\documentclass[aps,11pt,tightenlines,onecolumn,pra,citeautoscript,amsmath,amssymb,floatfix,nofootinbib,superscriptaddress,longbibliography]{revtex4-2}
\pdfoutput=1
\usepackage[utf8]{inputenc}

\usepackage{amsmath}
\usepackage{amssymb}
\usepackage{bbm}
\usepackage{amsthm, thm-restate}

\usepackage{graphicx}
\newcommand{\caphead}[1]{{\bf #1}}
\usepackage[table,usenames,dvipsnames]{xcolor}
\usepackage{tikz,pgfplots}\pgfplotsset{compat=1.18} 

\definecolor{verylightgray}{rgb}{0.95, 0.95, 0.95}
\definecolor{slightlyblue}{rgb}{0.95, 0.95, 1.0}

\newcommand{\matlabcolor}[0]{\color{verylightgray}}
\newcommand{\cppcolor}[0]{\color{slightlyblue}}

\usepackage{multirow}
\usepackage{dcolumn}
\newcolumntype{d}[1]{D{.}{.}{#1}} 

\usepackage{listings}
\lstnewenvironment{code:matlab}[1][]{
\lstset{language=MATLAB, backgroundcolor=\matlabcolor{}, rulecolor=\matlabcolor{}, 
          framextopmargin=0.4em, framexbottommargin=0.3em,#1}}{}

\lstnewenvironment{code:cpp}[1][]{
\lstset{language=C++, backgroundcolor=\cppcolor{}, rulecolor=\cppcolor{},
          framextopmargin=0.4em, framexbottommargin=0.3em,#1}}{}

\usepackage{natbib}
\usepackage[colorlinks]{hyperref}
\usepackage[capitalise]{cleveref}
\crefname{figure}{Figure}{figures}

\usepackage{enumerate}

\newcounter{lemmaN}

\newcounter{lemmaA}

\newcounter{colN}

\newcommand{\ket}[1]{|#1\rangle}
\newcommand{\bra}[1]{\langle #1|}

\DeclareMathOperator{\tr}{tr}
\newcommand{\id}{\mathbbm{1}}
\newcommand{\expt}[1]{\langle #1\rangle}

\newcommand{\inn}[2]{\langle #1, #2 \rangle}
\DeclareMathOperator{\st}{s.t.}

\newcommand{\psd}{\succeq}

\newcommand{\nats}{\mathbb{N}}
\newcommand{\ints}{\mathbb{Z}}
\newcommand{\nnints}{\mathbb{N}_0}
\newcommand{\reals}{\mathbb{R}}
\newcommand{\comp}{\mathbb{C}}
\newcommand{\hilb}[0]{\mathcal{H}}
\newcommand{\operators}[1]{\mathcal{L}\!\left(#1\right)}

\newcommand{\conj}[1]{{#1}^{*}}

\newcommand{\Csa}[1]{\mathbf{H}_{#1}(\mathbb{C})}

\newcommand{\rehead}[0]{\mathrm{Re}}
\newcommand{\imhead}[0]{\mathrm{Im}}
\newcommand{\re}[1]{\rehead\!\left(#1\right)}
\newcommand{\im}[1]{\imhead\!\left(#1\right)}

\newcommand{\inlineheading}[1]{\textbf{{#1}}}

\newcommand{\code}[1]{\texttt{\detokenize{#1}}}
\newcommand{\cpp}[0]{\mbox{C++}}
\newcommand{\matlab}[0]{\mbox{MATLAB}}
\newcommand{\cmake}[0]{\mbox{CMake}}

\newcommand{\moment}[0]{\emph{Moment}}
\newcommand{\version}[0]{\code{0.9.0}}
\newcommand{\cpplib}[0]{\code{lib_moment}}
\newcommand{\mex}[0]{\code{mex}}
\newcommand{\mtk}[0]{\code{mtk}}

\newcommand{\repo}[0]{\href{https://github.com/ajpgarner/moment/}{git respository}}
\newcommand{\eigen}[0]{\href{https://eigen.tuxfamily.org/}{\code{eigen}}}
\newcommand{\gtest}[0]{\href{https://github.com/google/googletest}{\code{googletest}}}
\newcommand{\cvx}[0]{\href{http://cvxr.com/cvx}{CVX}}
\newcommand{\yalmip}[0]{\href{https://yalmip.github.io/}{YALMIP}}

\makeatletter
\def\tocdepth@fullmunge{%
\let\l@section@saved\l@section
\let\l@section\@gobble@tw@
\let\l@subsection@saved\l@subsection
\let\l@subsection\@gobble@tw@
\let\l@subsubsection@saved\l@subsection
\let\l@subsubsection\@gobble@tw@
}%
\def\tocdepth@fullrestore{%
\let\l@section\l@section@saved
\let\l@subsection\l@subsection@saved
\let\l@subsubsection\l@subsubsection@saved
}%
\newcommand{\hidetoc}[0]{\addtocontents{toc}{\string\tocdepth@fullmunge}}
\newcommand{\restoretoc}[0]{\addtocontents{toc}{\string\tocdepth@fullrestore}}
\makeatother

\newcommand{\IQOQI}{Institute for Quantum Optics and Quantum Information,\\ Austrian Academy of Sciences, Boltzmanngasse 3, A-1090 Vienna, Austria}
\newcommand{\vd}{Departamento de Física Teórica, Atómica y Óptica, Universidad de Valladolid, 47011 Valladolid, Spain}

\begin{document}
\title{Introducing {\em Moment}:\\A toolkit for semi-definite programming with moment matrices}

\author{Andrew J.\ P.\ Garner}
\affiliation{\IQOQI{}}

\author{Mateus Araújo}
\affiliation{\vd{}}

\date{June 21, 2024;\quad Software version: \version{}.}

\begin{abstract}
Non-commutative polynomial optimization is a powerful technique with numerous applications in quantum nonlocality, quantum key distribution, causal inference, many-body physics, amongst others. 
The standard approach is to reduce such optimizations to a hierarchy of semi-definite programs, which can be solved numerically using well-understood interior-point methods.
A key, but computationally costly, step is the formulation of {\em moment matrices}, whose size (and hence cost) grows exponentially with the depth of the hierarchy.
It is therefore essential to have highly-optimized software to construct moment matrices.
Here, we introduce \moment{}: a toolkit that produces moment matrix relaxations from the specification of a non-commutative optimization problem. 
In order to obtain the absolute best performance, \moment{} is written in C++, and for convenience of use provides an interface via MATLAB.
We benchmark \moment{}'s performance, and see that it can be up to four orders of magnitude faster than current software with similar functionality.
\end{abstract}
\maketitle

\tableofcontents

\section{Introduction}
From analysts strategizing to maximize a business's profit, to scientists searching for the ground state energy of a complex solid-state system, numerical optimization is a critical use of computers.
Fundamentally, after modelling a system in terms of its essential parameters, one seeks the particular choice of parameters that extremize an {\em objective function} of these parameters, subject to constraints on which configurations of parameters are admissible.

{\em Convex optimization} problems~\cite{BoydV04} are those whose objective function is a convex function, and whose {\em feasible set} of parameters that satisfy the constraints is also convex~\cite{Rockafellar70}.
A particularly important type of convex optimization is {\em semi-definite programming}~\cite{VandenbergheB96}.
Here, the constraints take the form of imposing that a matrix, whose elements are affine functions of the optimization parameters, be positive semi-definite.
Semi-definite programs (SDPs) are computationally tractable with classical algorithms such as ellipsoid~\cite{GroetschelLS88, GaertnerM12} and interior methods~\cite{ForsgrenGW02}, and in practice enjoy wide software support (e.g.~\cite{Sturm99,mosek10,SCS}).

Moreover, semi-definite programming lends itself to many optimization problems that naturally arise in physics, since the set of quantum states can be represented with the convex cone of (complex) Hermitian positive semi-definite matrices~\cite{BengtssonZ06}.
Indeed, within quantum physics, semi-definite programming has been applied with substantial success to a myriad of topics, including 
 entanglement~\cite{Doherty2004, Doherty2005}
 nonlocality~\cite{Wehner06,NavascuesPA07,NavascuesPA08},
 quantum key distribution~\cite{BrownFF21di,Araujo22},
 many-body spin systems~\cite{Nakata2001,Kull2022,Wang2023},
 and causal inference~\cite{WolfeSF19,Ligthart2021}.
(For more examples, see the review of \citet{TavakoliPBA23}, or the book of \citet{SkrzypczykC23}).

A key technique in many of these scenarios, 
 building on the classical method of \citet{Lasserre01},
 is to relax convex {\em noncommuting polynomial optimization} (NPO) problems, whose objective functions and constraints are polynomial functions of quantum operators, into a hierarchy of SDPs~\cite{PironioNA10}.
In such SDPs, the positive semi-definite matrices involved are {\em moment matrices} and {\em localizing matrices}, whose elements are essentially expectation values of the involved quantum operators (and products thereof) when evaluated on some optimizing quantum state.
These SDPs can then be efficiently solved, allowing for the numerical solution of otherwise intractable problems.

\begin{figure}[tbh]
\begin{centering}
\includegraphics[width=0.625\textwidth]{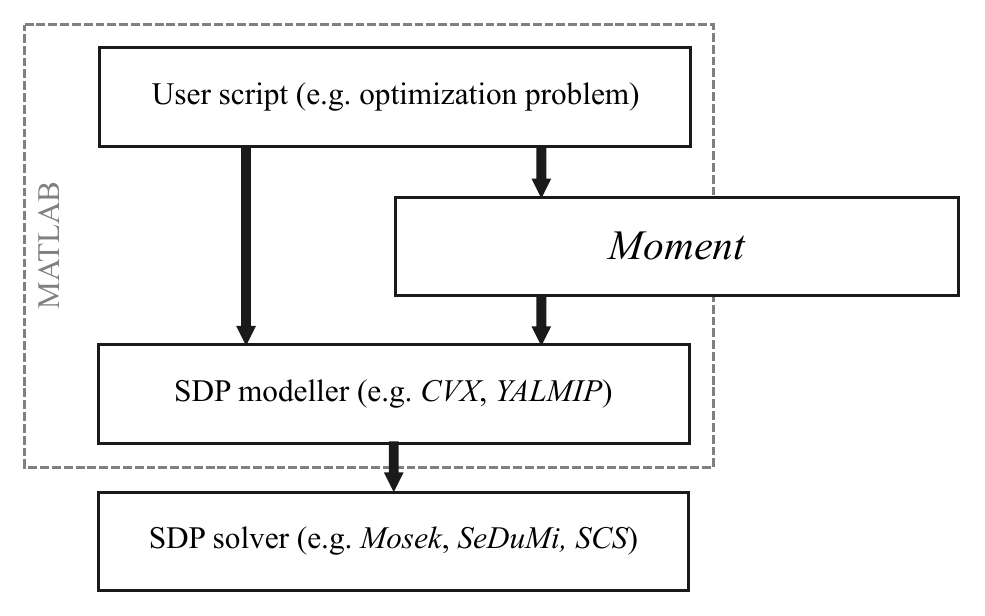}
\caption{%
\label{fig:overview}
\caphead{Schematic overview.}
\moment{} provides a suite of tools for use with MATLAB, to aid in the writing of semi-definite optimization problems involving moment matrices.
It has been designed to be used in conjunction with an SDP modeller (\cvx{}~\cite{cvx,GrantB08} or \yalmip{}~\cite{Lofberg04}), and a solver (e.g.\ Mosek~\cite{mosek10}, SeDuMi~\cite{Sturm99} or SCS~\cite{SCS}).
}
\end{centering}
\end{figure}

\moment{} is a suite of tools that enable the easy formulation and efficient generation of such SDPs.
Essentially, \moment{} generates the moment and localizing matrices of a problem from a description of the scenario's properties.
It also provides symbolic algebra support to handle formulating the associated objective functions and other constraints.
As sketched in \cref{fig:overview}, \moment{} is designed for use in a tool-chain with either \cvx{}~\cite{cvx,GrantB08} or \yalmip{}~\cite{Lofberg04} for modelling the SDPs,
 and an {\em SDP solver} (e.g.\ Mosek~\cite{mosek10}, SeDuMi~\cite{Sturm99} or SCS~\cite{SCS}) ultimately used to yield numerical results.
Although the interface of \moment{} is MATLAB, the majority of the computation is executed from a pre-compiled binary (whose source code is \cpp{}).

This document is organized as follows:
In \cref{sec:tech_intro}, we define the basic building blocks of semi-definite programs using moment matrices.
In \cref{sec:outline}, we outline the general function and components of \moment{}, introduced through a simple ``hello world'' optimization example from quantum nonlocality.
In \cref{sec:scenario}, we show in more depth how \moment{} can address problems in a variety of different scenarios: including quantum nonlocality (\cref{sec:localityscenario}), general noncommuting polynomial optimization (\cref{sec:algebraicscenario}), spin-$\frac{1}{2}$ lattices (\cref{sec:Pauli}), causal inference and inflation (\cref{sec:inflationscenario}); as well as two `utility' scenarios involving symmetrization (\cref{sec:symmetrizedscenario}), and the direct import of moment matrices generated using other software (\cref{sec:importedscenario}).
In \cref{sec:benchmark}, we measure and compare the performance of \moment{} to other software packages whose functionality overlaps with \moment{}.
Finally, in \cref{sec:discussion} we conclude with some remarks about the future of \moment{}.

\section{Technical introduction}
\label{sec:tech_intro}

In the following section, we will define the fundamental mathematical objects modelled by \moment{}.
The purpose of this document is to give an idea of the type of programs that can be composed using \moment{}.
As such, we will proceed in a ``bottom up'' manner.
Since \moment{} a toolkit that has been designed for broad application (i.e.\ may be used for tasks the authors have not hitherto considered): at this point we will put more emphasis on {\em what} tools are available, than {\em why} they are useful.
The latter, we hope, will be evident in the worked examples subsequently presented in \cref{sec:scenario}.

\subsection{Positive semi-definiteness}
Suppose we have a (possibly infinite-dimensional) Hilbert space $\hilb$,
 and let $\hat{M}\in\operators{\hilb}: \hilb \to \hilb$ be a self-adjoint operator acting on this space.
If $\bra{a}\hat{M}\ket{a} \geq 0$ for all $\ket{a}\in\hilb$, then $\hat{M}$ is said to be {\em positive semi-definite} (PSD). 
We write this as $M\psd 0$.
An operator is PSD if and only if its spectrum is entirely real and non-negative.
The set of PSD operators for a given Hilbert space is a convex cone: for any $A\psd0$ and $B\psd0$ and $\lambda\in[0,1]$ then $\lambda A + (1-\lambda) B \psd 0$, and for any $\mu\geq0$ and $C \psd 0$, $\mu C \psd 0$~\cite{AliprantisT07,Stormer13,FarautK94}.

\subsection{Semi-definite programs (SDPs)}
An SDP is an optimization of a linear function over the intersection of the convex cone of PSD matrices
 with a hyperplane. 
Writing $\Csa{m}$ as the set of $m\times m$ self-adjoint complex matrices (i.e.\ {\em Hermitian matrices}), and $\inn{\cdot}{\cdot}$ for the Hilbert-Schmidt inner product,
 then for some fixed matrix $C\in\Csa{m}$, a set of fixed matrices $A_i \in \Csa{m}$, and a fixed vector $\vec{b}$, every such optimization can be written~\cite{VandenbergheB96,TavakoliPBA23}:
\begin{align}
\label{eq:primalSDP}
\max_{X\in\Csa{m}} \quad & \inn{C}{X} \nonumber \\
\st \quad & \inn{A_i}{X} = b_i \quad \forall i, \nonumber \\
\mathrm{and}\quad &X \psd 0.
\end{align}
Here, $\inn{a}{b}$ denotes the Hilbert-Schmidt inner product $\inn{a}{b} := \tr\!\left(a^\dagger b\right)$.
By adding positive {\em slack variables}, inequality constraints can be formulated as equality constraints~\cite{BoydV04}.
A choice of $X$ that satisfies the constraints of the above program is known as {\em feasible}; and the SDP itself is said to be {\em feasible} if it admits of at least one feasible $X$.

The same set of $\{A_i\}$ $\{b_i\}$ and $C$ imply a {\em dual} program:
\begin{align}
\label{eq:dualSDP}
\min_{y \in\reals^{n}} \quad & \inn{b}{y} \nonumber \\
\st \quad & \sum_i A_i y_i - C \psd 0.
\end{align}
Problem~\eqref{eq:dualSDP} can be re-written in the form of Problem~\eqref{eq:primalSDP}, and so the dual program is also an SDP.
Every feasible point in the dual SDP lower bounds the optimum of the primal SDP, and every feasible point in the primal SDP upper bounds the optimum of the dual SDP -- a property known as {\em weak duality}.
For some special cases (e.g.\ {\em linear programs}, and {\em strictly feasible}\footnote{
A point $X$ is {\em strictly feasible} if it is in the interior of the PSD cone (i.e.\ $X$ is positive {\em definite}), and an SDP is strictly feasible if it contains at least one such point.}
SDPs~\cite{VandenbergheB96}) {\em strong duality} holds, where the optimum of the primal is exactly the optimum of the dual.

In practice, most interesting optimization problems will not arise naturally in the form of \cref{eq:primalSDP} or \cref{eq:dualSDP}.
It may sometimes be more natural to optimize over multiple smaller PSD cones, and some sets of constraints may be more naturally expressed as multiple half-spaces.
Indeed, a key function of an {\em SDP modeller} (such as \cvx{}~\cite{cvx,GrantB08} or \yalmip{}~\cite{Lofberg04}) 
 is to take an optimization problem with intuitively expressed objective functions and constraints,
 and translate them into a stricter, standardized input format, as required by the SDP solver.

The methods of solving SDPs are beyond the scope of this introduction (see instead, e.g.\ \cite{GroetschelLS88, GaertnerM12, ForsgrenGW02}).
It suffices for our purposes to treat a solver as black box software that takes an SDP as input,
 and returns a set of numerical values that define the solution, or otherwise signals that the SDP is infeasible.

\subsection{Expectation values and moments}
Consider a function $x\!\left(\lambda\right)$ over a random variable $\lambda$, distributed according to the probability distribution $P\!\left(\lambda\right)$ with support on the real axis $(-\infty, \infty)$.
The expectation value $\expt{x}$ of $x$ is:
\begin{align} 
\label{eq:classicalExpt}
\expt{x} := \int_{-\infty}^{\infty} P\!\left(\lambda\right) x\!\left(\lambda\right) \; d\lambda.
\end{align}
The classical moments $\mu_\alpha$ for $\alpha\in\nnints$ are then defined as the expectation values of nonnegative integer powers of $\lambda$, $\mu_\alpha := \expt{\lambda^\alpha}$.
These moments describe the shape of the probability distribution $P$. 
For instance, $\mu_0$ is the normalization, $\mu_1$ is the mean, $\mu_2$ relates to the variance, $\mu_3$ relates to the skew, etc.

Every probability distribution defines a sequence of moments, but not every sequence of real numbers can be interpreted as a sequence of moments defining a probability distribution.
Determining whether there exists a probability distribution to generate a given sequence is known as the {\em Hamburger moment problem}~\cite{ShohatT43}\footnote{Similarly, the {\em Stieltjes} and {\em Hausdorff} moment problems address this question for probability distributions over the half-closed interval $[0, \infty)$ and the closed interval $[0, 1]$ respectively~\cite{ShohatT43}.}.
To answer this one can examine the following Hankel matrix, $M_{ij} := \mu_{i+j}$, 
\begin{align}
\label{eq:classicalMM}
M := \begin{bmatrix}
\mu_0 & \mu_1 & \mu_2 & \cdots \\
\mu_1 & \mu_2 & \mu_3 & \cdots \\
\mu_2 & \mu_3 & \mu_4 & \cdots \\
\vdots & \vdots & \vdots & \ddots \\
\end{bmatrix}.
\end{align}
If and only if $M \psd 0$ will there exist a distribution $P$ with moments $\mu_i$ 

The concept of an expectation value can also be formulated within quantum theory (see, e.g.~\cite{NielsenC00}).
Rather than a general function, consider instead a linear operator $X \in \operators{\hilb}$ in some Hilbert space $\hilb$.
The analogue of a (normalized) probability distribution is played by a (trace-$1$) PSD {\em density operator} $\hat{\rho}$. 
This $\hat{\rho}$ defines a {\em state}\footnote{We will use the more precise meaning of the word `state' (i.e.\ a type of positive linear functional), as opposed to the common shorthand in quantum information theory, where the density operator is called a `quantum state'.
}%
$\rho: \operators{\hilb} \mapsto \comp$ that acts on any choice of operator $X\in\operators{\hilb}$ to give an expectation value:
\begin{align}
\label{eq:quantumExpt}
\rho\!\left(X\right) \equiv \expt{X}_{\rho} := \inn{\hat{\rho}}{X} := \tr\!\left(\hat{\rho} X \right).
\end{align}
If $X$ is Hermitian, then $\rho(X)\in\reals$, and if $X\psd 0$ then $\rho(X)\geq 0$.
We will omit the subscript $\rho$ from expectation values $\expt{\cdot}$ when it is obvious which state is being used.

For a Hermitian operator $X \in \operators{\hilb}$, consider the sequence of real-valued expectation values $\expt{\id}$, $\expt{X}$, $\expt{{X}^2}$, \ldots.
If  these values written as the matrix $M$ in \cref{eq:classicalMM} satisfy $M\psd0$,
 then they are also the moments of some classical probability distribution $P$.
Elementary quantum theory implies then that there is also a quantum density operator $\hat{\rho}_0$, diagonalizable in the same basis as $X$, that also yields the appropriate expectation values.
However, in general $\hat{\rho}_0$ is not the unique density matrix that yields these expectation values.
This is because there will be many other valid matrices $\hat{\rho}$ whose diagonal terms are the same as $\hat{\rho}_0$, but that differ in their off-diagonal terms.
These alternative states have all the same sequence generated by $X$; but there will be some other operator $Y$ (that does not commute with $X$), such that the sequence generated by $Y$ is different.

By analogy, we refer to the expectation value of some operator as a {\em moment} -- 
 and it is this that gives rise to our software's name.
The full, general, extension of the moment problem to quantum theory is not straightforward.
However, addressing whether a set of numbers can correspond to expectation values associated with a Hilbert space, a set of operators and a quantum state has lead to the development of an interesting subfield of quantum information (see \citet{TavakoliPBA23} and references within).

\subsection{Algebraic representation of operators}
In many optimizations problems 
 (particularly those whose solution methods descend from \citet{NavascuesPA07} and \citet{PironioNA10}),
 one seeks to optimize over a choice of (possibly infinite-dimensional) Hilbert space $\hilb$,
 a set of operators acting on $\hilb$, and a state of that space.

Early in the set-up of the problem (the point where \moment{} is designed to be used) $\hilb$ will be typically unknown.
As such, \moment{} does not store matrix representations of operators.
Instead, operators are handled in an abstract algebraic manner: 
 they are defined essentially by their multiplication rules, and behaviour under complex conjugation (i.e.\ properties that will be the same for {\em all} faithful representations).

Likewise, moments of operators are also handled abstractly:
 \moment{} determines whether a given operator (or product of operators) is Hermitian, anti-Hermitian, or neither, such that its  moment will respectively correspond to a real, imaginary, or complex scalar indeterminate when formulated into an SDP.

For a given setting, let the {\em alphabet} $\mathcal{X}:=\{x_1, \ldots, x_N\}$ be an ordered choice of $N\in\nats$ such operators (where $N$ is finite). 
It is taken that all $x_i$ act on the same, undetermined, Hilbert space.
The meaning of $\mathcal{X}$ could be physically motivated, e.g.\ corresponding to the set of measurements that might be made in a laboratory or by cryptographically-communicating parties.
Alternatively, $\mathcal{X}$ can also take on more abstract meanings (e.g. as terms in a non-commutative polynomial optimization problem).

In \moment{}, we restrict $\mathcal{X}$ to be explicitly closed under Hermitian conjugation such that $\conj{\mathcal{X}} := \{\conj{x_1}, \ldots, \conj{x_N}\}$ is (up to some permutation of elements) equal to $\mathcal{X}$.
This can be satisfied by taking all $x_i$ to be Hermitian operators -- but this is not a necessary condition, and \moment{} also supports the case where this does not hold.

\subsection{Indices, words, ordering, and dictionaries}
\label{sec:indices}
Since alphabets are finite ordered sets, we can associate each choice of element with an integer {\em index} $i\in {\ints}_N$ where $N=|\mathcal{X}|$.
In practice, these integer indices form \moment{}'s internal representation of operators.
One can then choose $L\in\nnints$ operators from $\mathcal{X}$ and label this choice by an $L$-dimensional integer {\em index vector} $\vec{i} \in {\ints_N}^L$.
The chosen operators can be composed to form a {\em word} $w_{\vec{i}} := x_{i_1}x_{i_2}\ldots x_{i_L}$ -- which itself would be an operator acting on the same Hilbert space as its constituents.
We define the word of length zero, associated with a zero-dimensional index $\vec{i}_0$, to be the identity operator $\id$.

Although every index vector identifies a unique operator, in general each operator is {\em not} uniquely identified by just one index vector.
As a simple example, consider the commuting set $\mathcal{X}_1 := \{x_1, x_2\}$ where $[x_1, x_2] = 0$.
The indices $(1, 2)$ and $(2, 1)$ describe $x_1 x_2$ and $x_2 x_1$ respectively, but $x_2 x_1 = x_1 x_2$, so these two distinct index vectors refer to the same operator.
Similarly, suppose we had the set $\mathcal{X}_2 := \{x_1\}$ with projective $x_1$ satisfying $x_1 x_1 = x_1$.
Then, {\em all} indices $(1)$, $(1, 1)$, $(1, 1, 1)$ etc.\ describe the same operator (with the exception of the empty string $\vec{i}_0$).

It is useful, then, to pick a canonical form of indices for each operator.
In \moment{}, we do this by assigning a total ordering to the index sequences known as {\em shortlex}:
 first, one orders index vectors by ascending dimension, and then by lexicographical order between vectors of a given length (i.e.\ first by lowest first index, tie-breaking by lowest second index, etc.).
The canonical choice of indices for a word is the {\em first} index vector in the shortlex sequence that matches the word.
For example, for a set with two (unrelated) operators, the first few index vectors (in order) are $i_0,\;(1),\;(2),\;(1,1),\;(1, 2),\;(2, 1),\;(2, 2),\;(1,1,1)$, \ldots.
In the above examples, for $\mathcal{X}_1$ we would prefer $(1, 2)$ over $(2, 1)$ to describe $x_1 x_2$; and for $\mathcal{X}_2$ we would prefer $(1)$ to describe $x_1$.

Finally, we define the {\em dictionary} $D(\mathcal{X}, L)$ of length $L$ as the shortlex--ordered set of words up to length $L$, where each included word is unique.
Clearly $D(\mathcal{X}, L_1) \subseteq D(\mathcal{X}, L_2)$ when $L_1 \leq L_2$,
 and $\lim_{L\to\infty} D(\mathcal{X}, L)$ is (an ordering of) the algebra generated by $\mathcal{X}$.
The conjugate dictionary $\conj{D}(\mathcal{X}, L)$ is defined as the set formed by taking every word in $D(\mathcal{X}, L)$ in order, and conjugating it\footnote{I.e.\ the elements of the conjugate dictionary are not directly ordered by shortlex, but rather by the shortlex of their conjugates.}.
As we have restricted ourselves to alphabets where $\mathcal{X} = \conj{\mathcal{X}}$,
 $D(\mathcal{X}, L)$ and $\conj{D}(\mathcal{X}, L)$ will consist of the same words in (generally) different order.
In the special case where $\mathcal{X}$ consists solely of commuting Hermitian operators, $D(\mathcal{X}, L) = \conj{D}(\mathcal{X}, L)$ exactly.

\subsection{Moment matrices}
\label{sec:introMM}
For an alphabet $\mathcal{X}$, the moment matrix level $L$ is the matrix $M_{ij}$ of expectation values associated with operators $\bar{w}_i w_j$ 
 where $\bar{w}_i$ is the $i^{\rm th}$ element of $\conj{D}(\mathcal{X}, L)$ and $w_j$ is the $j^{\rm th}$ element of $D(\mathcal{X}, L)$.

For example, for Hermitian operators $\mathcal{X} := \{x, y\}$, the moment matrix of level $0$ is:
\begin{align}
M(\mathcal{X}, 0) := \left[ \expt{\id}\right],
\end{align}
of level $1$ is:
\begin{align}
\label{eq:xymm}
M(\mathcal{X}, 1) := 
\begin{bmatrix}
\expt{\id} & \expt{x} & \expt{y} \\
\expt{x} & \expt{x^2} & \expt{xy} \\
\expt{y} & \expt{yx} & \expt{y^2}
\end{bmatrix},
\end{align}
and of level $2$ is:
\begin{align}
M(\mathcal{X}, 2) := 
\begin{bmatrix}
\expt{\id} & \expt{x} & \expt{y} & \expt{x^2} & \expt{xy} & \expt{yx} & \expt{y^2} \\
\expt{x} & \expt{x^2} & \expt{xy} & \expt{x^3} & \expt{x^2y} & \expt{xyx} & \expt{xy^2} \\
\expt{y} & \expt{yx} & \expt{y^2} & \expt{yx^2} & \expt{yxy} & \expt{y^2x} & \expt{y^3} \\
\expt{x^2} & \expt{x^3} & \expt{x^2y} & \expt{x^4} & \expt{x^3y} & \expt{x^2yx} & \expt{x^2y^2} \\
\expt{yx} & \expt{yx^2} & \expt{yxy} & \expt{yx^3} & \expt{yx^2y} & \expt{yxyx} & \expt{yxy^2} \\
\expt{xy} & \expt{xyx} & \expt{xy^2} & \expt{xyx^2} & \expt{xyxy} & \expt{xy^2x} & \expt{xy^3} \\
\expt{y^2} & \expt{y^2x} & \expt{y^3} & \expt{y^2x^2} & \expt{y^2xy} & \expt{y^3x} & \expt{y^4}
\end{bmatrix}.
\end{align}

By construction, moment matrices are Hermitian, and each contains every lower level moment matrix as a submatrix in its top left block.
The dimensions of a moment matrix are given by the number of elements in the associated dictionary, and this generally grows exponentially with $L$.
Because the first word in both $D$ and $\conj{D}$ is always $\id$, the top row of a moment matrix essentially lists the moments associated with elements of $D$ in order (likewise, the left-most column for $\conj{D}$).

It can be seen that for the trivial case of a single generating Hermitian operator $\mathcal{X}_0 := \{x\}$ with $x = \conj{x}$, 
 that the family of moment matrices will be Hankel matrices (i.e.\ corresponding to the $L\times L$ top-left block of \cref{eq:classicalMM}), motivating the usage of the expression {\em moment matrix}.

Recall that since \moment{} is used typically before the optimization is performed, such a matrix cannot be explicitly evaluated -- but instead must be handled symbolically.
In typical SDPs, a moment matrix is likely to play a role akin to ``$\sum_i A_i y_i \psd 0$'' in Problem~\eqref{eq:dualSDP}, where each $y_i$ is a moment (or the real or imaginary parts thereof), and each $A_i$ a Hermitian {\em basis element}.
A major task of \moment{} is to calculate $\{A_i\}_i$, and handle the indexing of $y_i$ in terms of the respective moments.

For example, the above $M(\mathcal{X}, 1)$ (\cref{eq:xymm}) is essentially expressed by \moment{} as:
\begin{align}
M(\mathcal{X}, 1) = \sum_{i=1}^5 a_i A_i + b_1 B_1,
\end{align}
where 
\begin{gather}
A_1 := \begin{bmatrix}
    1 & 0 & 0 \\ 0 & 0 & 0 \\ 0 & 0 & 0
\end{bmatrix},\;
A_2 := \begin{bmatrix}
    0 & 1 & 0 \\ 1 & 0 & 0 \\ 0 & 0 & 0
\end{bmatrix},\;
A_3 := \begin{bmatrix}
    0 & 0 & 1 \\ 0 & 0 & 0 \\ 1 & 0 & 0
\end{bmatrix},\;
A_4 := \begin{bmatrix}
    0 & 0 & 0 \\ 0 & 1 & 0 \\ 0 & 0 & 0
\end{bmatrix},\nonumber\\
A_5 := \begin{bmatrix}
    0 & 0 & 0 \\ 0 & 0 & 1 \\ 0 & 1 & 0
\end{bmatrix},\;
A_6 := \begin{bmatrix}
    0 & 0 & 0 \\ 0 & 0 & 0 \\ 0 & 0 & 1
\end{bmatrix},\;
B_1 := \begin{bmatrix}
    0 & 0 & 0 \\ 0 & 0 & i \\ 0 & -i & 0
\end{bmatrix}
\end{gather}
and
\begin{align}
a_1 := \expt{\id},\;
a_2 := \expt{x},\; 
a_3 := \expt{y},\;
a_4 := \expt{x^2},\;
a_5 := \re{\expt{xy}},\;
a_6 := \expt{y^2},\;
b_1 := \im{\expt{xy}}.
\end{align}

Here, $a_1, \ldots, a_5$, and $b_1$ are real scalar values that form the optimization variables -- finding a valid moment matrix amounts to seeking an appropriate choice of these parameters.
In this example enumerating the basis elements is pretty straightforward (the only subtle task being to determine that $\expt{x}$, $\expt{y}$, $\expt{xx}$ and $\expt{yy}$ are purely real, and that $\expt{yx} = \expt{xy}^*$).
However, for general settings where many non-trivial relations hold between operators, 
 the task can quickly become very involved 
 (see \cref{sec:benchmark} for an idea of the performance of \moment{} against other software on this task).

Finally, we remark on a notation choice here that persists throughout \moment{}: for any moment that could {\em a priori} be complex-valued (e.g.\ $\expt{xy}$), we have separated its real and imaginary parts and associated basis elements.
This serves several purposes.
First, there is a practical computational benefit: we can formulate a Hermitian matrix in the minimum set of real values, without having to store extraneous always-zero imaginary parts (e.g.\ $\im{\expt{x^2}}=0$).
This separation makes explicit what each (real-valued) degree of freedom in the choice of $\mathcal{M}$ is.
Another benefit arises for optimization problems that are symmetric under complex conjugation, such that any complex optimizer thereof implies the existence of a purely real optimizer with the same value of objective function (we shall discuss this explicitly later).
By splitting the imaginary components of the problem in such manner, it is possible to ``ignore'' the complex parts of a problem, where it will not affect the solution: allowing for the formulation of SDPs with fewer optimization parameters (and hence are quicker to solve).

\subsection{Localizing matrices}
\label{sec:introLM}
For an alphabet $\mathcal{X}$ and a word $v$ (called the {\em localizing word}) in the algebra generated by $\mathcal{X}$, 
 the level $L$ localizing matrix is the matrix formed by taking the expectation values of operators $\bar{w}_i v w_j$ where $\bar{w}_i$ is the $i^{\rm th}$ element of $\conj{D}(\mathcal{X}, L)$ and $w_j$ is the $j^{\rm th}$ element of $D(\mathcal{X}, L)$.
Clearly, the localizing matrices for word $v=\id$ are the moment matrices of the equivalent level.

For example, for non-commuting Hermitian operators $\mathcal{X} := \{x, y\}$, and a localizing word $x^2$ the localizing matrix of level $0$ is:
\begin{align}
L(\mathcal{X}, x^2, 0) := \left[ \expt{x^2}\right],
\end{align}
of level $1$ is:
\begin{align}
L(\mathcal{X}, x^2, 1) := 
\begin{bmatrix}
\expt{x^2} & \expt{x^3} & \expt{x^2y} \\
\expt{x^3} & \expt{x^4} & \expt{x^3y} \\
\expt{yx^2} & \expt{yx^3} & \expt{yx^2y}
\end{bmatrix},
\end{align}
and of level $2$ is:
\begin{align}
L(\mathcal{X}, x^2, 2) := 
\begin{bmatrix}
\expt{x^2} & \expt{x^3} & \expt{x^2y} & \expt{x^4} & \expt{x^3y} & \expt{x^2yx} & \expt{x^2y^2} \\
\expt{x^3} & \expt{x^4} & \expt{x^3y} & \expt{x^5} & \expt{x^4y} & \expt{x^3yx} & \expt{x^3y^2} \\
\expt{yx^2} & \expt{yx^3} & \expt{yx^2x} & \expt{yx^4} & \expt{yx^3y} & \expt{y^2x^3} & \expt{yx^2y^2} \\
\expt{x^4} & \expt{x^5} & \expt{x^4y} & \expt{x^6} & \expt{x^5y} & \expt{x^4yx} & \expt{x^4y^2} \\
\expt{yx^3} & \expt{yx^4} & \expt{yx^3y} & \expt{yx^4} & \expt{yx^4y} & \expt{yx^3yx} & \expt{yx^3y^2} \\
\expt{xyx^2} & \expt{xyx^3} & \expt{xyx^2y} & \expt{xyx^4} & \expt{xyx^3y} & \expt{xyx^2yx} & \expt{xyx^2y^2} \\
\expt{y^2x^2} & \expt{y^2x^3} & \expt{y^2x^2y} & \expt{y^2x^4} & \expt{y^2x^3y} & \expt{y^2x^2yx} & \expt{y^2x^2y^2}
\end{bmatrix}.
\end{align}

A localizing matrix is Hermitian if its localizing word is Hermitian.
Each localizing matrix includes all lower level localizing matrices of the same word in its top left block.

Localizing matrices can also be defined for scaled words and for polynomial expressions of operators in an obvious linear manner\footnote{This follows from the linearity of the state functional $\rho$.}.
For alphabet $\mathcal{X}$, level $n$, word $w$ and complex number $k$, $L(\mathcal{X}, kw, n) := kL(\mathcal{X}, w, n)$;
 likewise for alphabet $\mathcal{X}$, level $n$, and polynomial $p = \sum_i k_i w_i$ over words $\{w_i\}$ with complex coefficients $\{k_i\}$:
\begin{align}
L(\mathcal{X}, \sum_i k_i w_i, n) := \sum_i k_i\,L(\mathcal{X}, w_i, n)
\end{align}

As it does with moment matrices, \moment{} can generate basis elements for any localizing matrix, in terms of real parameters and complex basis matrices (themselves Hermitian, if $L$ is Hermitian).
Crucially, if the same moment appears as an element in multiple matrices in the specification of an optimization problem, it is mapped to a single optimization variable\footnote{Or, if the moment could be complex, it is mapped to the same pair of real optimization variables.}.

\clearpage
\section{Brief outline of using \moment{}}
\label{sec:outline}

\begin{figure}[tbh]
\begin{centering}
\includegraphics[width=0.95\textwidth]{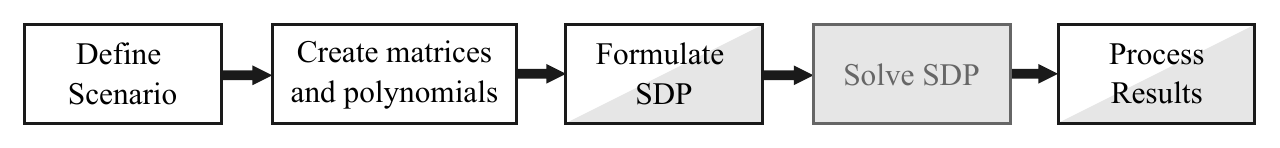}
\caption{%
\label{fig:Flow}
\caphead{Typical flowchart of a program using \moment{}.}
The third step requires the use of \cvx{} or \yalmip{} in addition to \moment{}, and the fourth step is handled entirely by external solver software.
}
\end{centering}
\end{figure}

\moment{} is a toolbox for \matlab{} to aid in the formulation of SDPs involving moment matrices.
A typical script that performs a computation aided by \moment{} follows the structure in \cref{fig:Flow}.
Full installation instructions are given in the \code{readme.md} of \moment{}'s \repo{}.

\subsection{Scenarios}
\label{sec:miniscenario}
All objects in \moment{} are conceptually part of a {\em scenario}.

A scenario is associated with exactly one {\em operator context} -- a description of the alphabet $\mathcal{X}$ in terms of a few abstract algebraic properties.
These include the number of fundamental operators, 
 along with a method for finding the canonical version of any index sequence (i.e.\ the {\em simplification} process),
 and a method for finding the conjugate of any index sequence.

A scenario also maintains a {\em symbol table} of ``known'' words -- that is, a list of words that have been seen in any generated dictionaries or matrices\footnote{This is necessary, because the number of words of a given length increases exponentially -- and some programs (especially those with localizing matrices) may want to consider only a subset of words of a particular length.}.
This table is used for the definition of the real SDP variables that will ultimately be passed to the SDP modeller. 
This list includes information such the conjugates of each sequence, as well as meta-information such as if the sequence is known to be Hermitian, anti-Hermitian, or neither (such that its associated moments will be respectively real, imaginary or complex).

In the \matlab{} toolbox of \moment{}, each scenario is an instance of a class inheritted from the class \code{MTKScenario}.
The various scenarios, descriptions of how their operator contexts are specified, and some worked examples are presented in depth in \cref{sec:scenario}.
For the purpose of this section, without elaboration let us define a scenario \code{scenario} associated with two Hermitian operators $x$ and $y$:
\begin{code:matlab}
scenario = AlgebraicScenario(2);
\end{code:matlab}
(Grey background fixed-width text denotes code to be executed in \matlab{}.)

\subsection{Objects: matrices and polynomials}
\label{sec:MomentObjects}
An {\em object} in \moment{} is a scalar, matrix, or tensor that represents one or more linear combinations of operators,
 that will ultimately be transformed into a scalar, matrix or tensor of moments when acted on by a state.
As previously mentioned, \moment{} determines where the same moment appears in multiple places (either within the same object, or in different objects within the same scenario), 
 and ensures that the same optimization variable(s) are used to represent it.

Objects can be created only after the scenario (and hence the relationship between operators) has been defined.
First, for scalar expressions, one can query the scenario to create the associated monomial object by index. For example, suppose we have a scenario with two operators $x$ and $y$. Then,
\begin{code:matlab}
xxy = scenario.get([1 1 2]);
\end{code:matlab}
creates the monomial \code{xxy} associated with operator sequence $xxy$ (and moment $\expt{xxy}$).

Often, it is useful to retrieve one monomial object for each element in the fundamental alphabet $\mathcal{X}$.
This can be done taking advantage of \matlab{}'s structured binding. For example,
\begin{code:matlab}
[x, y] = scenario.getAll();
\end{code:matlab}
will create $\code{x}$ and $\code{y}$ respectively associated with $x$ and $y$. 
(In general, one must provide the same number of variables as operators in the alphabet).
Alternatively it can be useful (e.g.\ if number of operators is parameterized) to return an array of such operators\footnote{
MATLAB can exhibit contextual behaviour depending on the number of output parameters of a function.
The \code{getAll} method will raise an error it is unsure how to interpret the number of outputs.
}:
\begin{code:matlab}
X = scenario.getAll();
\end{code:matlab}
where \code{X(1)} is $x$ and \code{X(2)} is $y$.

Scaled monomial objects and polynomials can be composed through arithmetic operations:
E.g,
\begin{code:matlab}
obj1 = 20 * x;
obj2 = x + 3*y;
obj3 = obj1 * obj2;
\end{code:matlab}
will respectively create $20x$, $x+3y$ and $20x^2+60xy$.

Almost every object in \moment{} represents an expression of operators that is later used to define moments, rather than directly representing the moments themselves.
As such \code{x*y} will result in a object that encodes the operator word $xy$, and subsequently yields the moment $\expt{xy}$, as opposed to yielding an object that represents the product of moments $\expt{x}\expt{y}$.

Scalar objects can be composed into vectors (and vectors into matrices, etc.) using \matlab{}'s concatenation syntax. 
For example,
\begin{code:matlab}
col_x_y = [x; y]
row_x_y = [x, y];
mat = [row_x_y; row_x_y]
\end{code:matlab}
creates a column vector, a row vector and a matrix respectively.

The primary function of \moment{} is the efficient generation of moment matrices.
To generate the moment matrix of level $n$, one executes:
\begin{code:matlab}
mm = scenario.MomentMatrix(n);
\end{code:matlab}

Meanwhile, to create a  localizing matrix, one first creates a monomial (or polynomial) object for the localizing word (or expression), and then executes the \code{LocalizingMatrix} method on that object.
For example, to make the level $2$ localizing matrix for $x$ (as \code{x}):
\begin{code:matlab}
lm_x = x.LocalizingMatrix(2);
\end{code:matlab}
or alternatively\footnote{
This second notation is useful due to limitations in how MATLAB handles subscripts (operator\ `\code{.}'), enabling a concise expression of complex queries such as \code{scenario.LocalizingMatrix(x+y, 1)}, while \code{(x+y).LocalizingMatrix(1)} would raise a syntax error.
}:
\begin{code:matlab}
lm_x = scenario.LocalizingMatrix(x, 2);
\end{code:matlab}

\subsection{Formulation of SDPs using objects from \moment{}}
\label{sec:formulateSDP}
Having created the necessary objects using \moment{}, one can proceed to assemble them into an SDP.
Here, one interfaces \moment{} with an SDP modeller (namely, one of \cvx{} or \yalmip{}).

\inlineheading{Declaring SDP variables.}
After the scenario is defined, and the objects of interest created, the next step is to create optimization variables (SDP variables) for the appropriate SDP modeller.
By default, \moment{} provides optimization variables for all moments in its symbol table.
Every moment $\expt{w}$ will be associated with up to two real variables $a_{\expt{w}}$ and $b_{\expt{w}}$ such that $\expt{w} = a_{\expt{w}}+i b_{\expt{w}}$.
If $w$ is Hermitian, then only $a_{\expt{w}}$ is generated such that $\expt{w}$ is purely real.
Likewise, if $w$ is anti-Hermitian ($\conj{w} = -w$), then only $b_{\expt{w}}$ will be generated such that $\expt{w}$ is purely imaginary.

As the architecture of \cvx{} and \yalmip{} are quite different, the syntax in \moment{} for generating these variables is necessarily different.
In both scripts below, for a moment scenario object \code{scenario}, vectors of optimization variables \code{a} and \code{b} are generated:
\begin{center}
\begin{tabular}{lcl}
Using \cvx{}: &\hspace{10px}& Using \yalmip{}:\\
\cellcolor{verylightgray} %
\begin{code:matlab}[linewidth=0.45\textwidth,backgroundcolor=,frame=]
cvx_begin sdp
    scenario.cvxVars('a', 'b');
    % ... 
cvx_end
\end{code:matlab}
&&
\cellcolor{verylightgray} %
\begin{code:matlab}[linewidth=0.45\textwidth,backgroundcolor=,frame=]
yalmip('clear'); 
[a, b] = scenario.yalmipVars();
\end{code:matlab} \\
\end{tabular}
\end{center}
In both modellers, this declaration should be towards the start of the SDP specification.
Moreover, once variables have been declared, no new \moment{} objects\footnote{
Technically, once the SDP variables are declared, there should be no new matrices, or invocations of \code{WordList} with the second parameter set to true.
Composition of monomials into polynomials is fine -- and \moment{} will raise an error if one attempts to use an object that contain moments that have not yet appeared in the symbol table (i.e.\ are not already in a moment matrix, localizing matrix or word list).
} should be generated until after the SDP is specified and solved -- as this has the potential to add new entries to the symbol table if a new moment is encountered, potentially invalidating the dimensions of \code{a},  \code{b} or both.

As previously mentioned, many SDPs of interest are symmetric under complex conjugation, implying that the optimal set includes solutions that are entirely real.
It is also possible to configure scenarios which can only produce Hermitian operators (e.g.\ the \code{PauliScenario} in \cref{sec:Pauli}).
Here, one can optimize only over real values and effectively ignore the imaginary part.
To define only the real components of moments, one would instead invoke \code{scenario.cvxVars('a');} or \code{a = scenario.yalmipVars();}.

Finally, \code{a(1)} always has the special meaning as the state's normalization, corresponding to the moment $\expt{\id}$.
In most cases, this value should be $1$ (and this should accordingly be imposed as a constraint in the SDP).
Beware that whenever a numeric constant appears in a polynomial, such as in $\expt{x} + 5$, moment will implicitly interpret this as $\expt{x} + 5\expt{\id}$, and so the evaluation of this expression will depend on \code{a(1)}.

\inlineheading{Creating modeller objects from moment objects.}
Once the optimization variables are declared, one can now create the polynomials and matrices that will be used by the modeller in the SDP (either as constraints or objective functions).
This is done by invoking the \code{Apply} method of the various \moment{} objects, with arguments \code{a} and \code{b} (or just \code{a}, for optimization restricted to real numbers).

For example, suppose \code{mm} is a moment matrix object, and \code{poly} is a polynomial objective, both generated by \moment{}.
To produce the modeller objects \code{MM} and \code{Poly}, we invoke:
\begin{code:matlab}
MM = mm.Apply(a, b);
Poly = poly.Apply(a, b);
\end{code:matlab} 

These objects can then be used as constraints, or objective functions, as per the instruction manuals of the respective modeller. (For objects depending only on real parts of moments, omit the argument \code{b}).

\inlineheading{A complete SDP.}
Tying the above together, to minimize some polynomial \code{poly} subject to a PSD moment matrix \code{mm}, both belonging to the scenario \code{scenario}, one would write:
\begin{center}
\begin{tabular}{lcl}
Using \cvx{}: &\hspace{10px}& Using \yalmip{}:\\
\cellcolor{verylightgray} %
\begin{code:matlab}[linewidth=0.45\textwidth,backgroundcolor=,frame=]
cvx_begin sdp
  scenario.cvxVars('a', 'b');
  MM = mm.Apply(a, b);
  Poly = poly.Apply(a, b);
  a(1) == 1;
  MM >= 0;
  minimize(Poly);
cvx_end
\end{code:matlab}
&&
\cellcolor{verylightgray} %
\begin{code:matlab}[linewidth=0.45\textwidth,backgroundcolor=,frame=]
yalmip('clear');
[a, b] = scenario.yalmipVars();
MM = mm.Apply(a, b);
Poly = poly.Apply(a, b);
constraints = [a(1) == 0; MM >= 0]
optimize(constraints, Poly);
\end{code:matlab} 
\end{tabular}
\end{center}

\inlineheading{A shortcut using \code{mtk_solve}.}
Although most non-trivial applications of \moment{} will likely require the fine-tuned configuration of the SDP,
 we provide the utility function \code{mtk_solve} for writing and solving simple SDPs.
In particular \code{mtk_solve} takes as its first argument either a single \moment{} matrix object (or a cell array of multiple such matrices)
 and as its second (optional) parameter a \moment{} polynomial object to use as an objective function.
When the second argument is supplied, the function formulates and attempts to solve an SDP that minimizes this objective function subject to all matrices in the first argument being PSD (and the normalization being $1$).
If the objective argument is omited, the function instead formulates a feasibility test of these same SDP constraints, returning true if a feasible solution could be found.

This allows for the formulation of a short ``hello world'' script for \moment{}, given as \code{chsh_terse.m}:
\begin{code:matlab}
scenario = LocalityScenario(2, 2, 2);
chsh = scenario.FCTensor([0 0 0; 0 1 1; 0 1 -1]);
result = mtk_solve(scenario.MomentMatrix(1), chsh)
\end{code:matlab}
which will, after some calculation, output $-2.8284$ (approximately $-2\sqrt{2}$).
The meaning of the first two lines (and the physical significance of what is being calculated) will be explained in \cref{sec:localityscenario}.
Essentially, this program minimizes a polynomial \code{chsh}, subject to the scenario's level $1$ moment matrix being PSD.

Although \code{mtk_solve} hides the \cvx{}/\yalmip{} stage from the user, at least one of the these modellers must be installed for the function to compute an answer.

\subsection{Imposing scalar constraints on moments}
\inlineheading{Direct implementation.}
\label{sec:eqconst}
As well as PSD constraints on matrices, SDPs often impose scalar constraints directly on the constituent moments.
The simplest way to impose such a constraint is during the modelling stage of the SDP, with syntax similar to the specification of objective functions.

For example, suppose we are formulating an SDP involving unknown operators $x$ and $y$, and want to impose $\expt{x}+\expt{y} \geq 3$.
We would first create a polynomial object: \code{cp = x + y - 3}.
Then, this polynomial can be translated to a \cvx{} or \yalmip{} object via the \code{Apply} method,
 and then interpreted as a constraint using the usual syntax of the respective modeller:
\begin{center}
\begin{tabular}{lcl}
Using \cvx{} (within \code{cvx_begin} block): &\hspace{10px}& Using \yalmip{}:\\
\cellcolor{verylightgray} %
\begin{code:matlab}[linewidth=0.45\textwidth,backgroundcolor=,frame=]
  cp.Apply(a, b) >= 0;
\end{code:matlab}
&&
\cellcolor{verylightgray} %
\begin{code:matlab}[linewidth=0.45\textwidth,backgroundcolor=,frame=]
constraints = [cp.Apply(a,b) >= 0]
\end{code:matlab} 
\end{tabular}
\end{center}
(Where \code{a} and \code{b} are formed as in \cref{sec:formulateSDP}.)

\inlineheading{Moment substitution rules.}
\moment{} provides a method by which an SDP's equality constraints can be used to eliminate some of the involved optimization variables.
As well as the obvious computational benefit of formulating a smaller SDP,
 this also can be used to solve a class of numerical issues arising from linearly dependent equality constraints.
Essentially, one rewrites each monomial and polynomial expression of moments in such a way as to assume the equality constraint holds.
For instance, suppose one has a moment matrix $M$ and a constraint $\expt{x} = \expt{y}$;
 then the simplification would be to replace $\expt{y}$ by $\expt{x}$ everywhere it appears in $M$,
  essentially reducing $M$'s number of basis elements.
Such substitutions are made at the level of moments, not at the level of the underlying operators 
 (the latter will be discussed in \cref{sec:algebraicscenario}).

\moment{} provides the \code{MTKMomentRulebook} object for collating and applying such substitutions (see also: \code{examples/moment_substitutions.m}).
Suppose we have a generic scenario with three unknown operators, $x$, $y$ and $z$:
\begin{code:matlab}
scenario = AlgebraicScenario(3);
[x,y,z] = scenario.getAll();
\end{code:matlab}
Then, a new rulebook (with name ``Example rulebook name'') can be initialized via:
\begin{code:matlab}
rulebook = scenario.MomentRulebook("Example rulebook name");
\end{code:matlab}

To enforce equality constraints, one lists polynomials that are implicitly equated with $0$.
For example  $\expt{xy} = i\expt{z}$ and $\expt{x} = \expt{y}$ are set by:
\begin{code:matlab}
rulebook.Add(x*y - 1i * z);
rulebook.Add(x - y);
\end{code:matlab}
or alternatively as a vector of polynomials (see \cref{sec:MomentObjects}):
\begin{code:matlab}
rulebook.Add([x*y - 1i * z; x - y]);
\end{code:matlab}
If there are many equality constraints to be added (e.g.\ because they have been algorithmically generated), the vector method will be significantly faster.

Once all rules have been specified, they can be applied to any \moment{} object.
For example, to apply the rules to moment matrix \code{mm}, one can either use the \code{Apply} method of the rulebook object with the target object as its argument: 
\begin{code:matlab}
sub_mm = rulebook.Apply(mm);
\end{code:matlab}
or equivalently one invokes the \code{ApplyRules} method of the target object with the rulebook as its argument:
\begin{code:matlab}
sub_mm = mm.ApplyRules(rulebook);
\end{code:matlab}
In the above example, this produces a new moment matrix \code{sub_mm} where all terms in $\expt{xy}$ have been replaced by $i\expt{z}$, all $\expt{yx}$ by $-i\expt{z}$ and all $\expt{y}$ by $\expt{x}$.

If a rulebook is applied to one object involved in an SDP, 
 chances are it should be applied to {\em all} involved objects, to avoid inconsistencies (e.g.\ $\expt{y}$ appearing in some objects but not others).

Finally, we remark that \moment{} internally stores the rules within a rulebook in a normalized {\em reduced} form, which ensures that if a rule matches a term and a substitution is made, no further substitutions by other rules in the set will be required on the newly-substituted part.
Conceptually, if one imagines the set of enforced constraints as a matrix acting to transform a vector whose basis is the real SDP variables, this is essentially a triangularization.
This reduction allows \moment{} to avoid a few potential problems that can arise when constraints involve shared terms (e.g.\ $\expt{z}=\expt{x}$ and $\expt{z}=\expt{y}$, which must also imply $\expt{y}=\expt{x}$),
 as well as consistently handle a few complicated edge cases where particular rules can affect whether a moment is complex- or real-valued.
This process also detects logically inconsistent set of polynomials constraints (namely, those that can be used to infer that $1 = 0$) before the solving process begins.
Finally, such a reduction typically also yields improved performance, because there are typically many elements on which the rules are applied (i.e.\ scaling with the number of elements in a moment matrix), amortizing the one-off cost (scaling with the number of rules) of computing this reduction at the point when the rules are defined.
Details of this reduction algorithm are in \cref{app:PolySubRule}.

\subsection{Retrieving and using numeric values after solving an SDP}
After a feasible problem has been solved,
 it is sometimes useful to extract properties of the solution beyond the objective function's value.
For example: one might have bounded the maximal violation of a Bell inequality, but wants also to know the individual probabilities that achieve this violation.
 
In \cvx{}, typically the variables \code{a} and \code{b} will have been automatically replaced with numbers reflecting the values they take in the solution.
In \yalmip{}, this replacement can be done manually with \code{a = value(a)} (and respectively for \code{b}).
Essentially, this produces a list of numbers corresponding to the real (and imaginary) components of the moments in the solution.

Once these numbers are available, it is possible to use the \code{Apply} method in any \moment{} object with the arguments \code{a} (and \code{b}) to evaluate the respective object numerically.
For example, suppose we had a polynomial \code{p}, then one could probe its value in the solution with \code{p_val = p.Apply(a,b)}.

\clearpage
\section{Applications of \moment{}}
\label{sec:scenario}
In this section, we shall look in depth at the types of scenario that can be implemented in \moment{}, through several worked examples.
Essentially, each scenario is a configurable framework describing the algebra underlying the moment matrices.
Some scenarios also provide additional features, that we will discuss below.

\subsection{Locality scenario}
\label{sec:localityscenario}
\subsubsection{Introduction.}
Semi-definite programming was arguably brought to the wider attention of the quantum information community to address questions about {\em quantum correlations}: statistics of measurements made by spatially-disjoint agents that cannot be accounted for by classical probability theory alone.
Typically, certification of non-classical behaviour is given by the violation of a Bell inequality~\cite{Bell64,BrunnerCPSW14}: a linear sum of expectation values of (joint) measurements made by the various agents in the scenario.

Among the most well-known of such inequalities is the CHSH~\cite{ClauserHSH69} inequality, applicable to two parties (Alice and Bob) who can each make a choice of two binary measurements ($A_0$, $A_1$ for Alice, $B_0$, $B_1$ for Bob).
If the outcomes of these measurements are given the numeric values $+1$ and $-1$, then one such inequality is:
\begin{align}
\label{eq:CHSH}
\expt{A_0 B_0} + \expt{A_0 B_1} + \expt{A_1 B_0} - \expt{A_1 B_1} \leq 2.
\end{align}
Measurement of statistics that break this bound imply that experiment cannot be accounted for by a local classical hidden variable model: a property known as {\em nonlocality}.
Famously, there are quantum systems for which the left-hand expression in \cref{eq:CHSH} takes the value of $2\sqrt{2}$: 
 for example, if the two parties share a maximally entangled Bell state of two qubits and make measurements $A_0 = \sigma_x$, $A_1 = \sigma_z$, $B_0 = \frac{1}{\sqrt{2}}\left( \sigma_x + \sigma_z \right)$ and $B_1 = \frac{1}{\sqrt{2}}\left(\sigma_x - \sigma_z\right)$, where $\sigma_x$ and $\sigma_z$ are Pauli matrices.
Tsirelson~\cite{Tsirelson87}, by way of operator algebra analysis, showed that $2\sqrt{2}$ is the maximum value that {\em any} quantum system (i.e.\ of any dimension, not just pairs of qubits) and choice of measurements can obtain in this scenario.

\Citet{Wehner06} showed that Tsirelson's bound can be obtained as the solution to a semi-definite program.
Shortly after, the more general {\em NPA hierarchy} was formulated~\cite{NavascuesPA07,NavascuesPA08}.
Here, the constraint that statistics are a quantum behaviour -- that is, are explicable by some quantum state and set of quantum measurements -- is relaxed onto a PSD constraint on the moment matrices formed (as per \cref{sec:introMM}) from the alphabet of operators associated with the various outcomes to the various measurements made by the disjoint agents.
By increasing the moment matrix level, one generates a hierarchy of increasingly strict necessary conditions for the quantum behaviour.
In the limit $\lim_{L\to\infty} M(\mathcal{X}, L) \psd 0$, one recovers the set of correlations produced by strategies involving commuting Hermitian operators.
Whether this is equivalent to the quantum set (in infinite dimensions), was a long-term open question known as Tsirelson's problem (e.g.\ \cite{JungeMPPSW11}), and \citet{JiNVWY20} claim the answer in the negative.

One can then posit a myriad of quantum optimisation problems~\cite{TavakoliPBA23} as SDPs,
 and the purpose of \moment{}'s locality scenario is to facilitate this.

\subsubsection{Implementation in \moment{}.}
We take the general approach that each operator corresponds to an {\em effect}: a type of operator associated with the outcome of a particular measurement, such that its expectation value gives the probability of that outcome occurring should the measurement be made.
In quantum theory, effects are generally POVMs (see, e.g.\ \cite{NielsenC00}),
 but here we can further take them to be {\em projectors} ($P^2 = P$).
This assumption is not a relaxation/constraint in settings where we have freedom of choice of the dimension of the underlying quantum system (as is the case if we want to optimise over all quantum systems), because of purification allows us to see POVM elements as projectors acting on a higher-dimensional system.

The locality scenario is specified by telling \moment{}:
\begin{enumerate}
\item How many disjoint parties are there (Alice, Bob, Charlie, etc.)?
\item How many measurements each agent can make?
\item How many outcomes each measurement of each agent has?
\end{enumerate}
The different agents can have different numbers of measurements, and each measurement can have a different number of outcomes.

From this information, Moment defines an alphabet $\mathcal{X}_{\rm locality}$ of operators, that satisfy the following algebraic rules:
\begin{enumerate}
\item All operators are Hermitian.
\item All operators are idempotent such that $x_i x_i = x_i$.
\item Every operator can be associated with exactly one party, and operators corresponding to different parties commute.
\item Operators within the same measurement are mutually exclusive, such that $x_i x_j = 0$ when $i\neq j$ and $x_i$ and $x_j$ are associated with different outcomes of the measurement.
\end{enumerate}

This allows an arbitrary operator sequence to be brought into canonical form by first sorting the operators by party (in a stable manner, so as to preserve the ordering of operators from the same party), and then within each party checking if a simplification can be made by applying the idempotent rule, or if two mutually exclusive operators are next to each other, implying that the entire string should be equal to zero.

To avoid adding non-linearly independent columns to moment matrices, no operators are generated for the final outcome of any measurement.
This is consistent with the scheme of \citet{CollinsG04}. 
The implicit outcomes are are given as $\id - \sum_{i\in M} x_i$ where $M$ are the set of indices labelling all other outcomes of that measurement.
Tools provided within \moment{} allow for the extraction and evaluation of these implicit moments, not just for single--party measurements but also where they appear as part of a joint measurement outcome.

\subsubsection{Example: (2,2,2)-Bell scenario.}
\label{sec:CHSH}
To return to the Ur--example, 
let us consider the CHSH scenario\footnote{All discussed examples may be found in the \repo{} in the \code{/matlab/examples} folder. This example in particular is \code{cvx_chsh.m} and \code{yalmip_chsh.m} for respective use with \cvx{} and \yalmip{}.}.
Recall, Alice and Bob, are each able to choose between two binary measurements.
From this, we define the following operators $\mathcal{X}_{\rm CHSH} = \{a_0, a_1, b_0, b_1\}$.
Here, $a_x$ is associated with the positive outcome to Alice's measurement of $A_x$ ($x = 0\mathrm{~or~}1$); a similarly $b_y$ are associated with Bob's measurement outcomes.

We want to computer Tsirelson's bound as an SDP:
\begin{align}
\mathrm{maximize}\quad &  p_{\rm CHSH}, \nonumber\\
\mathrm{s.t.}\quad& \expt{\id} = 1\quad~\mathrm{and}~\quad{}M\!\left(\mathcal{X}_{\rm CHSH}, L\right) \psd 0,
\end{align}
where $p_{\rm CHSH}$ is a polynomial expression of moments equivalent to the left-hand side expression of \cref{eq:CHSH}.
Since we seek the statistics associated with a normalized quantum state, we also impose the constraint $\expt{\id}=1$, as otherwise the problem will be unbounded.

To tackle this using \moment{}, we must first define the scenario.
As all parties have the same number of measurements and all measurements the same number of outcomes, we can use the following short syntax to define a \code{scenario} object\footnote{For alternative ways to define a locality scenario, one could also look at the examples \code{cvx_I3322.m} and \code{four_party.m}.}:
\begin{code:matlab}
scenario = LocalityScenario(2, 2, 2);
\end{code:matlab}

To generate the moment matrix level 1, we run:
\begin{code:matlab}
matrix = scenario.MomentMatrix(1);
\end{code:matlab}
which defines an object corresponding to the following moment matrix:
\begin{align}
\label{eq:chshMM}
M\!\left(\mathcal{X}_{\rm CHSH}, 1\right) := \begin{bmatrix}
\expt{\id} & \expt{a_0} & \expt{a_1} & \expt{b_0} & \expt{b_1} \\
\expt{a_0} & \expt{a_0} & \expt{a_0 a_1} & \expt{a_0 b_0} & \expt{a_0 b_1} \\
\expt{a_1} & \expt{a_1 a_0} & \expt{a_1} & \expt{a_1 b_0} & \expt{a_1 b_1} \\
\expt{b_0} & \expt{a_0 b_0} & \expt{a_1 b_0 } & \expt{b_0} & \expt{b_0 b_1} \\
\expt{b_1} & \expt{a_0 b_1} & \expt{a_1 b_1} & \expt{b_1 b_0} & \expt{b_1}
\end{bmatrix}.
\end{align}

The CHSH inequality (\cref{eq:CHSH}) has slightly more terms when expressed in Collins-Gisin notation,
 but \moment{} supplies various tools for composing the appropriate polynomial.
For example, 
\begin{code:matlab}
chsh_eq = scenario.FCTensor([0 0 0; 0 1 1; 0 1 -1]);
chsh_eq2 = scenario.CGTensor([[2 -4 0]; [-4 4 4]; [0 4 -4]]);
\end{code:matlab}
would both define the (same) equality via element-wise contraction with the {\em full--correlator} and {\em Collins-Gisin} tensors respectively.
One could alternatively achieve the same by:
\begin{code:matlab}
[A0, A1, B0, B1] = scenario.getMeasurements();
Corr00 = Correlator(A0, B0);
Corr01 = Correlator(A0, B1);
Corr10 = Correlator(A1, B0);
Corr11 = Correlator(A1, B1);
chsh_eq3 = Corr00 + Corr01 + Corr10 - Corr11;
\end{code:matlab}
with the advantage here that the individual correlators are now also exposed for subsequent use.

In any case, the polynomial that is created will have the explicit form:
\begin{equation}
\label{eq:CHSH:CG}
p_{\rm chsh} := 2 - 4 a_0 - 4 b_0 + 4 a_0 b_0 + 4 a_0 b_1 + 4 a_1 b_0 - 4 a_1 b_1.
\end{equation}

At this point, we have the objects necessary to formulate our SDP using our modeller of choice, and dispatch it to be solved. As per \cref{sec:formulateSDP}:
\vspace{0.25em}
\begin{center}
\begin{tabular}{lcl}
Using \cvx{} (see \code{cvx_chsh.m}): &\hspace{10px}& Using \yalmip{} (see \code{yalmip_chsh.m}):\\
\cellcolor{verylightgray} %
\begin{code:matlab}[linewidth=0.45\textwidth,backgroundcolor=,frame=]
cvx_begin sdp
    scenario.cvxVars('a');
    M = matrix.Apply(a);

    a(1) == 1;  
    M >= 0;

    objective = chsh_eq.Apply(a);
    maximize(objective);
cvx_end
\end{code:matlab}
&&\cellcolor{verylightgray} %
\begin{code:matlab}[linewidth=0.45\textwidth,backgroundcolor=,frame=]
yalmip('clear');
a = scenario.yalmipVars();
M = matrix.Apply(a);

constraints = [a(1) == 1, M >= 0];

objective = -chsh_eq.Apply(a);
optimize(constraints, objective); 
\end{code:matlab}
\end{tabular}
\end{center}

In both code snippets, the second line defines the vector of real optimization variables \code{a} that we solve over.
The next line defines the matrix \code{M}, as a cvx (resp.\ yalmip) object in terms of the variables $\code{a}$.
The middle section defines the constraints as per the modeller's preferred syntax.
Recall that the first element \code{a(1)} always refers to the normalization moment $\expt{\id}$.
Next, we define the objective function as a cvx (resp.\ yalmip) object (in terms of elements of $\code{a}$) that implements the polynomial $\vec{p}_{\rm chsh}$ as defined above.
The final lines instruct the modeller to parse the problem, and dispatch it to a solver.

In this particular example, the objective function only involves real-valued moments, and the contraints are symmetric under complex conjugation.
As such, in the interest of computational efficiency, we skipped the definition of the complex-basis SDP variables, and formulated the problem only in terms of the real elements \code{a}.

Once the SDP has been solved, one can query the modeller for the numerical values of \code{a}.
With cvx, \code{a} will be automatically replaced by the values from the solution; with yalmip, one needs to first write \code{a = value(a);}.
To get explicit numerical values associated with any \moment{} object, one uses the \code{Apply} method:
\begin{code:matlab}
chsh_eq.Apply(a)
matrix.Apply(a)
\end{code:matlab}
(The omission of \code{;} invites \matlab{} to output the value to console).

\subsubsection{Example: I3322 inequality.}
\label{sec:i3322}
Similarly to the CHSH scenario above, we can consider the case where Alice and Bob have {\em three} binary measurements to choose from.
The scenario, level 4 moment matrix and the ``I3322'' Bell functional~\cite{Froissart81,CollinsG04} can be generated via:
\begin{code:matlab}
i3322 = LocalityScenario(2, 3, 2);
moment_matrix = i3322.MomentMatrix(4);
i3322_eq = i3322.FCTensor([[0  -1 -1  0]
                           [-1 -1 -1 -1]
                           [-1 -1 -1  1]
                           [0  -1  1  0]]);
\end{code:matlab}
The above can be modelled and solved using an almost identical prescription to the previous example.
For a full implementation, see the \code{cvx_i3322.m} example script.

\subsubsection{Example: CGLMP inequality.}
\label{sec:cglmp}
The locality scenario can also handle measurements with more than two outputs. 
Consider the CGLMP scenario, where Alice and Bob have two measurements each with $k$ outputs~\cite{Collins2002,Acin2005}. For $k=3$ the scenario, level 2 moment matrix, and Bell functional can be generated via:
\begin{code:matlab}
CGLMP = LocalityScenario(2, 2, 3);
moment_matrix = CGLMP.MomentMatrix(2);
CGLMP_eq = CGLMP.CGTensor([[0  -1 -1  0  0]
                           [-1  1  1  0  1]
                           [-1  1  0  1  1]
                           [0   0  1  0 -1]
                           [0   1  1 -1 -1]]);
\end{code:matlab}
Here, we form the polynomial using the \code{CGTensor} method to contract with a Collins-Gisin tensor. 
Since the measurements are no longer binary, the full-correlator tensor is no longer well-defined.
For a full implementation, see the \code{yalmip_cglmp.m} example script.

\subsection{Algebraic scenario}
\label{sec:algebraicscenario}
\subsubsection{Introduction and rewrite rules.}
The algebraic scenario is a multipurpose scenario provided by \moment{}, geared for general non-commutative optimization problems.
In this scenario, one supplies the number of fundamental operators, whether they are Hermitian or not, 
 and then supplies a set of rules to impose equality constraints between strings of operators.
The generality of this scenario means it can also be used to implement the behaviour of the specialized locality (\cref{sec:localityscenario}) or Pauli spin (\cref{sec:Pauli}) scenarios, albeit with less efficiency.

To use an equality constraint to simplify a word in an automated manner,
 it must be {\em oriented} into as a {\em rewrite} (or {\em substitution}) rule.
In general, if one has the (monomial) equality $l_i = r_i$, the orientation will be $l_i \mapsto r_i$,
 where $l_i > r_i$ in shortlex order.
Then, to {\em reduce} (or {\em simplify}) word $w$, one searches for the substring $l_i$, and each time it appears, replaces that substring with $r_i$.

As an example, consider the equality constraint $aba=a$ on an algebra with two generating elements $a$ and $b$.
From this, we can infer infinitely many other identities, such as $aaba = aa$, $ababa = aba = a$, $baba = ba$. 
The associated rewrite rule is: $r: aba \mapsto a$;
 and by (repeated) find-and-replace this enforces all of the above identities to simplify words; for example $\underline{aba}ba \to \underline{aba} \to a$.

When there are multiple rules in a set, one must also consider {\em convergence} -- whether repeated systematic application of rules from the set will always reduce words to a standard form in a finite number of steps.
\moment{} provides an implementation of the Knuth-Bendix algorithm~\cite{KnuthB83} to test if a ruleset is convergent (or otherwise attempt to modify the ruleset to impose the same equality relations in a convergent manner).
Details of this algorithm (and a motivating example illustrating problems that can arise without convergence) are provided in \cref{app:MonoSubRule}.

\subsubsection{Example: Non-commuting polynomial optimization.}
Polynomial optimization concerns itself with finding the maximum of a (generally multivariate) polynomial expression, subject to a set of polynomial constraints.
The commuting case (i.e.\ for polynomials over real or complex numbers) was formulated as hierarchy of SDP relaxations by \citet{Lasserre01}.
Here, we consider the {\em non-commuting} extension presented by \citet{PironioNA10}, for the case of polynomials whose indeterminates are non-commuting operators.

To use their example, consider the consider the non-commuting polynomial optimization (NPO) problem:
\begin{align}
\label{eq:PNAExact}
\min_{\hilb,\, X_1, X_2 \in \operators{\hilb},\, \sigma}\quad 
& \sigma\left(X_1 X_2 + X_2 X_1\right), \nonumber\\
\mathrm{s.t.}\quad& X_1^2- X_1 = 0, \nonumber\\
\mathrm{and}\quad& -X_2^2 + X_2 + \frac{1}{2} \psd 0.
\end{align}

Using the prescription of \citet{PironioNA10}, we can relax this to a family of SDPs parameterized by moment-matrix level $m\in\nats$ and localizing-matrix level $l\leq m$.
\begin{align}
\min \quad & \expt{x_1 x_2} + \expt{x_2 x_1}, \nonumber \\
\st \quad & \mathcal{M}\!\left(\mathcal{X}, m\right) \psd 0, \nonumber \\
&\mathcal{L}\!\left(\mathcal{X}, -x_2^2 + x_2 + \frac{1}{2},  l\right) \psd 0, \nonumber \\
\mathrm{and}\quad& \expt{\id} = 1, \nonumber \\
\end{align}
where operators $\mathcal{X} := \{x_1, x_2\}$ are Hermitian, but subject to the rewrite rule $x_1 x_1 \mapsto x_1$.
This rewrite rule perfectly imposes the first constraint of \cref{eq:PNAExact};
 but the second constraint has been relaxed into a {\em localizing matrix} constraint.
To formulate this scenario in \moment{}, we will use the \code{AlgebraicScenario} object, to define two Hermitian operators and a single rule:
\begin{code:matlab}
scenario = AlgebraicScenario(["x1", "x2"], ...
                            'rules', {{[1, 1], [1]}}, ...
                            'hermitian', true);
\end{code:matlab}

We also need to formulate the polynomials for the constraint and for the objective function.
To do this, we first get the monomial objects associated with the fundamental alphabet:
\begin{code:matlab}
[x1, x2] = scenario.getAll();
I = scenario.id(); 
\end{code:matlab}
\noindent and then compose them:
\begin{code:matlab}
objective = x1 * x2 + x2 * x1;
constraint = -x2 * x2 + x2 + 0.5*I;
\end{code:matlab}

Let variables \code{mm_level} and \code{lm_level} respectively hold integer values $m$ and $l$.
To generate the two matrices we write:
\begin{code:matlab}
mm = scenario.MomentMatrix(mm_level);
lm = scenario.LocalizingMatrix(constraint, lm_level);
\end{code:matlab}

For $m=1$, $l=1$ the two matrices produced are:
\begin{align}
M\!\left(\mathcal{X}, 1\right) &=
\begin{bmatrix}
\expt{\id} & \expt{x_1} & \expt{x_2} \\ 
\expt{x_1} & \expt{x_1} & \expt{x_1 x_2} \\ 
\expt{x_2} & \expt{x_2 x_1} & \expt{x_2 x_2}
\end{bmatrix}, \\
L\!\left(\mathcal{X}, -x_2^2 + x_2 + 0.5, 1\right)  &= \nonumber\\
&\hspace{-9em}
\begin{small}
\begin{bmatrix}
\frac{1}{2}\expt{\id} + \expt{x_2} - \expt{x_2 x_2} & \frac{1}{2} \expt{x_1} + \expt{x_2 x_1} - \expt{x_2 x_2 x_1} & \frac{1}{2} \expt{x_2} + \expt{x_2 x_2} - \expt{x_2 x_2 x_2} \\ 
\frac{1}{2} \expt{x_1} + \expt{x_1 x_2} - \expt{x_1 x_2 x_2} & \frac{1}{2} \expt{x_1} + \expt{x_1 x_2 x_1} - \expt{x_1 x_2 x_2 x_1} & \frac{1}{2} \expt{x_1 x_2} + \expt{x_1 x_2 x_2} - \expt{x_1 x_2 x_2 x_2} \\ 
\frac{1}{2} \expt{x_2} + \expt{x_2 x_2} - \expt{x_2 x_2 x_2} & \frac{1}{2} \expt{x_2 x_1} + \expt{x_2 x_2 x_1} - \expt{x_2 x_2 x_2 x_1} & \frac{1}{2} \expt{x_2 x_2} + \expt{x_2 x_2 x_2} - \expt{x_2 x_2 x_2 x_2}
\end{bmatrix}
\end{small}.
\end{align}

We now have the objects necessary to formulate the SDP in our modeller of choice, and dispatch it to be solved:
\begin{center}
\begin{tabular}{lcl}
Using \cvx{} (see \code{cvx_polynomial.m}): &\hspace{10px}& Using \yalmip{} (see \code{yalmip_polynomial.m}):\\
\cellcolor{verylightgray} %
\begin{code:matlab}[linewidth=0.45\textwidth,backgroundcolor=,frame=]
cvx_begin sdp
    scenario.cvxVars('a', 'b');
    M = mm.Apply(a, b);
    L = lm.Apply(a, b);

    a(1) == 1;  
    M >= 0;
    L >= 0;

    Ob = objective.Apply(a, b);
    minimize(Ob);
cvx_end
\end{code:matlab}
&&\cellcolor{verylightgray} %
\begin{code:matlab}[linewidth=0.45\textwidth,backgroundcolor=,frame=]
[a, b] = scenario.yalmipVars();
M = mm.Apply(a, b);
L = lm.Apply(a, b);

constraints = [a(1) == 1];
constraints = [constraints, M >= 0];
constraints = [constraints, L >= 0];

Ob = objective.Apply(a, b);
optimize(constraints, Ob);
\end{code:matlab}
\end{tabular}
\end{center}

\subsubsection{Example: Brown--Fawzi--Fawzi calculation of conditional von Neumann entropy.}
Another more involved example arises from the field of device--independent quantum cryptography (see e.g.\ \cite{ScaraniBCDLP09,XuMZLP20}).

A crucial practical question in cryptography is ``how quickly can I establish a secure key?''
One way to quantify this is to consider the amount von Neumann entropy produced by the communicating party's cryptographic devices conditioned on the amount of {\em side information} that an eavesdropper has.
Optimising this, over all possible quantum devices in a particular scenario is a difficult problem,
 but \citet{BrownFF21di} provide a method that produces a family of values that converge on this conditional von Neumann entropy from below.

Using \moment{} and \yalmip{} in the example \code{brown_fawzi_fawzi.m}, we calculate a bound via this method for two devices (used by agent's Alice and Bob) achieving a minimal value of the CHSH function,
 using the ``speed up'' in {\em Remark 2.6.3} of \citet{BrownFF21di}.
Particularly, we will perform a numerical integration (via Gauss-Radau quadrature) of a univariate polynomial defined as the solution to a non-commuting polynomial optimization (further relaxed into an SDP).

For the numerical integration, the function \code{gauss_radau} determines the positions $t_i \in [0,1]$ of the sample points we must take, and their weights $w_i \in \reals$, such that the computed value $H$ will be
\begin{align}
H = \sum_i \dfrac{w_i}{t_i\log 2} f\left(t_i\right)
\end{align}
where each $f(t_i)$ is the solution to the following non-commuting polynomial optimization problem:
\begin{align}
\label{eq:BFFncp}
\inf_{\hilb{}, \psi, a_0, z_0, z_1} \quad & \bra{\psi} a_0\left(z_0 + \conj{z_0}+ \left(1-t_i\right)\conj{z_0}z_0\right) + t_i z_0 \conj{z_0}  \nonumber\\
& \qquad + (1-a_0)\left(z_1 + \conj{z_1}+ \left(1-t_i\right)\conj{z_1}z_1\right) + t_i z_1 \conj{z_1} \ket{\psi} \nonumber \\
\st \quad & \exists \; a_1, b_0, b_1 \mathrm{~where} \nonumber\\
&\quad \bra{\psi} - a_0 - b_0 + a_0 b_0 + a_0 b_1 + a_1 b_0 - a_1 b_1 \ket{\psi} \ge \alpha, \nonumber\\
\mathrm{and}&\quad [a_i, b_j] = [z_i, a_j] = [\conj{z_i}, a_j] = [z_i, b_j] = [\conj{z_i}, b_j] = 0, \quad \forall i, j \in \{0, 1\}, \nonumber\\
\mathrm{and}&\quad {a_0}^2 = a_0, \quad {a_1}^2 = a_1, \quad {b_0}^2 = b_0, \quad {b_1}^2 = b_1.
\end{align}
The astute reader might recognise the polynomial in the first constraint as an affine transformation of the CHSH inequality (\cref{eq:CHSH:CG}).
This arises because the general idea of the optimization is to maximize a key rate protocol subject to it being certifiably non-classical.

To relax this problem into a soluble SDP, we use the NPA hierarchy on the alphabet of operators $\mathcal{X}_{\rm bff} := \{a_0, a_1, b_0, b_1, z_0, \conj{z_0}, z_1, \conj{z_1}\}$
 satisfying the commutation and projection constraints in \cref{eq:BFFncp}.
Beyond the operators $\{a_0, a_1, b_0, b_1\}$ that also appear in the locality scenario, BFF's variational technique introduces two additional operators $z_0$ and $z_1$ that are neither projective nor Hermitian.

Declaring this alphabet, and imposing the operators relationships can be achieved in \moment{} by:
\begin{code:matlab}
scenario = AlgebraicScenario(["a0", "a1", "b0", "b1", "z0", "z1"], ...
                            'hermitian', false, 'normal', false);
rules = scenario.OperatorRulebook; 

for op = rules.OperatorNames(1:4) 
	rules.MakeHermitian(op);
	rules.MakeProjector(op);
	rules.AddCommutator("z0", op);
	rules.AddCommutator("z0*", op);
	rules.AddCommutator("z1", op);
	rules.AddCommutator("z1*", op);
end

rules.AddCommutator("b0", "a0");
rules.AddCommutator("b1", "a0");
rules.AddCommutator("b0", "a1");
rules.AddCommutator("b1", "a1");
\end{code:matlab}

Unlike in the previous example where we provided a simple rule into the constructor of the scenario, here we acquire a handle to the scenario's \code{OperatorRulebook} object.
This can be used to add the appropriate rules one at a time via the convenience methods \code{MakeHermitian}, \code{MakeProjector} and \code{AddCommutator}.
Beyond readability, this also allows for the programmatic construction of a ruleset -- for example, here many of the rules are defined via a \code{for} loop.
In the constructor of \code{AlgebraicScenario} we signal that (in general) the operators are not Hermitian, or even normal.
The Hermicity of specific operators can still be achieved through substitution rules (namely one of the form $\conj{x_i} \mapsto x_i$).
However, if all operators are Hermitian, setting the \code{hermitian} flag to true in the scenario's constructor will result in faster computations.

Since these rules essentially define the algebra of operators, it is necessary to define them {\em before} generating any objects (e.g.\ polynomials or moment matrices) associated with the scenario.

The objects required to define the SDP are generated within the \code{solve_bff_sdp} function by:
\begin{code:matlab}
mm = scenario.MomentMatrix(moment_matrix_level);
[a0, a1, b0, b1, z0, z1] = scenario.getAll();
chsh = - a0 - b0 + a0*b0 + a0*b1 + a1*b0 - a1*b1;
obj  = a0*(z0 + z0' + (1-t)*(z0'*z0)) + t*(z0*z0') + ...
      + z1 + z1' + (1-t)*(z1'*z1) + t*(z1*z1') ...    
      - a0*(z1 + z1' + (1-t)*(z1'*z1));
\end{code:matlab}
where \code{moment_matrix_level} and \code{t} are scalar variables passed into the function.
The \code{'} suffix operator in \matlab{} indicates Hermitian conjugation.
Relaxation into the NPA hierarchy requires the generation of a moment matrix \code{mm}.
Meanwhile the \code{chsh} and \code{obj} objects correspond respectively to
 the CHSH constraint and the objective function of \cref{eq:BFFncp}.

Even though \code{mm} is created within the function \code{solve_bff_sdp}, which is executed in a loop,
 the architecture of \moment{} avoids a much of a performance loss here: this matrix is generated once, and then cached in memory.
The \code{scenario.MomentMatrix} method returns a handle to this stored moment matrix, rather than explicitly copying out a block of data.

Modelling and solving of the SDP (here via \yalmip{}) follows the standard pattern:
\begin{code:matlab}
 a = scenario.yalmipVars();
 M = mm.Apply(a);
 constraints = [a(1) == 1;  M >= 0, chsh.yalmip(a) >= value_chsh];
 objective = obj.Apply(a);
 optimize(constraints, objective);
 val = value(objective);
\end{code:matlab}
Here, \code{val} is the return value of \code{solve_bff_sdp}, i.e.\ $f(t_i)$. (Again, we have used symmetry of the problem under conjugation to justify ignoring the imaginary parts of moments).

\subsection{`Pauli' spin-$\frac{1}{2}$ scenario}
\label{sec:Pauli}
\subsubsection{Introduction}

The {\em Pauli scenario} defines operators that act like the Pauli matrices ($\sigma_x$, $\sigma_y$ and $\sigma_z$) acting on a set of $N$ qubits.
These qubits can either be unstructured (i.e.\ ``glass''-like), or arranged in a one-dimensional chain, or two-dimensional lattice.
Such a qubit system is often used to model condensed-matter systems, and its semi-definite relaxation to bound properties such as ground state energy or magnetization that are otherwise untractable \cite{Kull2022,Wang2023}.

\subsubsection{Implementation}
\label{sec:PauliImpl}
For this particular scenario, an explicit matrix representation of the algebra is known ahead of time.
However, as is the case with all scenarios in \moment{}, the operators are still defined only by their algebraic properties.
Particular, one specifies the number of qubit spins $N$, and for each $N$ three Hermitian operators $X_i$, $Y_i$, and $Z_i$ are defined with multiplicative behaviour isomorphic to the Pauli matrices.
Namely, for all $i=1,\ldots, N$:
\begin{gather}
X_i X_i = Y_i Y_i = Z_i Z_i = \id \nonumber\\
X_i Y_i = i Z_i, \quad Y_i Z_i = i X_i, \quad Z_i X_i = iY_i \nonumber \\
Y_i X_i = -i Z_i, \quad Z_i Y_i = -i X_i, \quad X_i Z_i = -iY_i.
\end{gather}
Operators on different qubits commute, such that for all $i= 1,\ldots, N$ and $j\neq i$:
\begin{equation}
[X_i, X_j] = [X_i, Y_j] = [X_i, Z_j] = [Y_i, Y_j] = [Y_i, Z_j] = [Z_i, Z_j] = 0.
\end{equation}
This has the result that when multiplying (and subsequently simplifying) monomial strings of such Pauli operators, there is at most one operator associated with each qubit index.

For a Pauli scenario in \moment{}, one can define whether the qubits are conceptually arranged in a $1$-dimensional chain, or a $2$-dimensional lattice.
One can signal whether the system has {\em wrapping} such that the chain behaves topologically as if it were circle and the lattice as a torus.
One can also signal whether the setting has translational symmetry.

For instance, consider the following scenario definitions:
\begin{code:matlab}
chain = PauliScenario(6);
lattice = PauliScenario([3 2]);
cyclic_chain = PauliScenario(6, 'wrap', true);
symmetric_lattice = PauliScenario([3 2], 'wrap', true, 'symmetrized', true);
thermo_chain = PauliScenario(6, 'wrap', false, 'symmetrized', true);

\end{code:matlab}
Here, \code{chain} is a chain of $6$ qubits, whereas \code{lattice} is a $3\times 2$ lattice of qubits, neither with symmetries or wrapping.
Next, \code{cyclic_chain} is a chain of $6$ qubits such that qubit $6$ neighbours qubit $1$.
Similarly \code{symmetric_lattice} closes the $3\times2$ lattice into a torus.
The behaviour of the last case (\code{thermo_chain}) we shall discuss later.

The constructor for the chain without symmetry or wrapping provides the most `neutral' variant of the Pauli scenario, defining $N$ qubits without further structure (the chain-like behaviour only manifests if {\em nearest-neighbour} functions, defined in the subsequent section, are used).
This is appropriate for solving problems without a particularly symmetric topology (e.g.\ spin glasses, where the Hamiltonian couples qubits according to some arbitrary network).

\subsubsection{Nearest neighbours and wrapping}
\label{sec:NN}
The Pauli scenario allows for partial moment matrix levels by filtering the generating dictionary to only contain ``nearest--neighbour'' expressions.
Whether two qubits are deemed ``neighbouring'' depends on whether the scenario is a chain or a lattice, and whether wrapping is enabled.
For example, for an unwrapped chain, the only neighbour to qubit $1$ is qubit $2$; whereas for a wrapped chain qubit $1$ is also adjacent to qubit $N$.
In a lattice, each qubit can have up to $4$ neighbours.
For example, in a $r$ rows by $c$ columns lattice, qubit $k$ will be adjacent to qubits $k-1$, $k+1$ within the same column, as well as $k-r$ and $k+r$ in adjacent rows.
For a wrapping lattice, the left-most column is considered adjacent to the right-most column, and the top row is considered adjacent to the bottom row.

For chains, $m$-nearest neighbour filtering is possible, such that subsequent operators within a word must not act on qubits separated by an index greater than $m$.
For example, if $m=2$, expressions $X_1 Z_2$ and $X_1 Z_3$ are included, but $X_1 Z_4$ is not.
For expressions composed of three or more operators, nearest-neighbour filtering effectively restricts the maximum gap between indices of the operators (e.g.\ $X_1 Y_2 Z_3$ would be considered a nearest neighbour expression).
For lattices, only the $m=1$ (strictly nearest) case is supported.

Consider a scenario \code{chain} (wherein some scalar polynomial \code{p} has been previously defined).
For \code{MomentMatrix}, \code{LocalizingMatrix} and \code{WordList} functions, nearest neighbour variants are created by supplying an additional parameter:
\begin{code:matlab}
nn_mm = chain.MomentMatrix(2, 1);
nn_lm = chain.LocalizingMatrix(p, 2, 1);
nn_wl = chain.WordList(2, 1);
\end{code:matlab}
This respectively creates a level $2$ moment matrix, a level $2$ localizing matrix of $p$ and a list of all monomial expressions of length $2$, all restricted to nearest neighbours.
Beware: the nearest-neighbour restriction on matrices restricts only the terms in the generating dictionary (i.e.\ the top row of the matrix) -- the matrices themselves will in general still contain non-nearest neighbour expressions where they appear as the product of two nearest-neighbour-restricted terms.
To generate the usual, unrestricted, objects, one omits the nearest neighbour argument (or sets it to $0$).

\subsubsection{Symmetrization}
Sometimes, problems involving a spin system exhibit translational symmetry.
For the sake of discussion, consider an $N$-qubit cyclic Heisenberg spin chain with constant coupling constant $\vec{j} = (j_X, j_Y, j_Z)$, and constant external field $h$.
Its energy is given by the Hamiltonian:
\begin{equation}
\label{eq:HeisenbergSpinChain}
H = \sum_{i=1}^N j_X X_i X_{i+1} + j_Y Y_i Y_{i+1} + j_Z Z_i Z_{i+1} + h Z_i,
\end{equation}
where for notational brevity we take all addition in subscripts to be implicitly modulo $N$.
Clearly, substitutions of the form $X_i \to X_{i+k}$, $Y_i \to Y_{i+k}$ and $Z_i\to Z_{i+k}$ for all $i=1,\ldots, N$ do not change~$H$.

Suppose there is some feasible program with $\expt{H}$ as its objective, and without other constraints that break the problem's symmetry.
Then, for SDP relaxations of this problem, there will be a optimal solution whose moments also respect this symmetry. 
For example, $\expt{X_1} = \expt{X_2} = \ldots = \expt{X_N}$, but also $\expt{X_1 Z_2} = \expt{X_2 Z_2} = \ldots = \expt{X_N Z_1}$.
This allows the formulation of an SDP with (far!) fewer optimization parameters than the unsymmetrized equivalent.

To enable this in \moment{}, set the \code{'symmetrized'} parameter to \code{true} in the constructor of \code{PauliScenario} (as in \cref{sec:PauliImpl}).

This symmetrization should be understood as acting at the level of moments, not of operators.
As mentioned in \cref{sec:MomentObjects}, objects in \moment{} act as representations of operators up until the point they are applied to a state.
Thus, in a symmetric scenario when one has:
\begin{code:matlab}
x1 = scenario.sigmaX(1);
x2 = scenario.sigmaX(2);
p = x1 + x2;
q = 2*x1;
\end{code:matlab}
the objects \code{p} and \code{q} are distinct, even though they both produce a representation of the same moment $2\expt{X_1}$ when \code{.Apply(}\ldots\code{)} is used.
If \code{p} or \code{q} are used directly in objective function, or a scalar constraint, they will yield identical behaviour.
However, their distinction can be seen when they are multiplied by another object.
For instance,
\begin{code:matlab}
y1 = scenario.sigmaY(1);
p_res = p * y1;
q_res = q * y1;
\end{code:matlab}
produces objects \code{p_res} and \code{q_res} encoding respectively $iZ_1 + Y_1X_2$ and $2iZ_1$,
 whose moments (respectively $i \expt{Z_1} + \expt{Y_1 X_2}$ and $2i\expt{Z_1}$) are different, even with symmetrization applied to them.

The same behaviour applies consistently to matrix objects produced by \moment{}.
Continuing the above example,
\begin{code:matlab} 
mm = scenario.MomentMatrix(1);
x1_mm = x1 * mm;
\end{code:matlab}
produces a matrix by essentially premultiplying the unsymmetrized operators of \code{mm} by \code{x1}.
When \code{x1_mm} is ultimately used with a state, only then does the symmetrization take effect.
As such \code{x1_mm} will contain terms such as $X_1 Z_2$ on its top row (yielding moment $\expt{X_1 Z_2}$), even though \code{mm} will never produce the moment $\expt{Z_2}$ for any of its elements.

\inlineheading{Symmetrize method.}
The \code{PauliScenario} object defines a utility method \code{symmetrize}.
This acts on monomial and polynomial objects to produce a new object (typically polynomial) that obeys the scenario's symmetry,
 but is guaranteed to transform into the same moments as the input.
The quintessential use case is to define a Hamiltonian in terms of nearest neighbour interactions between the first and second qubit, and subsequently expand this to the whole chain.
For instance, take (with some real number $j$ in variable \code{j}):
\begin{code:matlab}
nn_H = j * x1 * x2
H = scenario.symmetrize(nn_H);
\end{code:matlab}
The object \code{nn_H} represents $jX_1 X_2$, whereas \code{H} represents $\frac{1}{N}\sum_{i=1}^N j X_i X_{i+1}$.
Both these objects evaluate to the identical moment $j \expt{X_1 X_2}$.

\inlineheading{Lattices and non-wrapping scenarios.}
Symmetrization can also be enabled for lattices.
This behaves in essentially the same manner as above, though for words longer than length one, it will differs in the exact set of moments identified as equivalent.
Symmeterization can also be implemented in scenarios without wrapping (this has uses in approximating the thermodynamic limit of large $N$, where an explicitly cyclic Hamiltonian may result in bad convergence due to frustration~\cite{AraujoKGVM23}).
Especially at small $N$, the set of identical moments will differ between these cases (e.g. for $N=2$, with wrapping enabled $\expt{X_1 Z_2} = \expt{Z_1 X_2}$, but these moments are different without wrapping).

\subsubsection{Example: Bounding the ground state energy of a Heisenberg spin chain.}
Let us now work through an example adapted from \citet{AraujoKGVM23}, whereby we wish to find the lower and upper bounds on the ground state energy of a Heisenberg spin chain with Hamiltonian $H$ as per \cref{eq:HeisenbergSpinChain} with $j_X = j_Y = j_Z = 0.25$ and $h = 0$.
The full version of this example is included with \moment{} as \code{yalmip_heisenberg.m}.

First, we set up the scenario and its objective Hamiltonian.
Let the variable \code{chain_length} contain the chain length $N$. Then, we make the scenario:
\begin{code:matlab}
scenario = PauliScenario(chain_length, 'wrap', true, 'symmetrized', true);
\end{code:matlab}
and with a three-element array \code{j = [0.25, 0.25, 0.25]}, corresponding to $j_X$, $j_Y$, $j_Z$ we set up the Hamiltonian using:
\begin{code:matlab} 
[X, Y, Z] = scenario.getAll();
base_H = j(1)*X(1)*X(2) + j(2)*Y(1)*Y(2) + j(3)*Z(1)*Z(2);
H = scenario.symmetrize(base_H);
\end{code:matlab}
As discussed above, the moments produced by applying a state to \code{base_H} and \code{H} will be the same; but they have crucially distinct behaviour when they are multiplied with other objects.

The lower energy bound can be computed as a straightforward minimization of the Hamiltonian as an objective function subject to a PSD moment matrix.
Let \code{mm_level} contain the maximum word length in the generating dictionary,
 and  \code{neighbours} contain a restriction of the maximum distance between subsequent qubits in any operator string (see \cref{sec:NN}).

Then, the moment matrix is generated by:
\begin{code:matlab}
raw_mm = scenario.MomentMatrix(mm_level, neighbours);
\end{code:matlab}

To compute the upper bound on the energy, we will maximize the Hamiltonian under the additional constraint that the optimizing state should be the ground state of the system (a so called {\em state--optimality constraint})\footnote{If we do not impose this, then maximization will simply yield an upper bounds on the {\em maximum} energy of the system.}.
\Citet{AraujoKGVM23} tell us we can impose this state optimality condition by imposing the following constraints:
For the optimizing state $\sigma^\star$ and for polynomials $p$ and $q$ (whose indeterminates are the Pauli operators acting on each qubit in the chain): 
\begin{equation}
\label{eq:GammaMatrix}
\sigma^\star \left( pHq - \frac{1}{2} \{pq, H\} \right) \geq 0,
\end{equation}
and for all polynomials $r$:
\begin{equation}
\label{eq:CommutatorConstraints}
\sigma^\star \left( [H, r] \right) = 0
\end{equation}
(where $\{x,y\}$ and $[x, y]$ respectively denote the anticommutator and commutator of $x$ with $y$).

We can translate and relax these constraints into the more familiar SDP language used in \moment{}.
Particularly, linearity of the functional $\sigma^\star$ allows us to consider instead a basis of monomials $p$, $q$ and $r$.
Moreover, we can impose a relaxed version of this by only considering monomials up to some fixed choice of maximum order\footnote{Here, since $N$ is finite, we could in principle enforce this condition on {\em every} monomial, since products of more than $N$ Pauli operators will always be equivalent to some product of $N$ or fewer operators. However, since the number of such monomials scales exponentially in $N$, computing this problem without this relaxation is impractical for all but the smallest values of $N$.}.

First, following \citet{PironioNA10}, if we had only $\sigma\left(pHq\right) \geq 0$ for all $p$ and $q$, this would be equivalent to imposing that the localizing matrix of $H$ is PSD.
However, the second term of this expression is less familiar -- but can likewise be expressed as a matrix of polynomial operator expressions produced by taking every operator component that defines the moment matrix, and calculating its anticommutator with $-\frac{1}{2}H$.
To produce this composite matrix expression (with an integer \code{lm_level} defined), one invokes:
\begin{code:matlab}
raw_lm_H = scenario.LocalizingMatrix(H, lm_level, neighbours);
raw_am_H = scenario.AnticommutatorMatrix(-0.5*H, lm_level, neighbours);
raw_gamma = raw_lm_H + raw_am_H;
\end{code:matlab}
Here, we use a new method \code{AnticommutatorMatrix}\footnote{A similar \code{CommutatorMatrix} method is also defined.}, and sum its result to the familiar localizing matrix of $H$.
\Cref{eq:GammaMatrix} is then imposed by constraining that the matrix of moments formed by applying a state to \code{raw_gamma} is PSD.
When we also include the commutator constraints, we must ignore the top-most row and left-most column of this matrix (because for all monomial $x$, $\expt{x H \id} - \expt{\{x, H\}} = \expt{[x, H]} = 0$, which would make the top-most row and left-most column zero, and hence the matrix would become trivially PSD).

Next, let us turn to \cref{eq:CommutatorConstraints}.
This is essentially a family of equality constraints.
With \code{so_level} containing an integer determining the maximum order of $r$,
 we generate a vector \code{linear} containing all such polynomials using:
\begin{code:matlab}
monomials = scenario.WordList(so_level, neighbours);
linear = commutator(monomials, H);
\end{code:matlab}
Again, the \code{neighbours} argument of \code{WordList} allows us to relax this list to neighbouring terms only.

Since \code{linear} are equality constriants, we can use these to define a set of moment substitution rules (potentially further reducing the number of real variables in the SDP we ultimately must solve).
We do this via:
\begin{code:matlab}
moment_rules = MomentRulebook(scenario, "Commutator constraints");
moment_rules.Add(linear, false);
mm = raw_mm.ApplyRules(moment_rules);
gamma = raw_gamma.ApplyRules(moment_rules);
\end{code:matlab}
This produces a matrices \code{mm} and \code{gamma}, which are guaranteed to enforce these constraints\footnote{
Typically, using \moment{} to reformulate equality constraints as substitutions, we should also transform the objective function (\code{H}) with the same rulebook. 
However, for this specific example, we know {\em a priori} that \code{H} cannot change under application of these rules.}

We now have all the objects to model the SDP.
For this example, we use \yalmip{}:
\begin{code:matlab}
a = scenario.yalmipVars(); 
M = mm.Apply(a);
G = gamma.Apply(a);
G = G(2:end, 2:end);

objective = H.Apply(a);
constraints = [a(1) == 1, M >= 0, G >= 0];

optimize(constraints, -objective);
upper_bound = value(objective);
\end{code:matlab}
This particular problem employs a mild numerical hack in the line \code{G = G(2:end, 2:end)}; which manually ignores the top-most row and left-most column of this matrix, for reasons mentioned above.

\subsection{Inflation scenario}
\label{sec:inflationscenario}
\subsubsection{Introduction.}
A common problem of classical causal inference in a network is to determine whether the statistics from a set of observables are compatible with an underlying causal structure~\cite{Pearl00} (specified in terms of connected {\em latent} variables).
The {\em inflation scenario} in \moment{} has been designed to address such questions, using the inflation technique of \citet{WolfeSF19}.

The idea key of inflation is to produce a new {\em inflated} network from the base network by making copies of the latent and observable variables.
Compatibility constraints on this inflated network can then be used to imply stricter compatibility constraints on the base network.
That is, in the language of SDPs, there are probability distributions that cannot be incorporated into a positive-semi-definite moment matrix of the inflated network,
 even though they can satisfy SDP constraints for moment matrices of the base network.
An example where this happens is presented later in this section.

\subsubsection{Implementation in \moment{}.}
Within \moment{}, a causal network is defined by its {\em observables}, and its latent hidden {\em sources}.
Each source can connect to one or more observable.
Each observable is specified to have a discrete finite number of outcomes, or can be set as continuous variable.
Finally, the desired level of inflation is set.

The inflation is performed according to the {\em web inflation} scheme in \citet{WolfeSF19} -- that is, for {\em inflation level} $N$, $N$ variants of each latent source are made; and then for every observable connected to $m$ latent sources, $N^m$ variants are made -- one for each combination of attached inflated sources.
For each observable variant, either $n-1$ projective Hermitian operators are generated if the observable corresponds to a measurement with $n$ outcomes; otherwise a single non-projective Hermitian operator is generated to handle the continuous variable case.

In addition to the simplification of operator strings into a canonical form,
 the inflation scenario brings two additional concepts: {\em moment simplification} and {\em factorization}.

The first takes into account the symmetries instrinsic to inflated networks,
 and equates {\em moments} that are equivalent up to a permutation of source variants,
 and thus replaces a moment by its equivalent with the lowest indices of sources.
For example, suppose we have a (trivial!) inflated scenario of two observables $X$ and $Y$ with common source $\lambda$.
Then, if we inflate by making $n$ copies of $\lambda$, we subsequently have inflated observables $\{X_i\}_{i=1,\ldots, n}$ and $\{Y_i\}_{i=1,\ldots, n}$.
Symmetry considerations allow us to replace every $\expt{X_i}$ with $\expt{X_1}$, or even $\expt{X_i Y_i} \mapsto \expt{X_1 Y_1}$.

The second concept, {\em factorization}, takes into account that when two observables do not share a latent hidden source, their moments factorize.
That is, suppose $X$ and $Y$ are observable variants with no common source, then $\expt{WV} = \expt{W}\expt{V}$.
This is particularly useful when applying fixed values (i.e.\ enforcing the statistics one wants to test) to the setting.
For example, if one knows $\expt{W} = 0.5$, then one can also make the substitution $\expt{WV}=0.5\expt{V}$.
\moment{} supplies tools for identifying and applying these additional substitutions.

These two concepts can interact, since inflation can ``break'' causal connections between some inflated variants of observables connected in the base scenario.
To return to the example of $X$ and $Y$ with a common source; although $\expt{X_i Y_i} = \expt{X_1 Y_1}$ does not factorize, $\expt{X_i Y_j} = \expt{X_i}\expt{Y_j} = \expt{X_1}\expt{Y_1}$ when $i\neq j$.

\subsubsection{Extended moment matrices.}
\moment{} provides an additional type of operator matrix in the inflation scenario, known as a (scalar) {\em extended matrix}~\cite{PozasKerstjensRRCCNA19},
 that for the imposition of relaxations of factorization relationships 

As a toy example, suppose $\mathcal{X} = \{W, V\}$.
The basic moment matrix generated is:
\begin{align}
\label{eq:underconstrainMM}
M(\mathcal{X}, 1) = \begin{bmatrix}
\expt{\id} & \expt{W} & \expt{V} \\
\expt{W} & \expt{W^2} & \expt{WV} \\
\expt{V} & \expt{WV} & \expt{V^2}
\end{bmatrix}
\end{align}
Suppose that $\expt{WV} = \expt{W}\expt{V}$ (and the explicit values of $\expt{W}$ and $\expt{V}$ are not determined).
Then, \cref{eq:underconstrainMM} is under-constrained  (i.e.\ relaxes this requirement) since all three moments could independently take values that do not satisfy $\expt{WV}=\expt{W}\expt{V}$.
Moreover, directly applying $\expt{WV}=\expt{W}\expt{V}$ as an extra constraint to the SDP usually will not work, as this non-linear constraint takes the problem outside the regime of {\em disciplined convex programming}~\cite{GrantB08} that \cvx{} and \yalmip{} rely on.

However, \citet{PozasKerstjensRRCCNA19} provide a technique to add a relaxation of this constraint that preserves the convexity of the problem.
Namely, one can add a column (and row) to the moment matrix representing an extra operator that is scalar multiple $k$ of the identity,
 where $k$ is one of the factors of the multiplicative constraint we wish to partially impose.

For instance, in the above example we could add $\expt{W}\id$.
Taking into account that $\expt{W}$ is just a number, such that $\expt{\expt{W}\id} = \expt{W}\expt{\id} = \expt{W}$,
 this results in the {\em extended moment matrix}:
\begin{align}
\label{eq:extendedMM}
E(\mathcal{X}, 1, \{\expt{W}\}) := \begin{bmatrix}
\expt{\id} & \expt{W} & \expt{V} & \expt{W} \\
\expt{W} & \expt{W^2} & \expt{WV} & \expt{W}^2 \\
\expt{V} & \expt{WV} & \expt{V^2} & \expt{W}\expt{V} \\
\expt{W} & \expt{W}^2 & \expt{W}\expt{V} & \expt{W}^2
\end{bmatrix}.
\end{align}
Since we know of the factorization relationship $\expt{WV} = \expt{W}\expt{V}$, we can then represent these two expressions by the same variable.
I.e.\ for this moment, there will be one basis element
\begin{align}
A_{\expt{WV}} := \begin{bmatrix}
0 & 0 & 0 & 0 \\
0 & 0 & 1 & 0 \\
0 & 1 & 0 & 1 \\
0 & 0 & 1 & 0
\end{bmatrix},
\end{align}
that covers where both expressions appear in the moment matrix.

\subsubsection{Example: Triangle scenario.}
Let us consider a conceptually-simple non-trivial example of an incompatible causal structure: the {\em triangle scenario}.
This is included with \moment{} as examples \code{cvx_inflation_triangle} and \code{yalmip_inflation_triangle} for the respective modeller.

A simple formulation consists of three binary observables (outcomes $0$ and $1$) each pair-wise connected by a hidden source.
We will attempt to refute the compatibility of this scenario with the tripartite measurement probabilities
\begin{align}
\label{eq:triP}
P(000) = P(111) = \frac{1}{2},
\end{align}
with all other outcome probabilities set to zero.

Let us motivate the usefulness of inflation, by showing analytically that the base network will not violate SDP constraints for any moment matrix level.
First, in the base scenario there are only three operators in the alphabet $\mathcal{X}_{\rm base}:=\{a, b, c\}$ (corresponding to the first outcome of each observable), such that once projection and commutation is taken into account, the longest unique string is $a b c$. 
It then follows that $M\left(\mathcal{X}_{\rm base}, n\right) = M\left(\mathcal{X}_{\rm base}, 3\right)$, for $n\geq3$.
Moreover, imposing \cref{eq:triP} sets the value of {\em every} moment to $0.5$, except for $\expt{\id} = 1$, and so we have the constant moment matrix:
\begin{align}
M(\mathcal{X}_{\rm base}, n\geq3) = 
\begin{bmatrix} 1 & 0.5 & \hdots & 0.5 \\
0.5 & 0.5 & \hdots & 0.5 \\
\vdots & \vdots & \ddots & \vdots \\
0.5 & 0.5 & \hdots & 0.5
\end{bmatrix} \psd 0,
\end{align}
where the positive-semi-definiteness can be directly verified.

Thus, to show the incompatibility of \cref{eq:triP} with the triangle scenario, we will inflate it to level $2$ (i.e.\ make two variant copies of each hidden source).
To define the causal network, and set the inflation level using \moment{}, we execute:
\begin{code:matlab}
triangle = InflationScenario(2,  [2, 2, 2], ...
                             {[1, 2], [2, 3], [1, 3]});
\end{code:matlab}
Here, the first argument is the inflation level (here, $2$).
The second argument defines the observables by their number of outcomes (here, three binary measurements).
The final argument defines the number of hidden sources by a list of observables they influence (here, a pairwise triangle configuration).

Since we are going to constrain a tripartite probability distribution, we will need the second (or higher) level moment matrix, to ensure that the necessary moments are in our SDP:
\begin{code:matlab}
moment_matrix = triangle.MomentMatrix(2);
\end{code:matlab}

Next we will create a set of moment substitution rules, which will impose the probability distribution (\cref{eq:triP}).
\begin{code:matlab}
[A, B, C] = triangle.getPrimaryVariants();
ABC = A * B * C;
distribution = ABC.Probability([0.5, 0, 0, 0, 0, 0, 0, 0.5]);
substitutions = triangle.MomentRulebook();
substitutions.Add(distribution);
\end{code:matlab}
The first line gets handles to variants of observables ($A$, $B$ and $C$) associated with the latent sources with the lowest indices (e.g. here $A_00$ instead of $A_{01}$, $A_{10}$ or $A_{11}$).
The second line creates the tripartite joint observable.
The third line creates an list of linear moment expressions that, if equated with zero, impose the supplied joint probability distribution.
 taking into account that most outcomes of this measurement are implicity defined, rather than corresponding directly to a single moment.
The fourth line creates a \code{MTKMomentRulebook} object \code{substitutions} that can be used to apply the substitutions to any matrices or polynomials (such as an objective function) within the scenario.
The final line registers the probability distribution with this rulebook.

Uniquely, in the {\em inflation scenario}, many more implicit rules are generated than are given by explicit list of polynomials.
This is to impose the distribution where it appears as a factor of another of another moment.
Indeed, here \code{distribution} defines $8$ polynomials, but the number of rules in \code{substitutions} is $74$.

To impose these constraints on our SDP, we will directly apply them to the moment matrix (effectively reducing the number of variables in the SDP we will ultimately solve):
\begin{code:matlab}    
subbed_matrix = substitutions.Apply(moment_matrix);
\end{code:matlab}
This creates a new matrix object \code{subbed_matrix}.

Finally, we execute a simple feasibility-test SDP (i.e.\ one with no objective function).
We could do this via the utility function \code{mtk_solve}:
\begin{code:matlab}
feasible = mtk_solve(subbed_matrix)
\end{code:matlab}
Alternatively, more fine-grained control of the SDP set-up can be done manually using \cvx{} or \yalmip{} as shown in the  \code{cvx_inflation_triangle.m} and \code{yalmip_inflation_triangle.m} examples respectively.

\subsection{Symmetrized scenario}
\label{sec:symmetrizedscenario}
\subsubsection{Introduction}
The symmetrized scenario is not directly a scenario in its own right, but rather a transformation that can be applied to any of the preceding three scenario types.
We follow the approach of \citet{IoannouR21}.
In particular, by exploiting symmetries in the polynomials that define an optimization problem, it is possible to map the SDP to another SDP with the same solution, but with a (potentially greatly) reduced number of scalar variables.
This has the benefit of reducing both computational time and memory usage of the solver, enabling the solution of problems that might have previously been beyond the computationally practical limit.

Consider the set of polynomials that define a problem's feasible solutions, and its objective.
The {\em symmetries} of the program are then the group $\mathcal{G}$ of transformations on the level of operators that leave this feasible set and objective unchanged.
If one has an optimal solution to such a problem, exchanging the operators according to any element of $\mathcal{G}$ results in another feasible solution that is also optimal.
Since the optimization problems are convex, any weighted combination of such solutions is also a solution.
One can then substitute the operators specifying of the problem with a weighted sum of transformed operators,
 and, after a change in variables, formulate a new program whose optimal value matches the original program.
For a good choice of substitution, the new program may have fewer SDP variables (and hence be computationally easier to solve) than the original program.

\subsubsection{Implementation in \moment{}}
\moment{} does not provide an implementation for the {\em discovery} of the symmetries in the polynomial problem specification. 
For this (short of calculating them by hand) one could automate the task through other software such as {\em SymDPoly}~\cite{SymDPoly}.
Rather, \moment{} provides the tools to apply a finite-sized symmetry group to the operators defining a problem, and to transform the elements defining the SDP (i.e.\ moment and localizing matrices, constraints, and objectives) into the new reduced variable set.

Denoting the operators of the program as $x_1, \ldots, x_N$, each generators of the symmetry group should be specified as a matrix that acts on the right on the row vector of operators $(\id, x_1, \ldots, x_N)$.
\moment{} then uses Dimino's algorithm~\cite{Butler91} to generate explicit matrix forms for each group element.
Such matrices can be seen as the representation of the symmetry group on operator strings of up to length $1$.
In order to practically apply these symmetries to a problem, their representation on longer words must be calculated.
The user specifies the maximum word length beforehand (e.g.\ for moment matrices of level $m$, one requires strings of length $2m$),
 and then \moment{} can generate the representation.

Once the larger representation is generated, the group average is calculated, to create a matrix $G$ which will ultimately be used for the reduction.
This is an arbitrary choice, but as seen in \citet{IoannouR21}, it can provide good results.
When matrix $G$ is not full rank, then there is the opportunity to make substitutions that reduce the number of SDP variables.
To find such substitutions, \moment{} calculates the {\em LU decomposition} (the implementation of this is provided by \eigen{}'s ``full pivot'' LU module~\cite{eigen}).
This decomposition effectively splits $G = P L U Q$ where $L$ is a lower-triangular matrix, and $U$ is an upper-triangular matrix, and $P$ and $Q$ are permutations.

The matrix $P L$ can then be interpreted as essentially a map from the list of monomial moments in the original problem, into a list of moments in the {\em reduced} variables.
The map that can be applied to moments of the original problem to map them into a basis of reduced variables is then given by $P L$.
Meanwhile, $U Q$ effectively gives the definition of each reduced variable as a function of moments of the original problem.

As a simple example, consider minimizing a two-operator program with objective $\expt{x}+\expt{y}$ subject to $\expt{x}+\expt{y}\geq1$, and the corresponding level $1$ moment matrix (\cref{eq:xymm}) being positive semi-definite.
Clearly, this problem is symmetric under exchange of $x$ with $y$, so the two element group $Z_2$ acting on $(1, x, y)$ is:
\begin{align}
\{
\id_3,  \quad
\begin{bmatrix}1 & 0 & 0 \\ 0 & 0 & 1 \\ 0 & 1 & 0\end{bmatrix}
\}.
\end{align}
To apply this symmetry to our problem with a level $1$ moment matrix, we need the length-$2$ representation, that acts on $(1, x, y, xx, xy, yx, yy)$:
\begin{align}
\{
\id_7, \quad
\begin{bmatrix}
1 & 0 & 0 & 0 & 0 & 0 & 0 \\
0 & 0 & 1 & 0 & 0 & 0 & 0 \\
0 & 1 & 0 & 0 & 0 & 0 & 0 \\
0 & 0 & 0 & 0 & 0 & 0 & 1 \\
0 & 0 & 0 & 0 & 0 & 1 & 0 \\
0 & 0 & 0 & 0 & 1 & 0 & 0 \\
0 & 0 & 0 & 1 & 0 & 0 & 0 
\end{bmatrix}
\},
\end{align}
such that the group average is
\begin{align}
G =  \begin{bmatrix}
1 & 0 & 0 & 0 & 0 & 0 & 0 \\
0 & \frac{1}{2} & \frac{1}{2} & 0 & 0 & 0 & 0 \\
0 & \frac{1}{2} & \frac{1}{2}& 0 & 0 & 0 & 0 \\
0 & 0 & 0 & \frac{1}{2} & 0 & 0 & \frac{1}{2} \\
0 & 0 & 0 & 0 & \frac{1}{2} & \frac{1}{2} & 0 \\
0 & 0 & 0 & 0 & \frac{1}{2} & \frac{1}{2}& 0 \\
0 & 0 & 0 & \frac{1}{2} & 0 & 0 & \frac{1}{2}
\end{bmatrix}.
\end{align}
This can be decomposed $G = P L U Q$:
\begin{align}
L = \begin{bmatrix}
1 & 0 & 0 & 0 & 0 & 0 & 0 \\
0 & 1 & 0 & 0 & 0 & 0 & 0 \\
0 & 1 & 0 & 0 & 0 & 0 & 0 \\
0 & 0 & 0 & 1 & 0 & 0 & 0 \\
0 & 0 & 0 & 0 & 1 & 0 & 0 \\
0 & 0 & 0 & 0 & 1 & 0 & 0 \\
0 & 0 & 0 & 1 & 0 & 0 & 0 
\end{bmatrix}, \quad
U = \begin{bmatrix}
1 & 0 & 0 & 0 & 0 & 0 & 0 \\
0 & \frac{1}{2} & \frac{1}{2} & 0 & 0 & 0 & 0 \\
0 & 0 & 0 & 0 & 0 & 0 & 0 \\
0 & 0 & 0 & \frac{1}{2} & 0 & 0 & \frac{1}{2} \\
0 & 0 & 0 & 0 & \frac{1}{2} & \frac{1}{2} & 0\\
0 & 0 & 0 & 0 & 0 & 0 & 0 \\
0 & 0 & 0 & 0 & 0 & 0 & 0 
\end{bmatrix}, \quad
P = Q = \id_7.
\end{align}
As $U$ has four non-zero rows, we can read this matrix as defining the following moments in a new system:
\begin{align}
\expt{\id} := \expt{\id}, \quad
\expt{Z_1} := \frac{1}{2}\left(\expt{x} + \expt{y}\right), \quad
\expt{Z_2} := \frac{1}{2}\left(\expt{xx} + \expt{yy}\right), \quad
\expt{Z_3} := \frac{1}{2}\left(\expt{xy} + \expt{yx}\right).
\end{align}
Meanwhile, $L$ shows how the original moments can be mapped to the symmetry--reduced moments:
\begin{align}
\expt{\id} \mapsto \expt{\id}, \;
\expt{x} \mapsto \expt{Z_1}, \;
\expt{y} \mapsto \expt{Z_1}, \;
\expt{xx} \mapsto \expt{Z_2}, \;
\expt{xy} \mapsto \expt{Z_3}, \;
\expt{yx} \mapsto \expt{Z_3}, \;
\expt{yy} \mapsto \expt{Z_2}.
\end{align}

After these substitutions are made, the reduced problem is then to minimize $\expt{Z_1}$ subject to $\expt{Z_1}\geq 1$, and
\begin{align}
\begin{bmatrix}
\expt{\id} & \expt{Z_1} & \expt{Z_1} \\
\expt{Z_1} & \expt{Z_2} & \expt{Z_3} \\
\expt{Z_1} & \expt{Z_3} & \expt{Z_2}
\end{bmatrix} \psd 0
\end{align}

Processes such as the above can be automated by \moment{}, as the following example demonstrates.

\subsubsection{Example: Symmetrized CHSH scenario}
We return to our most-solved example of the CHSH scenario, as described in \cref{sec:CHSH}, and show how it can be symmetrized.
First, we create the base scenario -- a locality scenario object representing the collection of operators and moments with no symmetry reduction applied.
This is the usual CHSH locality scenario of Alice and Bob, each with two binary measurements:
\begin{code:matlab}
chsh_scenario = LocalityScenario(2, 2, 2);
\end{code:matlab}

In particular, we will use the form of the CHSH inequality as in \cref{sec:CHSH}.
\begin{code:matlab}
CHSH_ineq = chsh_scenario.FCTensor([[0 0 0]; [0 1 1]; [0 1 -1]]);
\end{code:matlab}

The two generators of the $16$-element $D_4$ symmetry group on this polynomial (\cref{eq:CHSH:CG}) are given as a cell array:
\begin{code:matlab}
chsh_generators = {[[1 0 1 0 0];
                    [0 1 0 0 0];
                    [0 0 -1 0 0];
                    [0 0 0 0 1];
                    [0 0 0 1 0]], ...
                   [[1 0 0 0 0];
                    [0 0 0 1 0];
                    [0 0 0 0 1];
                    [0 1 0 0 0];
                    [0 0 1 0 0]]};
\end{code:matlab}

These can be seen as matrices acting from the right on the row vector of operators $(\id, a_0, a_1, b_0, b_1)$.
The first generator encodes the symmetry:
\begin{align}
\id \mapsto \id, \quad a_0 \mapsto a_0, \quad a_1 \mapsto \id - a_1, \quad  b_0 \mapsto b_1, \quad b_1 \mapsto b_0,
\end{align}
\noindent and the second:
\begin{align}
\id \mapsto \id, \quad a_0 \mapsto b_0, \quad a_1 \mapsto b_1, \quad  b_0 \mapsto a_0, \quad b_0 \mapsto b_1.
\end{align}

To set up the symmetrized scenario object, we invoke:
\begin{code:matlab}
sym_scenario = SymmetrizedScenario(chsh_scenario, chsh_generators, ...
                                   'word_length', 2*mm_level);
\end{code:matlab}
\noindent where \code{mm_level} is an integer denoting the maximum level of moment matrix we will consider generate.
This registers moments up to size \code{2*mm_level} in the base (unreduced) scenario, and determines the map from these moments onto moments in the symmetrized (reduced) scenario.
For level $1$, there is only one non-trivial reduced moment $\expt{y}$, which we can see by executing \code{sym_scenario.Symbols}:
\begin{align}
\expt{y}:= -0.25 \expt{a_0} - 0.25 \expt{b_0} + 0.25 \expt{a_0 b_0} + 0.25 \expt{a_0 b_1} + 0.25 \expt{a_1 b_0} - 0.25 \expt{b_1 b_1}.
\end{align}

To get the symmetry-reduced moment matrix, we query the symmetrized scenario object:
\begin{code:matlab}
sym_mm = sym_scenario.MomentMatrix(mm_level);
\end{code:matlab}
If the base (unreduced) moment matrix has not already been generated, it will be automatically generated at this point.
At level $1$, the unreduced matrix will be as per \cref{eq:chshMM}, and the symmetrized matrix is:
\begin{align}
\label{eq:chshMM:sym}
\mathcal{S}\!\left(M\!\left(\mathcal{X}, 1\right), \mathcal{Y}\right) := \begin{bmatrix}
\expt{\id} & 0.5\expt{\id} & 0.5\expt{\id} & 0.5\expt{\id} & 0.5\expt{\id} \\
0.5\expt{\id} & 0.5\expt{\id} & 0.25\expt{\id} & 0.375\expt{\id} + \expt{y} & 0.375\expt{\id} + \expt{y} \\
0.5\expt{\id} & 0.25\expt{\id} & \expt{a_1} & 0.375\expt{\id} + \expt{y} & 0.375\expt{\id} - \expt{y} \\
0.5\expt{\id} & 0.375\expt{\id} + \expt{y} & 0.375\expt{\id} + \expt{y} & 0.5\expt{\id} & 0.25\expt{\id} \\
0.5\expt{\id} & 0.375\expt{\id} + \expt{y} & 0.375\expt{\id} - \expt{y} & 0.25\expt{\id} & 0.5\expt{\id}
\end{bmatrix}.
\end{align}

To get the symmetry-reduced objective polynomial, we use the \code{Transform} method of the \code{SymmetrizedScenario} object:
\begin{code:matlab}
sym_CHSH_ineq = sym_scenario.Transform(CHSH_ineq);
\end{code:matlab}
This yields the polynomial:
\begin{equation}
\label{eq:CHSH:sym}
\tilde{p}_{\rm chsh} := 2 + 16\expt{y}.
\end{equation}

We now have the objects needed to model and solve the SDP.
\begin{center}
\begin{tabular}{lcl}
Using \cvx{} (see \code{cvx_chsh_symmetry.m}): &\hspace{10px}& Using \yalmip{} (see \code{yalmip_chsh_symmetry.m}):\\
\cellcolor{verylightgray} %
\begin{code:matlab}[linewidth=0.45\textwidth,backgroundcolor=,frame=]
cvx_begin sdp
    sym_scenario.cvxVars('a', 'b');
    M = sym_mm.cvx(a, b);
    a(1) == 1;
    M >= 0;
    Ob = sym_CHSH_ineq.cvx(a);
    maximize(Ob);
cvx_end
\end{code:matlab}
&&\cellcolor{verylightgray} %
\begin{code:matlab}[linewidth=0.45\textwidth,backgroundcolor=,frame=]
[a, b] = sym_scenario.yalmipVars();
M = sym_mm.yalmip(a, b);
constraints = [a(1) == 1];
constraints = [constraints, M >= 0];
Ob = sym_CHSH_ineq.yalmip(a, b);
optimize(constraints, Ob);
\end{code:matlab}
\end{tabular}
\end{center}
Both return the usual expected result of $2\sqrt{2}$.

\subsection{Imported scenario}
\label{sec:importedscenario}
The imported scenario is not a scenario by the strictest definition, as it does not contain any rules for manipulating operator strings, or directly generating moment matrices.
Instead, the imported scenario allows for the the manual input of matrices and polynomials that have been generated by other means,
 so that the other features of \moment{} (such as basis generation) can be used.

In an imported scenario, the moments themselves are the fundamental objects, as opposed to their underlying operators.
This means that features that rely on explicit algebraic relations (such as general multiplication) are unavailable.
Addition, subtraction, and complex conjugation still function.

\subsubsection{Importing moments}
In the imported scenario, strings are parsed into moments.
Let us write the $i^{\rm th}$ imported moment as $\expt{w_i}$, with the special cases $\expt{w_0}=0$ and $\expt{w_1}=\expt{\id}$.
Without further restriction, each moment is assumed to have a real and an imaginary part.

When parsing strings, if an integer along (e.g.\ \code{2}), or an integer prefixed with a hash (e.g.\ \#\code{2}) are supplied, the string is  interpreted as moment $\expt{w_2}$.
An asterisk suffix indicates the complex conjugate (e.g.\ \code{2*} parses to $\expt{w_2}^*$), and a minus sign prefix indicates negation (e.g.\ \code{-2} parses to $-\expt{w_2}$).
Arbitray real prefactors can be applied by writing them before the \code{#} (e.g.\ \code{0.5}\#\code{2} is $0.5\expt{w_2}$, \code{2.25}\#\code{3*} is $2.25\expt{w_3}^*$).
If a number with a decimal point is supplied, it is parsed into a constant, e.g.\ \code{0.25} becomes $0.25\expt{\id}$.

An imported scenario for complex moments is created:
\begin{code:matlab}
imported_scenario = ImportedScenario();
\end{code:matlab}
For a scenario with exclusively real-valued moments, one can invoke:
\begin{code:matlab}
real_scenario = ImportedScenario('real', true);
\end{code:matlab}

Suppose we define the $2\times 2$ array of strings
\begin{code:matlab}
m_str = ["1", "2";
       "3", "4"];
\end{code:matlab}

We can import this into moment with the \code{ImportMatrix} method:
\begin{code:matlab}
m = imported_scenario.ImportMatrix(m_str);
\end{code:matlab}
which will create a matrix representing the following moments:
\begin{align}
    \begin{bmatrix}
    \expt{\id} & \expt{w_2} \\ 
    \expt{w_3} & \expt{w_4}
    \end{bmatrix}.
\end{align}
Without further qualification, the moments $\expt{w_2}$, $\expt{w_3}$ and $\expt{w_4}$, will be complex, and the matrix will have no symmetry.

If it is known ahead of time that the imported matrix should be Hermitian, one could invoke:
\begin{code:matlab}
mH_str = ["1", "2";
          "2", "3"];
mH = imported_scenario.ImportHermitianMatrix(mH_str);
\end{code:matlab}
Here, because we have told \moment{} that the matrix is Hermitian, since exactly \code{"2"} appears in the top right and the transposed position in the bottom left, the imported moment $\expt{w_2}$ is inferred to be purely real (because $\expt{w_2}^* = \expt{w_2}$).

Similarly, for an entirely real scenario scenario, one could invoke:
\begin{code:matlab}
mS_str = ["1", "2";
          "2", "3"];
mS = real_scenario.ImportSymmetricMatrix(mS_str);
\end{code:matlab}

For real scenarios, the method \code{ImportSymmetricMatrix} and \code{ImportMatrix} are practically the same.
However, this function is subtly different for complex scenarios (since symmetric matrices and Hermitian matrices are not exactly the same, the former lacking a complex conjugation).
For instance, suppose (with \code{mH_str} defined as above) we call:
\begin{code:matlab}
mH2 = imported_scenario.ImportSymmetricMatrix(mH_str);
\end{code:matlab}
In this case, while $\expt{w_3}$ will be inferred to be real, no such inference can be made any more about $\expt{w_2}$ (since comparing the top-right and bottom-left simply yields the trivial identity $\expt{w_2}=\expt{w_2}$).

On the other hand, there are no side-effects to using the method \code{ImportHermitianMatrix} in a real scenario, since when all moments are a priori real, the concept of Hermitian and symmetric are identical.

\subsubsection{Example: Imported CHSH scenario}
Consider the following example input string array:
\begin{code:matlab}
mm_str = ["1", "2",  "3",  "4",   "5";
          "2", "2",  "6",  "7",   "8";
          "3", "6*", "3",  "9",   "10";
          "4", "7",  "9",  "4",   "11";
          "5", "8",  "10", "11*", "5"];
\end{code:matlab}
This is the level $1$ moment matrix of the CHSH scenario (\cref{eq:chshMM}), defined explicitly in terms of its unique moments.

To formulate an SDP using this, we first create the scenario object:
\begin{code:matlab}
imported_scenario = ImportedScenario();
\end{code:matlab}

Next, we use the \code{ImportHermitianMatrix} method to load the matrix into \moment{}:
\begin{code:matlab}
imported_mm = imported_scenario.ImportHermitianMatrix(mm_str);
\end{code:matlab}
\noindent Invocation of \code{disp(imported_scenario.Symbols)} reveals that \moment{} has automatically detected that all moments are Hermitian with the exception of ``6'' and ``11''.

Since no operators are explicitly defined in this scenario, 
 to specify a polynomial, we have to manually define it in terms of the imported symbol names.
This can be done by passing a string array into the \code{ImportPolynomial} method, with each element following the same notation as matrix in the matrix import.
We do this for the CHSH inequality (\cref{eq:CHSH:CG}) as follows:
\begin{code:matlab}
chsh_ineq = imported_scenario.ImportPolynomial(...
    ["2.0", "-4#2", "-4#4", "4#7", "4#8", "4#9", "-4#10"]);
\end{code:matlab}

\enlargethispage{2\baselineskip}
Once we have imported the moment matrix and objective function objects, we can model and solve the SDP in the usual way.
For brevity we use the \code{mtk_solve} utility function:
\begin{code:matlab}
mtk_solve(imported_mm, chsh_ineq)
\end{code:matlab}
This returns the expected result of $-2\sqrt{2}$.

\clearpage
\section{Performance benchmarks}
\label{sec:benchmark}
Although, to our knowledge, no software does exactly what \moment{} does, there are several packages available that provide overlapping functionality.
In this section, we will compare the performance of \moment{} across a variety of example problems against software that can be used for similar calculations.

The first such piece of software is {\em QETLAB}~\cite{qetlab} (Quantum Entanglement Theory Laboratory).
QETLAB is a MATLAB toolbox providing a wide range of tools for calculations involving quantum entanglement and questions of nonlocality.
Although QETLAB includes functions for other calculations, it has the most overlap with the {\em locality scenario} of \moment{}, providing functions such as \code{BellInequalityMax}, \code{BellInequalityMaxQubits} and \code{NPAHierarchy} that can optimize a Bell functional using the NPA Hierarchy.
For our tests, we use the main branch available on github~\cite{qetlabgit}, as (at the time of writing) the latest official release (v0.9) is $7$ years behind.

The second software we shall compare against is \code{ncpol2sdpa}~\cite{ncpol2sdpa} -- a python package for non-commutative polynomial optimization.
This software is the most directly comparable to \moment{}, in that it provides helpful tools and ``syntactic sugar'' for formulating a wide variety of optimization problems.
For testing, we again use Mosek as the coupled solver (using the python interface provided by Mosek themselves).
We also use an updated version of ncpol2spda~\cite{BrownNcpol2sdpa}, which incorporates several bug fixes.

For the inflation scenario, we shall test our software against \code{Inflation}~\cite{BoghiuWP23}.
This is a python library specifically for {\em quantum} inflation (currently, \moment{} is specialised to the classical case only),
 so the comparison is provided mainly for heuristic reference.

\inlineheading{Timing and other metrics.}
We will look at a few different timings for performance.
The first (and easiest to measure) is the {\em total time}: i.e.\ the time taken for the entire program to run, including the formulation, modelling and solving of the SDP.
The second (where available) is the {\em solve time} -- the time spent by the SDP solver.
Although this second metric does not measure directly the efficiency of \moment{}'s own code, 
 it provides an important reference time span, since almost all scientific applications of \moment{} will ultimately require solving an SDP.
The final time measure is the ``presolve'' time -- the time taken in the calculation by processes other than the SDP modeller and solver.
For \moment{}, this corresponds to the time spent on ``\moment{} code'' (e.g.\ generating moment matrices, translating objective functions, etc.), and is generally calculated as the total time excluding the solve time.

To minimize the influence of random system interruptions, we ran all programs ten times\footnote{Except for a few exceptionally slow cases, which we have explicitly marked in the tables.}, removed the fastest and slowest runs, then take the mean of the remaining times.
This procedure is done independently on the three different timings used.

Another metric we report is the number of constraints and variables in the generated SDP.
We retrieve this information from Mosek's presolve print-out, in particular using the ``scalarized'' column to report the number of SDP variables (otherwise, in many examples, the value will just be $1$ SDP matrix variable, and this does not reflect the increasing complexity of the programs).

\inlineheading{Benchmark system details.}
All benchmarks in this section were performed on a Dell Precision Tower 7910, equipped with a Intel Xeon E5-2623 processor with $4$ cores at $3.0$ GHz, $128$ GB of RAM, with Microsoft Windows 10 as its operating system.
We use \matlab{} version R2021a (9.10), and Python version 3.11.3 via Anaconda and the PyCharm IDE.
The \moment{} binary executables were  compiled with release--mode (``O3'') optimizations using LLVM/clang-cl version 16.0.5
.
SDPs were solved using Mosek version 10.0.40.

\subsection{Locality: Tsirelson bound for CHSH inequality via the NPA Hierarchy}

\begin{table}[bht]
\begin{center}
\begin{tabular}{|c|c|c|c|c|d{4.4}|d{4.4}|d{1.4}|}
\hline
\textbf{NPA}
&\textbf{Matrix}
&\multirow{2}{*}{\quad\textbf{Software}\quad}
&\multirow{2}{*}{\textbf{Variables}}
&\multirow{2}{*}{\textbf{Constraints}}
&\multicolumn{1}{c|}{\textbf{Total time}}
&\multicolumn{1}{c|}{\textbf{Set-up time}}
&\multicolumn{1}{c|}{\textbf{Solver time}}\\
\textbf{level}&\textbf{size}&&&
&\multicolumn{1}{c|}{\textbf{(seconds)}}
&\multicolumn{1}{c|}{\textbf{(seconds)}}
&\multicolumn{1}{c|}{\textbf{(seconds)}}\\
\hline
\multirow{2}{*}{5} & \multirow{2}{*}{61}  
& \moment{} & 1\,891 & 150 & 0.1927 & 0.04369 & 0.1485 \\
&& QETLAB  & 1\,891 & 150  & 3.078 & 2.965 & 0.112 \\
&& ncpol2sdpa & 1\,891 & 150  & 6.300 & 6.253 & 0.04701 \\
\hline
\multirow{2}{*}{6} & \multirow{2}{*}{85}  
& \moment{} & 3\,655 & 210 & 0.2392 & 0.06384 & 0.1754 \\
&& QETLAB & 3\,655 & 210 & 6.110 & 5.938 & 0.172 \\
&& ncpol2sdpa & 3\,655 & 210 & 22.71 & 22.61 &  0.08825 \\
\hline
\multirow{2}{*}{7} & \multirow{2}{*}{113}  
& \moment{} & 6\,441 & 280 & 0.3220 & 0.07307 & 0.2479 \\
&& QETLAB & 6\,441 & 280 & 11.48 & 11.26 & 0.2211 \\
&& ncpol2sdpa & 6\,441 & 280 & 83.47 & 83.28 & 0.1776 \\
\hline
\multirow{2}{*}{8} & \multirow{2}{*}{145}  
& \moment{} & 10\,585 & 360 & 0.4506 & 0.08595 & 0.3642 \\
&& QETLAB & 10\,585 & 360 & 21.41 & 21.06 & 0.3465 \\
&& ncpol2sdpa & 10\,585 & 360 & 269.6 & 269.3 & 0.3275  \\
\hline
\multirow{2}{*}{9} & \multirow{2}{*}{181}  
& \moment{} & 16\,471 & 450 & 0.6751 & 0.1048 & 0.5705 \\
&& QETLAB & 16\,471 & 450 & 41.25 & 40.63 & 0.6137 \\
&& ncpol2sdpa & 16\,471 & 450 & 943.8 & 943.2 & 0.6066  \\
\hline
\multirow{2}{*}{10} & \multirow{2}{*}{221}  
& \moment{} & 24\,531 & 550 & 1.092 & 0.1321 & 0.9604 \\
&& QETLAB & 24\,531 & 550 & 99.00 & 97.87 & 1.114 \\
&& ncpol2sdpa & 24\,531 & 550 & 3\,583 & 3\,582 & 1.103 \\
\hline
\multirow{2}{*}{11} & \multirow{2}{*}{265}  
& \moment{} & 35\,245 & 660 & 1.625 & 0.1817 & 1.441 \\
&& QETLAB & 35\,245 & 660 & 276.2 & 274.6 & 1.575 \\
\hline
\multirow{2}{*}{12} & \multirow{2}{*}{313}  
& \moment{} & 49\,141 & 780 & 2.310 & 0.2333 & 2.076 \\
&& QETLAB & 49\,141 & 780 & 1\,071 & 1\,069 & 2.266 \\
\hline
\multirow{2}{*}{13} & \multirow{2}{*}{365}
& \moment{} & 66\,795 & 910 & 2.859 & 0.3274 & 2.532 \\
&& QETLAB & 66\,795 & 910 & 4\,653 & 4\,650 & 2.854 \\
\hline
\multirow{1}{*}{14} & \multirow{1}{*}{421}
& \moment{} & 88\,831 & 1\,050 & 3.919 & 0.4182 & 3.501 \\
\hline
\multirow{1}{*}{15} & \multirow{1}{*}{481}
& \moment{} & 115\,921 & 1\,200 & 5.686 & 0.5376 & 5.146 \\
\hline
\multirow{1}{*}{16} & \multirow{1}{*}{545}
& \moment{} & 148\,785 & 1\,359 & 9.936 & 0.6983 & 9.236 \\
\hline
\end{tabular}
\end{center}
\caption{
\label{table:CHSH}
\caphead{CHSH benchmark results.}
}
\end{table}

\begin{figure}[h!]
\centering
\begin{tikzpicture}
\begin{axis}[%
    scale only axis,
    ymode = log,
    xmin=5,
    xmax=16,
    grid=major,
    xlabel={NPA Hierarchy Level},
    ylabel = {Total time (s)},
    axis background/.style={fill=white},
    legend style={at={(0.975,0.225)},legend cell align=left, align=left, draw=white!15!black}
]
\addplot [blue, mark color=blue, mark=*] table[col sep=space]{
  5.0     0.1927
  6.0     0.2392
  7.0     0.3220
  8.0     0.4506
  9.0     0.6751
 10.0     1.092
 11.0     1.625
 12.0     2.310
 13.0     2.859 
 14.0     3.919
 15.0     5.686 
 16.0     9.936
 }; \addlegendentry{Moment}
\addplot [olive, mark color=olive, mark=square*] table[col sep=space]{
  5.0      3.078 
  6.0      6.110 
  7.0     11.48
  8.0     21.41
  9.0     41.25
 10.0     99.00
 11.0    276.2 
 12.0   1071
 13.0   4653
 }; \addlegendentry{QETLAB}
\addplot [red, mark color=red, mark=triangle*, mark size=3] table[col sep=space]{
  5.0      6.300
  6.0     22.71 
  7.0     83.47
  8.0    269.6
  9.0    943.8
 10.0   3583
 }; \addlegendentry{ncpol2sdpa}
\end{axis}
\end{tikzpicture}
\caption{\caphead{CHSH benchmark results (total time).}
Total run time in seconds (logarithmic scale) as a function of NPA Hierarchy level for solving the Tsirelson bound on the CHSH inequality. Corresponding data is in \cref{table:CHSH}.
}
 \label{fig:CHSH}
\end{figure}
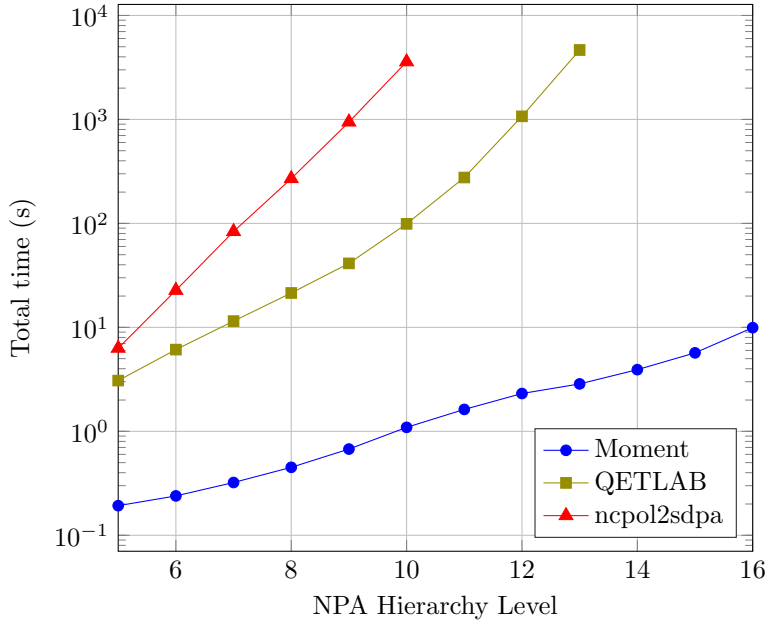 

In \cref{table:CHSH} and \cref{fig:CHSH}, we compare the performance of calculating the Tsirelson bound on the CHSH inequality via the NPA hierarchy.

For \moment{}, we will use code similar to that in the first example of \cref{sec:localityscenario}.
For QETLAB, we will solve this function using the \code{BellInequalityMax} function.
For ncpol2sdpa, we follow the example for the documentation -- namely, construct a \code{Probability} class, define the objective using the \code{define_objective_with_I} function (analogue to \moment{}'s \code{CGTensor} function), and generate the SDP with the \code{SdpRelaxation} class.
Since QETLAB uses \cvx{}, for direct comparison we will use \cvx{} with \moment{} (in both cases, specifically version 2.2).
In all runs, the expected value of approximately $2\sqrt{2}$ is returned.

First, the number of SDP variables and constraints reported by Mosek is the same with \moment{} as it is with QETLAB and ncpol2sdpa.
Likewise, the solve time is approximately equal between the packages.
This suggests, after all pre-processing is finished, essentially the same SDP is being solved by Mosek.
There is a slight advantage (about 50-100\ ms) for ncpol2sdpa, that is more significant for the lower levels.
This may be due to slight differences in how the timing is reported between Python and MATLAB, or a minor advantage in Mosek's python interface over its MATLAB one.

Where there is a huge performance advantage for \moment{} over QETLAB and ncpol2sdpa is in the set-up stage -- crucially, in generating the moment matrix.
This is a {\em diverging} performance difference (i.e.\ not just a constant factor, as might be expected from using a compiled language like \cpp{} instead of an interpreted language like \matlab{}), demonstrating an algorithmic advantage in using \moment{}.
In particular, the largest practical moment matrix to generate with ncpol2sdpa was level $10$ taking almost an hour, which \moment{} could generate in $\frac{1}{10}$ of a second;
and with QETLAB, the largest practical moment matrix was level $13$ taking an hour and seventeen minutes, which \moment{} could generate in $0.3$ seconds.
Level $16$ is the maximum supported by \moment{}, as with $4$ fundamental operators, words of length $32$ are the longest that can be hashed by a $64$-bit integer. 
Nonetheless, even at this level, the moment matrix was produced in a moderately quick time of about $0.7$ seconds.
 
We conclude that for \moment{}, {\em solving} the SDP is the bottleneck -- whereas for the other software, generating the SDP relaxation takes the vast majority of computational time.

\subsection{Locality: Tsirelson bound for I3322 inequality via the NPA Hierarchy}

\begin{table}
\begin{center}
\begin{tabular}{|c|c|c|c|c|d{5.6}|d{3.5}|d{3.5}|}
\hline
\textbf{NPA}
&\textbf{Matrix}
&\multirow{2}{*}{\quad\textbf{Software}\quad}
&\multirow{2}{*}{\textbf{Variables}}
&\multirow{2}{*}{\textbf{Constraints}}
&\multicolumn{1}{c|}{\textbf{Total time}}
&\multicolumn{1}{c|}{\textbf{Set-up time}}
&\multicolumn{1}{c|}{\textbf{Solver time}}\\
\textbf{level}&\textbf{size}&&&
&\multicolumn{1}{c|}{\textbf{(seconds)}}
&\multicolumn{1}{c|}{\textbf{(seconds)}}
&\multicolumn{1}{c|}{\textbf{(seconds)}}\\
\hline
\multirow{3}{*}{1} & \multirow{3}{*}{7}  
& \moment{} & 28 & 7 & 0.2002 & 0.08678 & 0.1135 \\
&& QETLAB & 28 & 7 & 0.1742 & 0.08717 & 0.08712 \\
&& ncpol2sdpa & 27 & 21 & 0.03373 & 0.01929 & 0.01444 \\
\hline
\multirow{3}{*}{2} & \multirow{3}{*}{28}  
& \moment{} & 406 & 153 & 0.215 & 0.08593 & 0.1291 \\
&& QETLAB & 406 & 153 & 0.5764 & 0.4769 & 0.09959 \\
&& ncpol2sdpa & 406 & 153 & 0.4573 & 0.4289 & 0.02841 \\
\hline
\multirow{3}{*}{3} & \multirow{3}{*}{88}  
& \moment{} & 3\,916 & 867 & 0.5735 & 0.1452 & 0.4283 \\
&& QETLAB & 3\,916 & 867 & 5.614 & 5.230 & 0.3837 \\
&& ncpol2sdpa & 3\,915 & 867 & 9.883 & 9.498 & 0.3849 \\
\hline
\multirow{3}{*}{4} & \multirow{3}{*}{244}  
& \moment{} & 29\,890 & 4\,491 & 8.158 & 0.3903 & 7.759 \\
&& QETLAB & 29\,890 & 4\,491 & 68.01 & 59.60 & 8.416 \\
&& ncpol2sdpa & 29\,890 & 4\,491 & 162.1 & 151.9 & 10.19 \\
\hline
\multirow{3}{*}{5} & \multirow{3}{*}{628}  
& \moment{} &197\,506 & 22\,179 & 406.9 & 2.177 & 404.8 \\
&& QETLAB & 197\,506 & 22\,179 & 2\,282 & 1\,837 & 444.9 \\
&& ncpol2sdpa & 197\,506 & 22\,179 & 2\,294 & 1\,873 & 420.5 \\
\hline
6 & 1\,540 & \moment{} & 1\,186\,570  & 106\,084  & \multicolumn{1}{c|}{---} & 15.26 & \multicolumn{1}{c|}{---} \\
\hline
7 & 3\,652 & \moment{} & --- & --- & \multicolumn{1}{c|}{---} & 270.9 & \multicolumn{1}{c|}{---} \\
\hline
\end{tabular}
\end{center}
\caption{
\label{table:I3322}
\caphead{I3322 benchmark results.}
For levels 6 and 7, we measure the set-up time only, and for \moment{} only.
This is because  the other software did not complete the computation within a reasonable time-frame.
Indeed, level 6 is at the very limit of what is practically solvable with Mosek on our benchmark system, and level 7 exceeded the system's available resources.
}
\end{table}

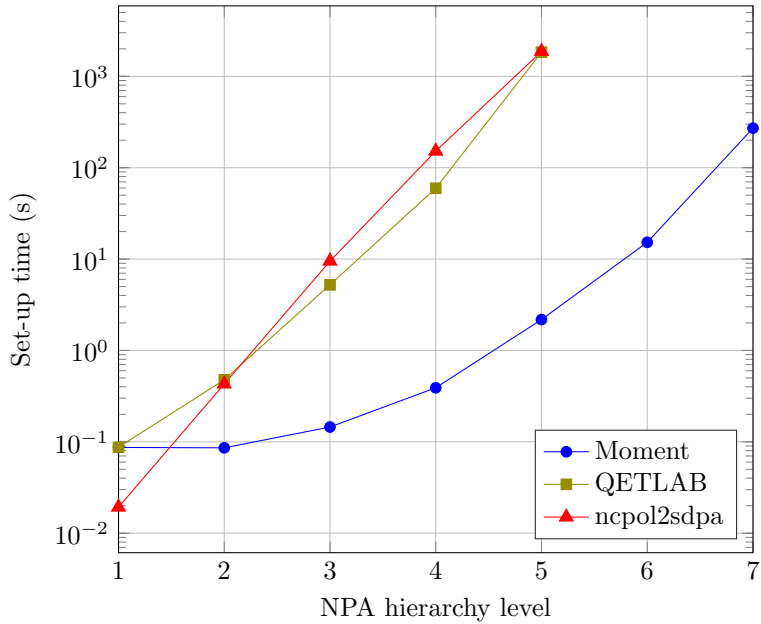
\begin{figure}[h!]
\centering
\begin{tikzpicture}
\begin{axis}[%
    scale only axis,
    ymode = log,
    xmin=1,
    xmax=7,
    grid=major,
    xlabel={NPA hierarchy level},
    ylabel = {Set-up time (s)},
    axis background/.style={fill=white},
    legend style={at={(0.975,0.225)},legend cell align=left, align=left, draw=white!15!black}
]
\addplot [blue, mark color=blue, mark=*] table[col sep=space]{
  1.0     0.08678
  2.0     0.08593
  3.0     0.1452
  4.0     0.3903
  5.0     2.177
  6.0    15.26
  7.0   270.9
}; \addlegendentry{Moment}
\addplot [olive, mark color=olive, mark=square*] table[col sep=space]{
  1.0     0.08717
  2.0     0.4769
  3.0     5.230
  4.0    59.60
  5.0   1837 
}; \addlegendentry{QETLAB}
\addplot [red, mark color=red, mark=triangle*, mark size=3] table[col sep=space]{
  1.0      0.01929
  2.0      0.4289
  3.0      9.498
  4.0   151.9 
  5.0  1873 
 }; \addlegendentry{ncpol2sdpa}
\end{axis}
\end{tikzpicture}
\caption{\caphead{I3322 benchmark results (set-up time).}
Problem set-up time in seconds (logarithmic scale) as a function of NPA Hierarchy level for generating an SDP to bound the Tsirelson bound for the I3322 inequality. Corresponding data is in \cref{table:I3322}.
}
 \label{fig:I3322}
\end{figure} 

Let us consider the I3322 Bell inequality (as previously discussed in \cref{sec:i3322}).
Here, the size of the moment matrix (and hence the complexity of the problem) grows exponentially with the hierarchy level. 
The results, comparing \moment{}, QETLAB and ncpol2sdpa are presented in \cref{table:I3322} and \cref{fig:I3322}.

For NPA level $2$ and higher, the same number of SDP variables and constraints are obtained, suggesting that the same SDP is being generated by all three softwares.
An anomaly occurred with the level $1$ matrix of ncpol2sdpa, causing it to have {\em more} constraints (here \moment{} and QETLAB are still in accordance).
In this case, the numerical answers differed by about $10^{-8}$ (on a value close to $5.5$).

We remark that for $N=1$, ncpol2sdpa was about $0.06$ seconds faster than both \moment{} and QETLAB. Since this is a very simple problem, this might due to factors such as a relatively faster speed of interpreting python vs.\ interpreting MATLAB.

With every library except \moment{}, the set-up time is the dominant portion of total computational time.
Conversely, \moment{}'s set-up time is considerably shorter; for example, at level $5$, the problem set-up time accounted for only $0.5\%$ of the total computational time.

\subsection{Algebraic: Non-commuting polynomial optimization via the PNA Hierarchy}
We now test an algebraic example with localizing matrices  -- namely, the first example from \citet{PironioNA10}, discussed in \cref{sec:algebraicscenario} (\cref{eq:PNAExact}).

We will contrast \moment{}'s performance against the implementation of the same SDP relaxation in the \code{readme.md} of ncpol2spda~\cite{BrownNcpol2sdpa} (also described in Section 5 of \citet{ncpol2sdpa}).
Probing the output of the \code{SdpRelaxation} class \code{write_to_file} function, we see that ncpol2sdpa produces a localizing matrix of $M-1$ for moment matrix level $M$ (there is no option to configure this in ncpol2spda), so we do the same with \moment{}.

\begin{figure}[h!]
\centering
\begin{tikzpicture}
\begin{axis}[%
    scale only axis,
    ymode = log,
    xmin=1,
    xmax=10,
    ymin=0.001,
    ymax=1000,
    grid=major,
    xlabel={NPA hierarchy level},
    ylabel = {Set-up time (s)},
    axis background/.style={fill=white},
    legend style={at={(0.975,0.225)},legend cell align=left, align=left, draw=white!15!black}
]
\addplot [blue, mark color=blue, mark=*] table[col sep=space]{
  1.0     0.02791
  2.0     0.02775 
  3.0     0.02913 
  4.0     0.03143
  5.0     0.03773 
  6.0     0.05447
  7.0     0.1142
  8.0     0.276
  9.0     0.6742 
 10.0     1.979
}; \addlegendentry{Moment}
\addplot [red, mark color=red, mark=triangle*, mark size=3] table[col sep=space]{
  1.0      0.004885
  2.0      0.008567 
  3.0      0.02447 
  4.0      0.08730 
  5.0      0.6103
  6.0      1.940
  7.0     11.71
  8.0     34.68 
  9.0     99.56
 10.0    286.7
 }; \addlegendentry{ncpol2sdpa}
\end{axis}
\end{tikzpicture}
\caption{\caphead{PNA non-commutative benchmark results (set-up time).}
Problem set-up time in seconds (logarithmic scale) as a function of NPA Hierarchy level for generating an SDP relaxation of \cref{eq:PNAExact}. Corresponding data is in \cref{table:PNA}.
}
 \label{fig:PNA}
\end{figure}
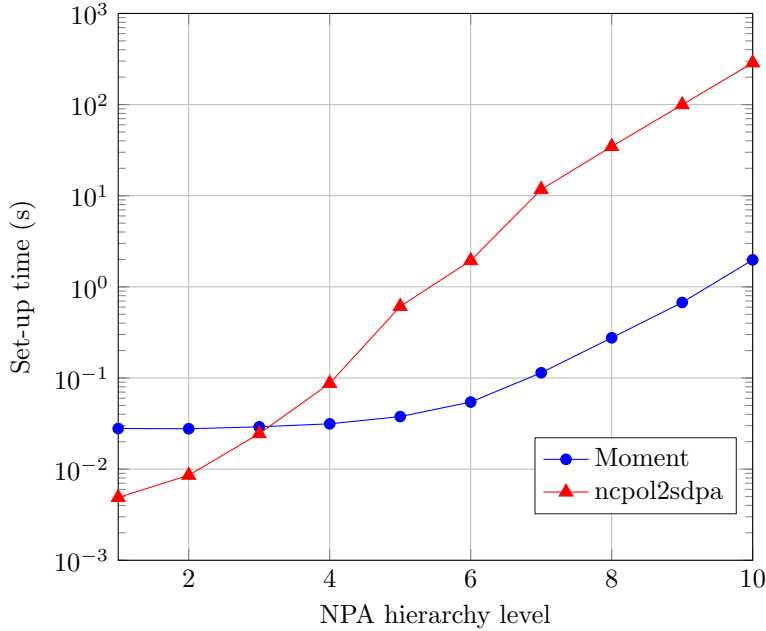

The results of this benchmark are presented in \cref{table:PNA} and \cref{fig:PNA}.

\begin{table}[hbt]
\begin{center}
\begin{tabular}{|c|c|c|c|c|d{3.5}|d{3.6}|d{3.5}|}
\hline
\textbf{PNA}
&\textbf{Matrix}
&\multirow{2}{*}{\quad\textbf{Software}\quad}
&\multirow{2}{*}{\textbf{Variables}}
&\multirow{2}{*}{\textbf{Constraints}}
&\multicolumn{1}{c|}{\textbf{Total time}}
&\multicolumn{1}{c|}{\textbf{Set-up time}}
&\multicolumn{1}{c|}{\textbf{Solver time}}\\
\textbf{level}&\textbf{size}&&&
&\multicolumn{1}{c|}{\textbf{(seconds)}}
&\multicolumn{1}{c|}{\textbf{(seconds)}}
&\multicolumn{1}{c|}{\textbf{(seconds)}}\\
\hline
\multirow{2}{*}{1} & \multirow{2}{*}{3}  
& \moment{} & 9 & 5 & 0.113 & 0.02791 & 0.08475 \\
&& ncpol2sdpa & 10 & 4 & 0.0322 & 0.004885 & 0.02752 \\
\hline
\multirow{2}{*}{2} & \multirow{2}{*}{6}  
& \moment{} & 29 & 14 & 0.1132 & 0.02775 & 0.08543 \\
&& ncpol2sdpa & 45 & 13  & 0.05956 & 0.008567 & 0.05100 \\
\hline
\multirow{2}{*}{3} & \multirow{2}{*}{11}  
& \moment{} & 89 & 35 & 0.1160 & 0.02913 & 0.08651 \\
&& ncpol2sdpa & 153 & 34  & 0.1163 & 0.02447 & 0.09031 \\
\hline
\multirow{2}{*}{4} & \multirow{2}{*}{19}  
& \moment{} & 258 & 86 & 0.1242 & 0.03143 & 0.09221 \\
&& ncpol2sdpa & 465 & 85 & 0.2844 & 0.08730 & 0.1964 \\
\hline
\multirow{2}{*}{5} & \multirow{2}{*}{32}  
& \moment{} & 720 & 213 & 0.1414 & 0.03773 & 0.1035 \\
&& ncpol2sdpa & 1\,326 &  212 & 1.099 & 0.6103 & 0.4883 \\
\hline
\multirow{2}{*}{6} & \multirow{2}{*}{53}  
& \moment{} & 1\,961 & 535  & 0.2082 & 0.05447 & 0.1524 \\
&& ncpol2sdpa & 3\,655 & 534 & 3.293 & 1.940 & 1.411 \\
\hline
\multirow{2}{*}{7} & \multirow{2}{*}{87}  
& \moment{} & 5\,261 & 1\,361 & 0.6400 & 0.1142 & 0.5258 \\
&& ncpol2sdpa & 9\,870 & 1\,360 & 16.00 & 11.71 & 4.286 \\
\hline
\multirow{2}{*}{8} & \multirow{2}{*}{142}  
& \moment{} & 13\,983 & 3\,496 & 3.625 & 0.276 & 3.347 \\
&& ncpol2sdpa & 26\,335 & 3\,495 & 51.80 & 34.68 & 17.11 \\
\hline
\multirow{2}{*}{9} & \multirow{2}{*}{231}  
& \moment{} & 36\,951 & 9\,041 & 31.20 & 0.6742 & 30.52 \\
&& ncpol2sdpa & 69\,751 & 9\,040 & 136.7 & 99.56 & 37.17 \\
\hline
\multirow{2}{*}{10} & \multirow{2}{*}{375}  
& \moment{} & 97\,298 & 23\,486 & 471.6 & 1.979 & 469.6 \\
&& ncpol2sdpa & 183\,921 & 23\,485 & 796.6 & 286.7 & 510.2 \\
\hline
\end{tabular}
\end{center}
\caption{
\label{table:PNA}
\caphead{PNA non-commutative benchmark results.}
Matrix size refers to the size of the moment matrix; the localizing matrix is the size of moment matrix of the level below (with ``level $0$'' being $1\times1$).
}
\end{table}

Throughout there are approximately twice as many scalar variables for ncpol2spda than for \moment{}.
This is likely because ncpol2sdpa is also generating the complex basis elements in its relaxation, whereas with \moment{} it is possible to exploit the symmetry of the problem and ignore the complex terms before passing to \yalmip{}.
While this explains the discrepancy in {\em solve time} between the two, there is no major advantage from this for \moment{} during its set-up time, as the full moment and localizing matrices must be generated in any case.

For the low levels of Hierarchy ($M\leq 3$), ncpol2sda has an advantage of a few milliseconds.
These calculations are so quick that this may be a constant factor advantage of Python over MATLAB, 
 or some other small difference in overheads from required libraries.
However, for larger matrix sizes \moment{} starts to exhibit significantly favorable performance (taking around $2$ seconds to calculate level $10$, which takes ncpol2sdpa almost five minutes).

\subsection{Algebraic: Brown--Fawzi--Fawzi entropy calculation.}

Let us measure the performance of conditional Entropy calculations~\cite{BrownFF21di}, as per \cref{sec:algebraicscenario}.
We will compare \moment{}'s implementation against Brown's implementation using ncpol2sdpa~\cite{BrownDIRates}. 
In particular, to be directly comparable, we use a modified version of test case in the script \code{chsh_local.py} with the parameters \code{M = 4} and \code{KEEP_M = 0}. This results in 7 SDPs (for 8 sample points in the Gauss-Radau quadrature, where the final point is bounded trivially). 
Both \moment{}'s and Brown's implementation generate one moment matrix, and then cycle through different objective functions to calculate the final value.
We also set the \code{extramonomials} flag in the \code{get_relaxation} method to zero, so as to directly compare performance for a given NPA level\footnote{These monomials, which appear as terms in the objective function, are generated in the level 2 moment matrix, but not the level 1 moment matrix.}.
For the solve time, we use the sum of solution times reported by ncpol2sdpa; and the presolve time the constitutes the remaining portion of the total time.

\begin{table}[bht]
\begin{center}
\begin{tabular}{|c|c|c|c|c|d{3.3}|d{4.4}|d{3.3}|}
\hline
\textbf{NPA}
&\textbf{Matrix}
&\multirow{2}{*}{\quad\textbf{Software}\quad}
&\multirow{2}{*}{\textbf{Variables}}
&\multirow{2}{*}{\textbf{Constraints}}
&\multicolumn{1}{c|}{\textbf{Total time}}
&\multicolumn{1}{c|}{\textbf{Set-up time}}
&\multicolumn{1}{c|}{\textbf{Solver time}}\\
\textbf{level}&\textbf{size}&&&
&\multicolumn{1}{c|}{\textbf{(seconds)}}
&\multicolumn{1}{c|}{\textbf{(seconds)}}
&\multicolumn{1}{c|}{\textbf{(seconds)}}\\
\hline
\multirow{2}{*}{2} & \multirow{2}{*}{49}  
& \moment{} & 1\,225 & 499  & 2.695 & 0.8397 & 1.858 \\
&& ncpol2sdpa & 1\,275 & 498 & 3.189 & 1.921 & 1.269  \\
\hline
\multirow{2}{*}{3} & \multirow{2}{*}{221}  
& \moment{} & 24\,531 & 7\,729 & 217.4 & 1.162 & 215.3 \\
&& ncpol2sdpa & 24\,753 & 7\,728 & 336.5 & 79.50 & 256.9  \\
\hline
\multirow{2}{*}{4} & \multirow{2}{*}{925}  
& \moment{} & 428\,275 & 122\,017 & \multicolumn{1}{c|}{---}  & 9.791 & \multicolumn{1}{c|}{---}  \\
&& ncpol2sdpa & --- & --- & \multicolumn{1}{c|}{---} & 2\,587.9 &  \multicolumn{1}{c|}{---} \\
\hline
\end{tabular}
\end{center}
\caption{
\label{table:BFF}
\caphead{Brown-Fawzi-Fawzi entropy calculation benchmark.}
The NPA level $4$ case requires more memory to solve than our benchmark system has, so only the set-up time is included.
}
\end{table}

The results are reported in \cref{table:BFF}.
Again, we can observe  \moment{}'s performance advantage over ncpol2sdpa in the setup times, with the speed-up becoming considerably noticeable for the larger matrices (with about $70\times$ speed advantage for level $3$ and about $250\times$ speed advantage at level $4$).

\subsection{Inflation: Classical 4-outcome triangle}
Let us now test the inflation scenario against the python package \code{Inflation}~\cite{BoghiuWP23}.
In particular, we shall run a feasibility test of the classical triangle scenario with four outcomes for each observable.
We shall impose the distribution $P(000)=P(111)=P(222)=P(333)=\frac{1}{4}$ on the joint measurement of these observables.
These statistics should be provably infeasible after inflation (levels $2$ and $3$), but are feasible for the uninflated moment matrices.
Because the imposed probability distribution is tripartite, we start at level $2$ of the NPA hierarchy (the level $1$ moment matrix does not contain all the terms required to impose to the probability distribution).

For \moment{}, we execute code similar to that in \cref{sec:inflationscenario}. 
For \code{Inflation}, we use code similar to the example in their \code{readme.md}, but making sure to pass the commuting flag to the \code{InflationSDP} object.

\begin{table}[bht]
\begin{center}
\begin{tabular}{|c|c|c|c|c|c|d{4.4}|d{7.4}|d{4.6}|}
\hline
\textbf{Inflation}
&\textbf{NPA}
&\textbf{Matrix}
&\multirow{2}{*}{\quad\textbf{Software}\quad}
&\multirow{2}{*}{\textbf{Variables}}
&\multirow{2}{*}{\textbf{Constraints}}
&\multicolumn{1}{c|}{\textbf{Total time}}
&\multicolumn{1}{c|}{\textbf{Set-up time}}
&\multicolumn{1}{c|}{\textbf{Solver time}}\\
\textbf{level}&\textbf{level}&\textbf{size}&&&
&\multicolumn{1}{c|}{\textbf{(seconds)}}
&\multicolumn{1}{c|}{\textbf{(seconds)}}
&\multicolumn{1}{c|}{\textbf{(seconds)}}\\
\hline
\multirow{2}{*}{1} & \multirow{2}{*}{2} & \multirow{2}{*}{37}  
& \moment{} & 703 & 0 & 0.2559 & 0.1289 & 0.1271 \\
&&& Inflation & 704 & 0 & 0.7367 & 0.7334 & 0.003410 \\
\hline
\multirow{2}{*}{1} & \multirow{2}{*}{3} & \multirow{2}{*}{64}  
& \moment{} & 2080 & 0 & 0.2172 & 0.1014 & 0.1159 \\
&&& Inflation & 2081 & 0 & 1.892 & 1.883 & 0.009575 \\
\hline
\multirow{2}{*}{2} & \multirow{2}{*}{2} & \multirow{2}{*}{631}  
& \moment{} & 199\,396 & 8\,073 & 24.02 & 1.555 & 22.44 \\
&&& Inflation & 204\,724  & 5\,328 & 121.2 & 99.45 & 21.73 \\
\hline
\multirow{2}{*}{2} & \multirow{2}{*}{3} & \multirow{2}{*}{6\,571}  
& \moment{} & 21\,592\,306  & 330\,534 & \multicolumn{1}{c|}{---}  & 385.6 & \multicolumn{1}{c|}{---} \\
&&& Inflation & --- & --- & \multicolumn{1}{c|}{---} & 22\,400.+ & \multicolumn{1}{c|}{---} \\
\hline
\multirow{2}{*}{3} & \multirow{2}{*}{2} & \multirow{2}{*}{3\,241}  
& \moment{} & 5\,253\,661 & 15\,282 & 1\,673 & 21.75 & 1\,651 \\
&&& ~Inflation$^{*}$ & 5\,266\,105 & 12\,444  & 7\,589 & 2\,436 & 5\,153 \\
\hline
\end{tabular}
\end{center}
\caption{
\label{table:Inflation}
\caphead{$4$-outcome triangle inflation scenario benchmark.}
Inflation level $2$ NPA level $3$ is too memory--intensive for the benchmark machine to solve with Mosek. 
Moreover, with \code{Inflation}, a memory bottleneck was reached for these settings after around $22\,400$ seconds, before the final stage of the set-up completed, in the \code{solveSDP_MosekFUSION} function -- though we anticipate on hardware with more RAM, this stage might complete.
}
\end{table}

The results of the benchmark are in \cref{table:Inflation}.
The difference in the number of variables between \code{Inflation} and \moment{} comes from additional scalar variables introduced in \code{Inflation}'s formulation of the SDP not present in \moment{}. The scalarized number of SDP variables (and the size of the moment matrix, etc.) was the same between the two packages, for all inflation and NPA levels.

Unlike most of the other software we tested, \code{Inflation} can generate SDPs in a shorter time than it takes to solve them.
As such, it can surely solve every inflation problem that \moment{} can (and, due to its specialization, can also solve other {\em quantum} inflation problems that are currently not supported in \moment{}).
Nonetheless, for classical problems \moment{} is faster -- in the larger cases by two orders of magnitude.

\section{Discussion and outlook}
\label{sec:discussion}
In this article, we have introduced \moment{} and shown how it can be used to formulate SDPs for various optimization problems in quantum information theory and beyond.
We have measured its performance, and seen that at the limits of modern hardware, it removes the computational bottleneck in the set-up stage of SDPs for many scientific calculations.
In comparison with similar software, speed-ups of up to $4$ orders of magnitude were observed, allowing the formulation of SDPs that would otherwise take hours (or not complete at all) to be computed in seconds.
Moreover, \moment{} has been formulated as a general purpose library, making minimal assumptions about the form of the optimization problems to which it can be applied.
For these reasons, we are pleased to present \moment{} to the general scientific community.

Although \moment{} can already be used to address complicated problems~\cite{AraujoKGVM23,AraujoGN24},
 we anticipate that a few general features and optimizations can still be made -- and that the nature of scientific research will always invite specializations (i.e.\ new ``scenarios'') for particularly tricky tasks.
Additional features could include partial NPA hierarchy levels (e.g.\ for matrices like `2+AAB' in the locality setting, other general groupings of operators, or functionality akin to the \code{extramonomials} parameter in ncpol2sdpa).
Another optimization would be to improve memory management for large symbolic matrices (preliminary profiling indicates that up to $30-40\%$ of runtime is spent {\em releasing} piece-wise allocated memory for large matrices, suggesting a significant benefit in switching to arena memory allocation~\cite{Hanson90}).

Currently, \moment{} is used to formulate optimizations where the moments are the SDP's variables.
However, some problems are better solved via the dual of this: optimizing over polynomial expressions whose indeterminates are moments (usually explicitly in a ``sum of squares'' form), where the coefficients of these polynomials are the SDP's variables.
An extension of \moment{} might then be used to facilitate the generation of such dual programs  from \moment{} (noting that \moment{} can already calculate the algebraic equivalences between differently expressed moments).

Solving an SDP with \moment{} currently requires the use of an SDP modeller (\cvx{} or \yalmip{}).
The richness of \moment{}'s features means the role of these modellers appears somewhat minimal,
 and it is thus natural to ask whether dependency on this software could be eliminated entirely.
Our conclusion, however, is that this would be undesirable.
Namely, the SDP modellers we employ essentially function as MATLAB ``drivers'' for the solver that ultimately provides numerical solutions to the problems.
Thus, as it is now, \moment{} is entirely agnostic of the solver used.
If we eliminated modeller dependency, we would need to add support separately for every single solver that one would want to use with moment.
This would add many more dependencies, and would violate the programming principle of separation of concerns.

Finally, we remark that our choice of MATLAB for this project was born out of our familiarity with this language, and its wide usage among quantum physicists in the context of semi-definite programming (especially with \yalmip{} or \cvx{}).
However, our extensive usage and deep inspection of  MATLAB during development of this software has often lead us to question whether we should have adopted another language (e.g.\ for performance, for better/more consistent language features, or simply to avoid relying on proprietary software).
Natural alternative languages would be {\em Python} or {\em Julia}~\cite{Julia17}.
While porting \moment{} to these languages would not be trivial,
 we remark that a significant core part\footnote{Namely the library \code{lib_moment}, which contains all the algorithms and a significant part of the scenario architecture.} of the \cpp{} source code is entirely independent of MATLAB, and could be used as a significantly advanced starting point for future adaptations.

\hidetoc
\section*{Note on version}
This document has been written for \moment{} version \version{}, which may be found with pre-compiled binaries at 
\url{https://github.com/ajpgarner/moment/releases/tag/v0.9.0-beta}.
An up-to-date version of the software (with compilation and installation instructions) may be found at \url{https://github.com/ajpgarner/moment/}.

\section*{Acknowledgments}
We are grateful for comments from and conversations with Paolo~Abiuso, Peter~Brown, Jonathan~Clark, Jan~Mandrysch, Miguel~Navascu\'es, David~Trillo, and Mirjam~Weilenmann.
We acknowledge funding from the FWF stand-alone project P 35509-N. 
The research of M.A. was supported by the European Union--Next Generation UE/MICIU/Plan de Recuperación, Transformación y Resiliencia/Junta de Castilla y León.

\restoretoc
\bibliography{moment}

\clearpage
\appendix
\section*{Appendix}
\hidetoc

\section{Algorithms}
\label{sec:algorithm}

%
%
\subsection{Equality constraints and polynomial substitution rules on moments}
\label{app:PolySubRule}

As described in \cref{sec:eqconst}, \moment{} provides a method for imposing polynomial equality constraints on moments, by way of rewriting expressions such that the constraints are satisfied.
A rewrite rule is essentially a ``find and replace'' instruction consisting of a LHS pattern that is checked against an expression, and if a match is found, that part of the expression is replaced by the RHS.

First, we remark on the ordering of moments and of polynomials thereof.
For two moments, their symbol order is almost shortlex order on the underlying operator strings, and in the special case where all moments are real-valued, it is exactly this.
In the general complex case, rather than evaluating $a \leq_{\rm shortlex} b$, we first evaluate $a \leq_{\rm sym} b := \min(a, \conj{a}) \leq_{\rm shortlex} \min(b, \conj{b})$ (where minimization is also by shortlex comparison), and only then tie-break by shortlex.
The effect of this is that moments that are complex conjugates of each other are consecutive in the ordering, with the lower-shortlex-valued one appearing first.
We also define the empty polynomial $0$ such that $0 \leq_{\rm sym} a$ for all $a$.

The expectation values of polynomials of operators, which are essentially linear combinations of moments, can then be written in a normalized form, in descending order, with like terms gathered.
Two polynomials can then be (partially) ordered by first applying $\leq_{\rm sym}$ to the first term, and tie-breaking with the second term, and so forth\footnote{A stricter order could be imposed by also tie-breaking monomials on the same indeterminates by some arbitrary ordering on their coefficient, e.g.\ first the real part, and then the imaginary part. The algorithms in \moment{} do not require this level of ordering, so we do not implement it.}.

\inlineheading{Orientation of polynomials.}
Let us remark on the {\em orientation} polynomials into rewrite rules.
Consider a finite length polynomial $p= c_1 \expt{m_1} + \ldots + c_N \expt{m_N}$, whose coefficients are $c_i \in \comp{}\setminus 0$, and whose indeterminates are the moments $\{m_i\}_i$; and let $p$ be written in the normalized order described above.
Then, the oriented rule is
\begin{equation}
\label{eq:orient}
    r: \expt{m_1} \mapsto -\frac{1}{c_1} \sum_{j=2}^N c_j \expt{m_j} 
\end{equation}

A polynomial $q$ is {\em reduced} by a rule $r$ by comparing the moment of each term in the $q$ to the monomial defining the LHS of the rule.
If these are the same, a new polynomial is produced by substituting the LHS with the RHS (taking into account the coefficient in front of the matched term in $q$).
A subtlety occurs in scenarios where complex (or imaginary) moments can appear: namely, each term in the polynomial also compared with the complex conjugate moment of the rule's LHS, and if matched, the rule's RHS is substituted into the expression.
Essentially, a single moment substitution rule $r$ enforces both a constraint and the complex conjugate of that constraint.

The reader who has read too much of this document and thought about it for too long will identify that there will be a problem here with orientation if $\expt{m_2} = \expt{m_1}^*$.
As a pathological example: consider $p = \expt{x^*} - 2\expt{x} - 1$, which na\"ively orients to $r: \expt{x^*} \mapsto 2\expt{x} + 1$.
Suppose we apply $r$ to the monomial $q = \expt{x^*}$.
This will yield $2\expt{x}+1$, which we can apply $r$ to again (in conjugate form) to yield $4\expt{x^*} + 3$, and then again to yield $8\expt{x} + 7$, and so on, ad infinitum.
On the other hand, there is nothing mathematically wrong with the constraint $\expt{x^*} = 2\expt{x} + 1$; indeed, by considering the linear system of equations of this and its conjugate $\expt{x} = 2\expt{x^*} + 1$, we can quickly deduce that $\expt{x} = \expt{x^*} = -\frac{2}{3}$.

Resolving this tension via such straightforward linear algebra works in most cases.
Consider polynomial $p$ with $\expt{m_1}=\expt{m_2}^*$, 
 written as $p = c_1 \expt{w_1} + c_2 \expt{\conj{w_1}} + c_3 \expt{w_3} + \ldots$.
Its conjugate can be written $\conj{p} = \conj{c_2} \expt{w_1} + \conj{c_1} \expt{\conj{w_1}} + \conj{c_3}\expt{\conj{w_3}} + \ldots$.
Then, we can take the linear combination:
\begin{equation}
\label{eq:reorient}
 \frac{1}{c_1} p - \frac{1}{\conj{c_2}}\conj{p} = (1 - 1) \expt{m_1} + \left(\frac{c_2}{c_1} - \frac{\conj{c_1}}{\conj{c_2}}\right) \expt{m_2} + \sum_{i=3}^N \left(\frac{c_i}{c_1}\expt{w_i} + \frac{\conj{c_i}}{\conj{c_2}}\expt{\conj{w_i}}\right),
\end{equation}
and orient the resulting polynomial whose leading term is now in $\expt{m_2}$ into the rule $r$:
\begin{equation}
\label{eq:reorientation}
r: \expt{m_2} \mapsto - \dfrac{c_1 \conj{c_2}}{c_2\conj{c_2} - c_1 \conj{c_1}} \sum_{i=3}^N \left(\frac{c_i}{c_1}\expt{w_i} + \frac{\conj{c_i}}{\conj{c_2}}\expt{\conj{w_i}}\right).
\end{equation}
We call polynomials like $p$, where this procedure works, {\em reorientable}.

However, examining the denominator of the prefactor in \cref{eq:reorientation}, we see there remains a final problematic case; namely where $\expt{m_1} = \expt{m_2^*}$ but also $|c_1| = |c_2|$.
An example of a polynomial in this class is $p' = \frac{1}{2}\expt{x} + \frac{1}{2}\expt{x^*} - 1$.
Indeed, fully orienting this polynomial into a substitution rule for $\expt{x}$ {\em should} be impossible, 
 because this constraint does not in fact constrain the two real degrees of freedom in $\expt{x}$.
On the other hand, $p'$ is still meaningful as a constraint: here, explicitly mandating that $\re{\expt{x}} = 1$, while leaving $\im{\expt{x}}$ entirely unconstrained.
For this reason, we call $p'$ of this form {\em non-orientable}, and must handle such polynomials as special cases.

\inlineheading{Non-orientable polynomials and partial rules\footnote{
See method \code{MomentRule::resolve_nonorientable_rule} in \code{/lib_moment/symbolic/rules/moment_rule.cpp}.
}%
.}
To resolve the case of non-orientable polynomials, for all $\delta\in(-\pi,\pi]$, we can define a pair of functions $K_\delta: \comp \to \reals$ and $J_\delta: \comp \to \reals$:
\begin{align}
K_\delta(x) & := \frac{1}{2}\left( e^{-i\delta} x + e^{i\delta} \conj{x} \right) \\
J_\delta(x) & := \frac{1}{2}\left( -i e^{-i\delta} x + i e^{i\delta} \conj{x} \right)
\end{align}
such that
\begin{align}
e^{-i\delta} x = K_\delta(x) + i J_\delta(x).
\end{align}
These real-valued functions project onto a pair of orthogonal axes in the complex plane characterized by the angle $\delta$ (when $\delta=0$, $K_0(x) \equiv \re{x}$ projects onto the real axis, and $J_0(x) \equiv \im{x}$ onto the imaginary axis). (The real-valuedness may be directly confirmed.)

The general non-orientable polynomial $p$ has the form $p = k e^{ia} \expt{X} + k e^{ib} \expt{\conj{X}} + q$ for $k\in \reals \setminus 0$, $a, b \in \reals$, and polynomial $q <_{\rm sym} \expt{X}$.
To impose $p=0$ as a rule, we must identify the constrained axis of $\expt{X}$ in the complex plane.
Rearranging $p=0$:
\begin{align}
\frac{1}{2}\left(e^{i\frac{(a-b)}{2}} \expt{X} + e^{i\frac{(b-a)}{2}}\expt{\conj{X}}\right) = -e^{-i\left(\frac{a+b}{2}\right)} \frac{1}{2k} q,
\end{align}
which has the form
\begin{align}
\label{eq:KdeltaConstrain}
K_\delta\left(\expt{X}\right) = -e^{-i\left(\frac{a+b}{2}\right)} \frac{1}{2k} q,
\end{align}
with $\delta = \frac{1}{2}(a-b)$.

The LHS of \cref{eq:KdeltaConstrain} is real by construction, but the RHS might be complex.
Thus we must also consider the implied constraint
 \begin{align}
 \label{eq:splitrule}
     q' := \im{-e^{-i\left(\frac{a+b}{2}\right)} \frac{1}{2k} q} = 0.
 \end{align}
Indeed, if $q'$ is not manifestly identically zero, then $p$ has implied another non-trivial constraint that must be handled  when assembling the complete set of rules\footnote{Some rules like $\re{\expt{X}}=i$ will be obviously logically inconsistent, corresponding to a bad input polynomial. However, if $q$ is not a scalar, then this splitting can result in two meaningful rules; e.g.\ $\re{Y} = \expt{X}$ implies $\re{Y} = \re{\expt{X}}$, but also that $\im{X}=0$.}.
Splitting off such additional rules cannot result in an infinite loop, since $q' <_{\rm sym} \expt{X}$ by construction (and there is a lowest order polynomial, $0$).

It remains to handle the real part of \cref{eq:KdeltaConstrain},
 that is: $K_\delta\left(\expt{X}\right) = \re{-e^{-i\left(\frac{a+b}{2}\right)} \frac{1}{2k} q}$.
In this form, we can see explicitly how the projection of $\expt{X}$ onto one axis of the complex plane is constrained (while the orthogonal axis is free).
To implement this as a rewrite, we form the {\em partial rule}:
\begin{align}
\label{eq:nonorient}
\expt{X} & \mapsto i e^{i\delta} J_{\delta}(\expt{X}) + e^{i\delta} K_{\delta}(\expt{X}) \nonumber\\
& = \frac{1}{2}\left(\expt{X} - e^{-2i\delta} \expt{\conj{X}}\right) + e^{i\delta} \re{-e^{-i\left(\frac{a+b}{2}\right)} \frac{2}{k} q}.
\end{align}

Clearly, this is not orientated, as the RHS contains the terms $\expt{X}$ and $\expt{\conj{X}}$.
However, in this form, the rule now has the useful property of {\em idempotence} -- if we directly apply \cref{eq:nonorient} to its own RHS (matching $\expt{X}$ and $\expt{\conj{X}}$) and gather like terms, then the RHS is unchanged.
This means this partial rule can still be useful for reducing polynomial expressions.

Let us consider the motivating example constraint of the previous subsection, $p_2 = \frac{1}{2}\expt{x} + \frac{1}{2}\expt{x^*} - 1 = 0$.
In this subsection's notation, $a=b=0$, $k=\frac{1}{2}$ and $q=-1$.
Thus, $\delta=0$, and according to \cref{eq:nonorient} the associated rewrite rule is:
\begin{align}
\expt{x} \mapsto \frac{1}{2}\left(\expt{x}-\expt{\conj{x}}\right) + 1,
\end{align}
which could also be written as $\expt{x} \mapsto i\im{\expt{x}} + 1$.
Thus, this rule has the intuitively expected behaviour of substituting the real part of $\expt{x}$ with $+1$, while leaving the imaginary part as a free variable.
(In this example, trivially $\im{-1}=0$, so there is not a non-trivial second constraint of the form \cref{eq:splitrule}.)

\inlineheading{Reducing rules.}
Suppose we have two rules $r_1: \expt{z}\to\expt{y}$ and $r_2: \expt{y}\to\expt{x}$.
If we had a polynomial that contained a term in $\expt{z}$, we would have to iteratively apply rule $r_1$ and then rule $r_2$ to the substituted substring to get our desired final term in $\expt{x}$.
As alluded to in \cref{sec:eqconst}, this adds a scaling factor to the time taken to reduce any objects (such as moment matrices).
On the other hand, if our rules were $r_1': \expt{z} \mapsto \expt{x}$ and $r_2': \expt{y} \mapsto \expt{x}$, exactly the same equality constraints are imposed, but we have the advantage that we only have to match each term in the input polynomial against one rule in the ruleset, and once substitution is done, we do not have to further consider the replacement terms (except to order and group them at the end, to put the transformed polynomial into normalized form).

More extremely, suppose we had constraints $\expt{z} = \expt{y}$ and $\expt{z} = \expt{x}$ (with ordering $z>y>x$).
Na\"ive orientation would yield $\tilde{r}_1: \expt{z}\to\expt{y}$ and $\tilde{r}_2: \expt{z}\to\expt{x}$.
This has clear problems: it is ambiguous whether to match $\expt{z}$ against $\tilde{r}_1$ or $\tilde{r}_2$, and moreover, $\expt{y}$ is not matched by either expression, even though it should be replaced with $\expt{x}$.
Again, the above $r'_1: \expt{z}\to\expt{x}$ and $r'_2: \expt{y}\to\expt{x}$ avoids both these problems.

Thus, to ensure consistency in a set of rules, we must have a {\em reduced} set:
one where the RHS of every rule in the set is idempotent under application of every other rule in the set, and where each LHS is strictly unique (that is, unique, even if $\expt{x}$ and $\expt{\conj{x}}$ are also identified).
By this definition, we see the avoidance of ambiguity (and the computational advantage) can persist also for partial rules (of the form \cref{eq:nonorient} with reduced $q$).
It is thus possible to use essentially the same systematic logic for reduction by oriented, re-oriented, and partial rules.

\inlineheading{The algorithm for producing a reduced ruleset\footnote{
Corresponding to method \code{MomentRulebook::complete} in \code{/lib_moment/symbolic/rules/moment_rulebook.cpp}.
}%
.}
Let $\mathcal{P} := \{p_i\}_{i=1,\ldots, N}$ be a list of normalized polynomials, whose equation with zero defines the desired equality constraints.
To produce the reduced set of rules $\mathcal{R}$, we use the following algorithm:
\begin{enumerate}
    \item \label{step:sort} Sort $\mathcal{P}$, so that the first element is the lowest according to the ordering discussed above\footnote{This step is not essential for correctness, but at the cost of an $O(N\log N)$ sort, we can mostly avoid an $O(N^2)$ scaling arising from step \ref{step:fixrules}.}.
    \item Prepare an empty ruleset $\mathcal{R}$.
    \item \label{step:PolyLoop} If $\mathcal{P}$ is empty, exit the algorithm reporting success.
    \item Otherwise, pop (remove for consideration) the first polynomial $p$ from the front of list $\mathcal{P}$.
    \item Reduce $p$ by every rule in $\mathcal{R}$ to form polynomial $p'$.
    \item If $p'$ is zero, then $p$ is redundant, so it can be discarded. In this case, loop back to step \ref{step:PolyLoop}.
    \item Else if $p_i'$ is a non-zero scalar value (i.e. $k\expt{I}$ for $k\neq0$), then the set of input polynomials implies a logically inconsistent set of constraints. In this case, exit the algorithm reporting failure.
    \item Otherwise, consider the leading terms of $p_i'$:
        \begin{enumerate}
            \item If $p_i'$ has only one term, or if the indeterminate of the second term is not the complex conjugate of the first term, form the oriented rule $r$ according to \cref{eq:orient}.
            \item Else, if the indeterminate of the second term is the complex conjugate of that in the first term, but the coefficients $|c_1|\neq|c_2|$, then form the re-oriented rule $r$ according to \cref{eq:reorientation}
            \item Otherwise, the first term is the complex conjugate of the second term up to a complex phase. Form the partial rule $r$ according to \cref{eq:nonorient}.
            If the partial rule implies a second split rule (\cref{eq:splitrule}) that is not identically zero, insert the split polynomial $q'$ into the list $\mathcal{P}$ (ideally, in order).
        \end{enumerate}
    \item \label{step:fixrules} For each rule $r' \in \mathcal{R}$ whose LHS is ordered after the LHS of $r$, attempt to reduce the RHS of $r'$ by $r$ (replacing the RHS with the reduction, where it matches).
    \item Insert rule $r$ into the set $\mathcal{R}$, and loop back to step \ref{step:PolyLoop}.
\end{enumerate}

In the best-case scenario, no polynomial is non-orientable, and, due to the ordering in step \ref{step:sort}, each $r$ is essentially appended to $\mathcal{R}$ (without the need to test for reduction of any other rules in $\mathcal{R}$) and the algorithm's scaling is bottle-necked by the sort (i.e.\ $O(N\log N)$ for $N$ input polynomials).
In the theoretical worst-case scenario, every polynomial is non-orientable, and moreover always results in a split rule that has to be inserted at the front of $\mathcal{P}$, leading to a scaling of $O(N^2)$.
Thus far, we have found in our example uses that the favourable case is more typical.

This algorithm is used in \citet{AraujoGN24}, where \moment{} is employed in the generation of around $4\,500$ polynomial constraints. 
There, the runtime of the reduction algorithm was of the order of $7$ seconds.

%
%
\subsection{Monomial rewrite rules}
\label{app:MonoSubRule}
\inlineheading{Motivating example.}
Recall that a ruleset is convergent if repeated application of its members can reduce a word to its standard form in a finite number of steps.
Take as an example, the algebra of $a, b, c$ with the ruleset $\mathcal{R} := \{r_1 := ab \mapsto a, r_2 := bc \mapsto b\}$.
Suppose we wish to reduce the string $abc$.
If we start by using $r_1$, we transform $\underline{ab}c\mapsto ac$, to which there are no further matches.
Alternatively we can match with $r_2$ to transform $a\underline{bc}\mapsto ab$ and subsequently match with $r_1$ to give $\underline{ab}\to a$.
As such, the resultant ``simplified'' string ($ac$ or $a$) is dependent on the order in which the rules were applied, and as such $\mathcal{R}$ is said to be {\em non-convergent}\footnote{Another way a ruleset can be non-convergent is if it contains an infinite cycle of rules (e.g. $b\mapsto a$, $a\mapsto b$). However, this can be avoided by construction, by systematically imposing a consistent orientation (here, shortlex) and omitting trivial rules (such as $a\mapsto a$).%
}.
This is a problem, since we can see that clearly $ac$ and $a$ are the same (because they both are equal to $abc$), and in the context of generating SDPs, this will lead to incorrect answers arising from implicit relaxations of constraints (ultimately allowing for what should be the same moment to instead take different values where it has been `spelled' differently).

In our example, we could correct this by appending an additional rule $r_3: ac \mapsto a$ to form $\mathcal{R}' := \mathcal{R}\cup r_3$, such that $abc\mapsto a$ regardless of the order of rule application.
This fixes the problem for $abc$, but how can we be sure that there are no other strings on which $\mathcal{R}'$ will encounter a similar problem?

The solution employed in \moment{} is to attempt {\em completion} via the {\em Knuth-Bendix algorithm}~\cite{KnuthB83}.
Such an algorithm, when it runs to completion, determines whether a ruleset is convergent, or otherwise provide a convergent ruleset implementing the same algebraic relations.
Unfortunately, telling ahead of time whether a general ruleset can {\em ever} be completed by this algorithm is equivalent to the halting problem~\cite{Post47}, and hence uncomputable.
To mitigate this in \moment{}, one specifies a heuristic parameter on the number of attempted new rules one is willing to add to the set before it gives up (and issues a warning).
Our experience has been that completing a ruleset is a bit of dark art.
In particular, one quirk of the algorithm is that whether the set completes or not is affected by the ordering that determines the orientation of equalities into rules.
As such, if a ruleset does not seem to complete, one might try declaring the fundamental operators in a different order.
Indeed, our default choice of representing non-Hermitian operators by interleaving indices (i.e.\ $a$, $\conj{a}$, $b$, $\conj{b}$, \ldots instead of $a$, $b$, \ldots, $\conj{a}$, $\conj{b}$, \ldots) came about because we typically found this more likely to successfully complete when we had rulesets that impose unitary behaviour (e.g.\ $\conj{a} a \mapsto a\conj{a} \mapsto \id$) alongside commutation relations.

\inlineheading{The algorithm.}
Suppose we begin with an oriented ruleset $\mathcal{R}$.
To complete this set, we do the following:
\begin{enumerate}
\item Loop through each ordered pair of rules $r_i, r_j \in \mathcal{R}$ (with $i\neq j$). \label{step:pairloop}
\item Compare the LHS of $r_i$ for {\em string overlap} with the LHS of $r_j$: 
 that is, test if the suffix of $r_i$'s LHS has any non-zero overlap with the prefix of $r_j$'s LHS.
If there is no overlap, move to the next pair in the loop (step \ref{step:pairloop}).
\item {\bf Composition.} For each overlapping pair $r_i$, $r_j$, compose the shortest new string $w$ that begins with the LHS of $r_i$ and ends with the LHS of $r_j$.
Apply rule $r_i$ to the prefix of $w$ to make string $w_i$, and rule $r_j$ to the suffix of $w$ to make string $w_j$.
\item Repeatedly apply every rule in $\mathcal{R}$ to $w_i$ until the reduced string $w_i'$ is formed that matches no other rules.
Likewise do the same with $w_j$ to form $w_j'$. \label{step:simplify}
\item If $w_i'$ and $w_j'$ are identical, then the overlap is confluent, so move to the next pair in the iteration.
\item {\bf Orientation.} If $w_i'$ and $w_j'$ are non-identical,
 equate them, and then orient (via shortlex) to form a new rule $r': w_i' \mapsto w_j'$ or $r': w_j' \mapsto w_i'$.
\item {\bf Reduction.} Loop over every other rule $r_k\in\mathcal{R}$: \label{step:reduceloop}
\begin{enumerate}
    \item Attempt to match $r'$ to $r_k$'s RHS\footnote{The simplification in step \ref{step:simplify} ensures that $r'$ will never match $r_k$'s LHS.}, $w_k$; if there is no match, move to the next rule in the loop of \ref{step:reduceloop}.
    \item If there is a match, apply $r'$ to $w_k$ and then subsequently attempt to apply $r'$ and every rule in $\mathcal{R}$ repeatedly, until no further matches are found, forming the reduced string $w_k'$.
    \item If $w_k'$ is equal to the LHS of $r_k$, then delete rule $r_k$ from the set (it is now redundant) and move on to the next $r_k$ in the loop of step \ref{step:reduceloop}.
    Otherwise, replace $r_k$ with a new rule with the same LHS as $r_k$ but with $w_k'$ as its RHS.
\end{enumerate}
\item Append rule $r'$ to the set $\mathcal{R}$; and {\em restart} the loop over pairs (step \ref{step:pairloop}).  \label{step:restart}
\item If the loop completes, every pair of rules either does not overlap or is confluent, and the ruleset is hence convergent.
\end{enumerate}

In deviation from the Knuth-Bendix algorithm, to dodge the issue of never-ending loops, in \moment{} we amend step \ref{step:restart} as follows: Count how many restarts have been performed, and if this exceeds a predetermined threshold, exit with a failure warning; otherwise restart the loop.
Without this amendment, there is the possibility that the loop never terminates.

The source code for this implementing is in \code{operator_rule.cpp}/\code{.h} (determining how a single rule reduces an string) and \code{operator_rulebook.cpp}/\code{.h} (determining how an entire ruleset acts to reduces a string, and the Knuth-Bendix algorithm) in the folder \code{cpp/scenarios/algebraic/}.

As a debugging feature, \moment{} provides an option for logged completion. 
This is done by manually invoking the \code{Complete} method on an \code{AlgebraicScenario} 
 (or \code{OperatorRulebook} object) with the second argument set to \code{true}.
This produces a verbose log of the manipulations performed on the rules.
An simple example of this is given as \code{examples/complete.m}.

\section{Overview of \moment{} components}
\label{app:Components}

\subsection{MATLAB components}
\label{app:MATLAB}
We hope the guide earlier in this document is sufficient for using \moment{} in MATLAB, but provide here a little extra information for power users, and those who wish to make modifications.
We have tried our best to document the code, so that the builtin \code{doc} and \code{help} functions from MATLAB may be of some assistance. 

Essentially, most of \moment{}'s MATLAB code is syntactic sugar for a mex function \code{mtk}, which is implemented as the resultant compiled binary from the \cpp{} code.
If we have done our job correctly, there should be almost no need for an end-user to {\em directly} call the \code{mtk} function (except maybe to check it has compiled correctly).

\subsubsection{File structure}
The MATLAB interface of \moment{} has the following directory structure:
\begin{enumerate}
\item \code{/matlab/} -- The MATLAB root, containing the compiled binary \code{mtk} and the installation script \code{mtk_install.m}.
\item \code{/matlab/classes} -- The majority of MATLAB code, defining handles for the various objects needed to formulate SDPs using moment.
\item \code{/matlab/examples} -- Example scripts (including many of the worked examples in \cref{sec:scenario}).
\item \code{/matlab/functions} -- Stand-alone utility functions, that are not associated with any particular class.
\item \code{/matlab/tests} -- MATLAB unit test suites. 
These can be run individually through MATLAB's editor, or detected and executed as a whole using the \code{mtk_test} function.
\end{enumerate}

Within the \code{/matlab/classes} folder, folders beginning with \code{@} are classes, whose methods have been split across multiple files.
Meanwhile, folders beginning with \code{+} contain one or more minor classes associated with particular scenarios.

\subsubsection{Class hierarchy}
\begin{figure}[tbh]
\begin{centering}
\includegraphics[width=0.85\textwidth]{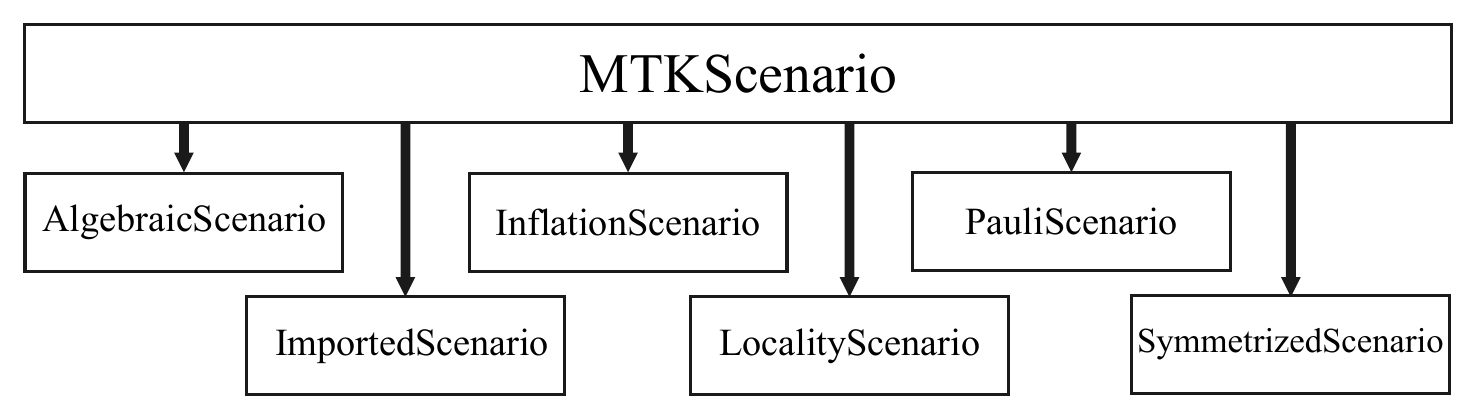}
\caption{%
\label{fig:MTKScenario}
\caphead{Inheritance diagram for \code{MTKScenario}.}
}
\end{centering}
\end{figure}

\begin{figure}[tbh]
\begin{centering}
\includegraphics[width=0.825\textwidth]{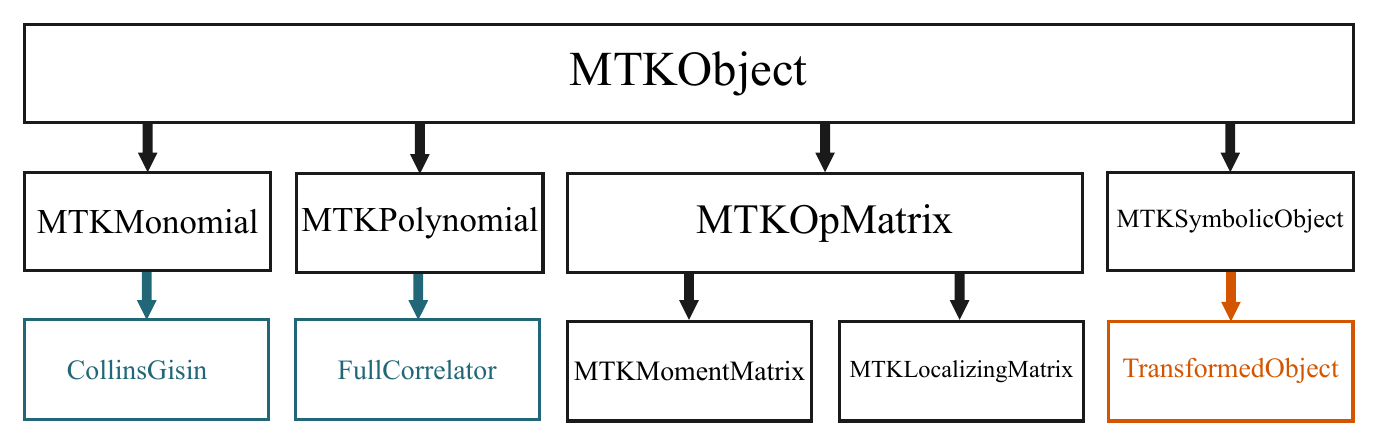}
\caption{%
\label{fig:MTKObject}
\caphead{Inheritance diagram for \code{MTKObject}.}
\code{MTKObject} and its derived classes are the building blocks for expressing SDPs with \moment{}.
Blue classes are only in the locality and inflation scenarios, and the orange class only in the symmetrized scenario.
}
\end{centering}
\end{figure}

The MATLAB interface of \moment{} predominantly follows an object-oriented paradigm (in so far as is supported by MATLAB).
Particularly, there are two important class hierarchies: \code{MTKScenario} and its descendants (\cref{fig:MTKScenario}) and \code{MTKObject} and its descendants  (\cref{fig:MTKObject}).
Most classes, especially when they would otherwise have very generic names such as \code{Polynomial} or \code{Object} have had their names prefaced with \code{MTK} (standing for \moment{} tool kit).
We adopt this convention because MATLAB has almost no support for namespaces.

The class \code{MTKScenario} handles scenarios, as introduced in \cref{sec:miniscenario} and elaborated upon in  \cref{sec:scenario}. 
Essentially, each instance of a class derived from \code{MTKScenario} defines a set of operators, a table of known moments, and a list of generated matrices; as well as methods to create certain specific polynomials or monomials in some of the scenarios.
Most of the large data involved in \moment{} is not stored as a variable within MATLAB, but rather is handled through the \cpp{} side.
As such, these classes are thin-wrappers and forwarders to functions that are implemented in \cpplib{}, via the compiled binary \mtk{}.
Reference counting is employed such that when the last reference to a particular scenario is removed in MATLAB, the corresponding memory in \cpp{} is also freed.

The class \code{MTKObject} is the base class of any object that might be associated with a scenario, and at some point in its life be interpreted as (a collection of) moments (be they scalar, vector, matrix, or general tensor).
This includes monomial and polynomial objects, as might appear in objectives and constraints.
An important subclass of this is \code{MTKOpMatrix} (standing for operator matrix), which includes in particular \code{MTKMomentMatrix} and \code{MTKLocalizingMatrix} associated with the obvious respective objects.

For matrices, \moment{} avoids storing information in MATLAB--managed memory directly. 
As such, these classes are somewhat thin handles to the respective data as stored and allocated via the \cpp{} code.
Some properties, like \code{.SequenceStrings} and \code{.SymbolStrings} are generated in a deferred manner (i.e.\ when they are first requested). This is because  forming a human-readable string representation of a moment matrix takes considerably more computational time than just generating the matrix for use in an SDP,
 and usually this information is useful only when debugging an SDP, and not when running in production.

\subsubsection{Unloading moment.}
It is possible to manually unload \moment{} completely, and free all objects from memory, without having to restart MATLAB. 
To do this, execute:
\begin{code:matlab}
clear
clear mtk
\end{code:matlab}
The first command will free {\em all} variables from the workspace (so be sure to export any required data first!)
This process also allows one to edit/replace the compiled \code{mtk} executable file, releasing any locks that MATLAB places on the file -- and so is an essential step if one wants to make changes to \moment{}'s binary while MATLAB is running.

\subsection{Overview of \cpp{} components}
The \cpp{} code defines several compilation targets: the core library \cpplib{}, the \mex{} function \mtk{}, a suite of \cpp{} unit tests, and several standalone stress tests.

The source is written to be compiled and linked according to the \texttt{c++20} standard, which is supported by most modern compilers (we tested with \texttt{clang 15.0.1}, \texttt{gcc 12.3}, and \texttt{msvc 19.36.32537}).
The \code{CMakeLists.txt} files are tested to work with \cmake{} version \texttt{3.22}.

Two dependencies are included as submodules of the git repository: \eigen{} and \gtest{}.
\eigen{} is necessary to compile \cpplib{} depends on this.
\gtest{} should be included in order to compile the \cpp{} unit tests.
To compile the \mex{} function \mtk{}, \matlab{} version 9.4 or later (2018a) must be locatable by \cmake{} via the `\code{Matlab}' \cmake{} package.

It is possible to compile just \cpplib{} completely independently of any \matlab{} / \mex{} code.
This could be done either as the starting point of an interface for \moment{} in another language (i.e.\ by writing an analogue of \mtk{} and/or the \matlab{} library),
 or directly embedding \moment{} features for use in another application (e.g.\ to solve some compiled computation, and directly translating objects from \moment{} into a format that can be passed into an SDP solver).

\subsubsection{The \cpp{} library \texttt{lib\_moment}}
The subfolder \code{/cpp/lib_moment} defines the build target \cpplib{} resulting in production of a library of the same name.
The majority of the functional, scientific code of \moment{} is found within this library.
Although a full description of this is way beyond this document, we briefly outline its file structure:
\begin{itemize}
\item \code{/dictionary} -- relates to generating dictionary objects. Of particular note here is the class \code{OperatorSequence}, which handles the storage of (canonically expressed) operator words.
\item \code{/matrix} -- relates to moment matrices, and how they can be converted to bases.
\item \code{/matrix/operator_matrix} -- code related specifically to the generation of matrices from dictionaries of operator sequences.
\item \code{/matrix_system} -- relates to the storage (and indexing) of moment matrices. Essentially, each instance of \code{MatrixSystem} is the underlying object tracked by each instance of \code{MTKScenario} in the MATLAB code.
\item \code{/multithreading} -- utility functions relating to multi-threading support.
\item \code{/probability} -- specific functions shared between locality and inflation scenarios, for converting between Collins-Gisin notation, full correlator tensors, and measurement outcome probabilities.
\item \code{/scenarios} -- each subfolder of this contains methods specific to a particular scenario.  The base class \code{Context} here (descended from in each subfolder) contains the methods for simplifying strings of operators into their canonical form.
\item \code{/symbolic} -- handles objects that are expressed as linear combinations of moments. Also contains the \code{SymbolTable} class, which is responsible for assigning a unique ID to each moment encountered within a scenario, and tracking properties such as whether it has imaginary components.
The \code{Monomial} and \code{Polynomial} classes in here are somewhat akin to their respective \code{MTKMonomial} and \code{MTKPolynomial} MATLAB objects.
\item \code{/symbolic/rules} -- relates to the substitution rules described in \cref{app:PolySubRule}.
\item \code{/tensor} -- contains templates for generic objects containing sub-objects that might be accessed either by a multi-dimensional index or a singular offset.
\item \code{/utilities} -- defines general purpose functions and classes, not unique to \moment{}.
\end{itemize}

Essentially, one can think of \code{MatrixSystem} as \moment{}'s main container class, that is associated with a single \code{Context} defining the relationship between operators, and a single \code{SymbolTable} enumerating the moments encountered so far within that system.
Together, these form what we have conceptualized as a scenario in this document, and what is wrapped by the MATLAB object \code{MTKScenario}.
Typically, the \cpp{} classes \code{MatrixSystem} and \code{Context} will be overloaded (in the subfolders of \code{/scenarios}), so as to inject the unique behaviour required for each unique scenario.
This follows a predictable naming structure (e.g. \code{LocalityContext} for the \code{Context} associated with the locality scenario).

\subsubsection{The \mex{} function \mtk{}}
The subfolder \code{/cpp/mex_functions} defines the build target \code{moment_mex} resulting in production of the \mtk{} binary.

\mex{} stands for {\em MATLAB executable},
 and is the name shared by two distinct architectures produced by Mathworks (with great detriment to the quality of search engine results): the first being for C, the second for \cpp{}.
\moment{} uses the \cpp{} architecture.
The required headers for this interface are included within a \matlab{} installation in the \code{extern} folder.
However, it is not necessary to manually add this folder to the include path,
 instead prefering to use the \cmake{} package \code{Matlab} which provides the \code{matlab_add_mex} command for defining a compile target that generates a \mex{} function.
Our experience is that this has been surprisingly robust, generating functioning binaries with \texttt{clang 15.0.1}, \texttt{gcc 12.3} and \texttt{msvc 19.36.32537} on Windows, and  \texttt{gcc 12.3} in linux.
We warn, however, that MathWorks officially supports a much more limited number of older compilers~\cite{MATLABCompilers}.

Most of \mtk{} is defined within the namespaces \code{Moment::mex} and \code{Moment::mex::functions}.

\inlineheading{mex\_entry.cpp---}%
At its heart, a \mex{} function is a dynamically linked library, that is loaded the first time it is used within a MATLAB script.
The entry point is defined by a class \code{::MexFunction} that publically inherits from \code{matlab::mex::Function}.
In \mtk{}, this is implemented as \code{mex_entry.cpp}.
This class defines three methods:
\begin{enumerate}
\item \code{MexFunction()} \emph{(the constructor)} -- called the {\em first} time \mtk{} is executed from \matlab{}.
\item \code{~MexFunction()} \emph{(the destructor)} -- called when \matlab{} closes, or when \code{clear mtk} is executed.
\item \code{operator()} -- the main function call, supplying the arguments passed to the particular invocation of \mtk{}, and handles to the output arrays that \matlab{} expects the function to populate.
\end{enumerate}
For a simple \mex{} program the last function could execute the entirety of the desired functionality.
However, \moment{} is a complex library of multiple functions, with storage managed by \cpp{} (cf.\ storing as an array in \matlab{}) that needs to persist between subsequent calls.
As such, we implement effectively a switch-board from this one \mex{} entry point to the many different \mtk{} functions in the library.
Here, \code{operator()} effectively just forwards these arguments to an instance of the \code{MexMain} class.

\inlineheading{mex\_main.cpp/.h---}%
The \code{MexMain} class can be thought of as the aforementioned switch-board operator.
It handles the overall flow of a function call, and (after querying for environmental variables in the constructor) is entered through its \code{operator()} function.
The flow of handling a function call is as follows:
\begin{enumerate}
\item Identify from the first argument to \mtk{} which \mtk{} function (see \cref{app:mtk-functions}) has been requested -- or yield an exception (via \matlab{}'s error reporting system) if the function name is unrecognized.
\item Construct the relevent specialization to \code{MTKFunction} identified in step 1.
\item Check if the number of remaining inputs is within the range expected by the specific \mtk{} function, as well as identifying and trimming any named parameters or named flags from the inputs. These cleaned inputs are provided as a \code{SortedInputs} structure.
\item Check if the number of outputs requested is within the range expected by the specific \mtk{} function.
\item Provide the cleaned inputs, and output handles to the \code{MTKFunction}'s call operator, to perform the requested computation.
\item Do any logging, if enabled (if an exception is thrown in the previous steps, and logging is enabled, this failure is also reported).
\end{enumerate}
 
\inlineheading{function\_list.cpp/.h---}%
This file contains integer tokens \code{MEXEntryPointID} that uniquely enumerate the different \mtk{} functions supported.
The function \code{make_str_to_entrypoint_map} in the anonymous namespace of \code{function_list.cpp} defines the string identifier that is invoked as the first argument to \mtk{} in \matlab{}.
The function \code{make_mtk_function} constructs, switching on a supplied \code{MTKEntryPointID}, the relevent specialization of \code{MTKFunction}.

\inlineheading{mtk\_function.cpp/.h---}%
This class defines the abstract interface \code{MTKFunction}.
Particularly, the constructors of the classes derived from \code{MTKFunction} should specify the minimum and maximum number of expected inputs and outputs, as well as the names of any named parameters and flags.
It can also specify what sets of flags (or parameters) are mutually exclusive (i.e.\ cannot be simultaneously supplied by the user without error).
The function \code{operator()} then should be overloaded by children of \code{MTKFunction} to do the actual computation on the inputs, and populate the outputs.

In practice, almost every \mtk{} function uses the template \code{ParameterizedMTKFunction} (defined in the same files).
This template effectively mixes in an intermediate step between the cursary input handling done in \code{MTKMain} and the actual computation, by allowing (as a template argument) the specification of a class derived from \code{SortedInputs}.
Then, when the function is executed, the \code{SortedInputs} is first parsed into the derivative input class (with the possibility of raising an error if the inputs are invalid), which allows for the parsing of abstract MATLAB arrays into standardized \cpp{} data types (e.g.\ the first input to \code{moment_matrix} is the ID of the matrix system the matrix is to be created for. \code{SortedInputs} would have this as a \code{matlab::data::Array} object, but before we can act upon it, we would want this object as a single unsigned integer.).
This parsed input is then passed to an overloaded \code{operator()} function, to do the computation.

\inlineheading{storage\_manager.h and environmental\_variables.cpp/.h---}%
This \code{StorageManager} class manages the common storage shared between calls to \mtk{} in a thread-safe manner.
Currently this consists of the systems of matrices linked to scenarios,
 the logger (if active), and the environmental variables.
This latter basically manages a few settings that are applied across every function call (e.g.\ turning on/off multithreading during matrix generation).

\inlineheading{Folder: \code{functions}---}%
Every pair of \code{.cpp} and \code{.h} files in this folder (and its subfolders) defines one \mtk{} function by providing an implementatoin of \code{MTKFunction}.
Most also provide a specialization of \code{SortedInputs} corresponding to the inputs expected by that particular function.
See \cref{app:mtk-functions} for a list of these functions.

\inlineheading{Folder: \code{import}---}%
This folder contains functions and classes for parsing \matlab{} data objects (typically \code{matlab::data::Array}) into classes defined in \cpplib{}.

\inlineheading{Folder: \code{export}---}%
This folder contains functions and classes for exporting classes defined in \cpplib{} into MATLAB data objects (i.e.\ derivates of \code{matlab::data::Array}).

\inlineheading{Folder: \code{eigen}---}%
This folder contains functions for converting numerical \matlab{} data objects (i.e.\ \code{matlab::data::Array} or derivates thereof)
 into \eigen{} library matrices; and vice versa.
Unfortunately (but understandably), the raw storage of sparse matrices is not exposed by \matlab{}, so many of these functions involve a copying data.
Thankfully, this does not need to be done in a tight loop, and in practice is a tiny fraction of the runtime of \moment{} (and an even smaller fraction of the runtime of producing and solving an SDP).

\inlineheading{Folder: \code{utilities}---}%
This folder contains general boilerplate code for interfacing between \matlab{} and \cpp{} -- in particular, the import and export of standard \cpp{} types from \matlab{} data types -- as well as few wrappers to \matlab{} engine functions (primarily, for writing output to console, and reporting warnings and  errors\footnote{In principle \matlab{} might catch exceptions directly thrown by \cpp{} code, but the faith this would require in the ABI's stability is beyond most \cpp{} developers.}).

\section{Hints on extending moment}
\moment{} aims to be useful across a wide-range of SDP applications,
 but we cannot anticipate every single desired feature ahead of time.
However, we have designed \moment{} in a modular manner, to facilitate the addition of new features.
The following section provide various references for extension.
Programming such extensions will likely require a high degree of familiarity with \cpp{}, \matlab{}, or both.

\subsection{Adding a new function to \mtk{}}

Suppose we want to add a new function with the name `\code{your_function}' to \mtk{}.
We will do this via \code{ParameterizedMTKFunction} (for an implementation without this, see \code{mex_functions/functions} \code{/debug/version.cpp} and its associated header).

In practice, one should use an existing function as a starting reference (e.g.\ \code{make_representation} is a fairly straight-forward example).
The steps to take are thus:
\begin{enumerate}
\item To the folder \code{mex_functions/functions/} (or a subfolder thereof), add a pair of files \code{your_function.cpp} and \code{your_function.h}.
\item Add \code{functions/your_function.cpp} to \code{mex_functions/CMakeLists.txt} within the \\\code{matlab_add_mex} function.
\item In the header file \code{your_function.h} define a class \code{YourFunctionParams} in the namespace \code{Moment::mex::functions} inheritting publicly from \code{SortedInputs}.
This class must define the following constructor:
\begin{code:cpp}
explicit YourFunctionParams(SortedInputs&& raw_inputs);
\end{code:cpp}
\item In the source file \code{your_function.cpp} this constructor should be implemented. 
The call to the parent construct must forward \code{raw_inputs}:
\begin{code:cpp}
YourFunctionParams::YourFunctionParams(SortedInputs&& raw_inputs)
 : SortedInputs(std::move(raw_inputs)) {
    /* your code */
}
\end{code:cpp}
The code you add to this function should effectively parse the member \code{this->inputs} (which is defacto an array of \code{matlab::data::Array})
 into the various member variables defined in \code{YourFunctionParams}.
\item In the same file provide a class definition for \code{YourFunction}.
This class will publicly inherit from \code{MTKFunction} via a \code{ParameterizedMTKFunction} specialization:
\begin{code:cpp}
class YourFunction
  : public ParameterizedMTKFunction<YourFunctionParams, 
                                    MTKEntryPointID::YourFunction>
\end{code:cpp}
This class should provide a constructor,
 and must declare the call operator:
\begin{code:cpp}
void operator()(IOArgumentRange output, 
                YourFunctionParams &input) override;
\end{code:cpp}
\item Define the constructor and call operator in \code{your_function.cpp}.
The constructor is an opportunity to set the minimum and maximum number of inputs/outputs handled, as well as to register any named parameters or flags.
The call operator is the actual functionality of your function; and should ultimately write objects to the \code{output} parameter supplied.
\item To \code{/mex_function/function_list.h} add \code{YourFunction,} within the enum \code{MTKEntryPointID} alphabetically.
\item In \code{/mex_function/function_list.cpp} to \code{make_str_to_entrypoint_map} at the appropriate alphabetical point, add the line:
\begin{code:cpp}
output.emplace("your_function", MTKEntryPointID::YourFunction);
\end{code:cpp}
\item In \code{/mex_function/function_list.cpp} to \code{make_mtk_function} in the switch, add the following case at the appropriate alphabetical point:
\begin{code:cpp}
case functions::MTKEntryPointID::YourFunction:
  the_function = 
     std::make_unique<functions::YourFunction>(engine, storageManager);
  break;
\end{code:cpp}
\end{enumerate}

\subsection{Adding another modeller for use with \moment{} in \matlab{}}
We chose to initially support \cvx{} and \yalmip{} as, at the time of writing, they are the dominant SDP modelling toolkits for \matlab{}.
Should \moment{} support be requried for another (possibly new) modeller, this can be achieved solely via modifications to the \matlab{} code (i.e.\ without explicit need to modify \cpplib{} and \mtk{}).
In practice, one should take our yalmip implementation as a starting point.

To add support for a new a modeller, take the following steps:
\begin{enumerate}
\item Add a new section to \code{MTKMatrixSystem.m} labelled by the name of the new modelling toolkit. Within this section, add a new appropriately labelled \code{methods} block and within this define a function \texttt{\underline{modeller}Vars} (where `modeller' is replaced with the name of the new toolkit).
This function should, ideally\footnote{CVX does not do this; but CVX eschews the conventions of functional programming in order to provide a ``natural'' syntax for expressing SDPs in matlab code. As a developer, this was exactly as fun to work with as it sounds.} detect the number of outputs, such the first output are the SDP variables associated with the real basis elements, and the second (if provided) as the variables associated with imaginary basis elements.
The expected number of such elements may be found from (integer) member properties \code{RealVarCount} and \code{ImaginaryVarCount}.
\item To folder \code{@MTKScenario} add a new file \texttt{\underline{modeller}Vars.m} to define a function in \code{MTKScenario} with the same name as the function in \code{MTKMatrixSystem.m}.
This function should effectively forward to the function in the matrix system class. The simplest implementation would be:
\begin{code:matlab}
[varargout{1:nargout}] = obj.System.modellerVars();
\end{code:matlab}
\item If the SDP variables are \matlab{} classes that define meaningful overloads for \code{plus} (addition) and \code{mtimes} (matrix multiplication) and \code{reshape},
 then the \code{Apply} method can be used in the same way as currently implemented for cvx and yalmip.
Otherwise, to the folder \code{@MTKObject} add a new file \texttt{Apply\underline{Modeller}.m}, defining a method of the same name.
This function should accept as its inputs the objects produced by the \code{modellerVars} method (with second, imaginary, input optional),
 and as an output generate an object corresponding to the contraction of the supplied SDP variables with the coefficients associated with the object:
\begin{code:matlab}
expr = obj.ApplyModeller(real_basis, im_basis);
\end{code:matlab}
\end{enumerate}

\subsection{Write a completely different mex function for another purpose}
A large portion of the \mtk{} source code is boiler-plate for interfacing \cpp{} with \matlab{} via the \mex{} framework.
As open-source, free software, you are of course free (under the conditions of the GNU General Public License version 3.0) to use any or all of this code as the starting point for your own \mex{} functions.
This would surely be overkill for writing a single function (which, hypothetically, can be done as a single \cpp{} file) -- but may prove useful for writing a complicated suite, like \moment{}.

Files of particular general usefulness may found within the \code{cpp/mex_functions/eigen} and \code{cpp/mex_functions/utilities} folders,
 as well as the general architecture files  \code{mex_entry.cpp}, \code{mex_function.cpp}, and \code{mex_main.cpp}  in \code{cpp/mex_functions/}
 and some of the general-purpose utility classes in \code{cpp/lib_moment/utilities}.

\end{document}